\documentclass{article}
\usepackage[utf8]{inputenc}
\usepackage{graphicx}
\usepackage[portrait, margin=0.8in]{geometry}
\usepackage{hyperref}
\usepackage{authblk}
\usepackage{todonotes}
\usepackage{amsmath,amsthm,amssymb}
\usepackage{braket}
\usepackage{enumitem}
\usepackage{tabularx,booktabs,multirow}
\usepackage[sorting = none, backend = bibtex, style=numeric-comp]{biblatex}

\title{Scaling Whole-Chip QAOA for Higher-Order Ising Spin Glass Models on Heavy-Hex Graphs}

\author[1]{Elijah Pelofske\thanks{Email: epelofske@lanl.gov}}
\author[1]{Andreas Bärtschi}
\author[1]{Lukasz Cincio}
\author[1]{John Golden}
\author[1]{Stephan Eidenbenz}

\affil[1]{Los Alamos National Laboratory}

\date{\vspace{-6ex}}

\begin{document}
\maketitle

\begin{abstract}
We show through numerical simulation that the Quantum Approximate Optimization Algorithm (QAOA) for higher-order, random-coefficient, heavy-hex compatible spin glass Ising models has strong parameter concentration across problem sizes from $16$ up to $127$ qubits for $p=1$ up to $p=5$, which allows for computationally efficient parameter transfer of QAOA angles on instance sizes where exhaustive grid-search is prohibitive even for $p>1$. We use Matrix Product State (MPS) simulation at different bond dimensions to obtain confidence in these results, and we obtain the optimal solutions to these combinatorial optimization problems using CPLEX. In order to assess the ability of current noisy quantum hardware to exploit such parameter concentration, we execute short-depth QAOA circuits (with a CNOT depth of 6 per $p$, resulting in circuits which contain $1420$ two qubit gates for $127$ qubit $p=5$ QAOA) on ensembles of $100$ higher-order (cubic term) Ising models on noisy IBM quantum superconducting processors with $16, 27, 127$ qubits using QAOA angles learned from a single $16$-qubit instance using the JuliQAOA tool. We show that (i) the best quantum processors generally find lower energy solutions up to $p=3$ for 27 qubit systems and up to $p=2$ for 127 qubit systems and are overcome by noise at higher values of $p$, (ii) the best quantum processors find mean energies that are about a factor of two off from the noise-free numerical simulation results. Additional insights from our experiments are that large performance differences exist among different quantum processors even of the same generation and that dynamical decoupling significantly improves performance for some, but decreases performance for other quantum processors. Lastly we compute $p=1$ QAOA angle mean energy landscapes computed using up to a $414$ qubit IBM quantum computer, showing that the mean QAOA energy landscapes remain very similar as the problem size changes. 

\end{abstract}

\section{Introduction}
\label{section:introduction}

The Quantum Alternating Operator Ansatz (QAOA)~\cite{hadfield2019qaoa}, and the predecessor Quantum Approximate Optimization Algorithm~\cite{farhi2014qaoa, farhi2014qaoabounded}, is a quantum algorithm that is intended to be a heuristic solver of combinatorial optimization problems. QAOA is typically considered to be a variational hybrid quantum-classical algorithm because there is a set of parameters (usually called angles) that must be tuned in order for QAOA to perform well - and typically the standard tuning approach is to use a classical processor to perform iterative gradient descent learning on the QAOA angles, using the quantum computer to evaluate the expectation value of the algorithm at a different angles. The motivation for this approach, typically, is that because quantum computers are very difficult to engineer to have low error rate gate operations, current technologies have fairly high error rates - but by using variational algorithms, part of the computation can be off-loaded onto the classical part of the computation. Unfortunately, the task of learning good QAOA angles (and learning variational parameters for hybrid quantum-classical algorithms in general), is computationally hard and only made harder by the presence of noise in the quantum computation~\cite{bittel2021training, wang2021noiseinduced}. For these reasons, the suitability of QAOA for Noisy Intermediate-Scale Quantum (NISQ)~\cite{preskill2018quantum} computers is unclear, and is being actively studied using a variety of different approaches~\cite{shaydulin2023qaoa, pelofske2023high, harrigan2021quantum, sack2023largescale, weidenfeller2022scaling, shaydulin2023evidence, lotshaw2022scaling}. 

The Quantum Alternating Operator Ansatz consists of the following components:
an initial state $\ket{\psi}$,
a \texttt{phase separating} cost Hamiltonian $H_C$,
a \texttt{mixing} Hamiltonian $H_M$ (here the standard transverse field mixer $H_M = \sum_{i=1}^N \sigma^x_i$),
a number of \texttt{rounds} $p\geq 1$ to apply $H_C$ and $H_M$ (also referred to as the number of \emph{layers}), and
two real vectors of \texttt{angles} $\vec{\gamma} = (\gamma_1,...,\gamma_p)$ and $\vec{\beta} = (\beta_1,...,\beta_p)$, each with length $p$. Note that because we use the standard initial state, mixer, and phase separator, this algorithm is the original Quantum Approximate Optimization Algorithm -- in particular we do not use more complex mixers.

There exist a large number of QAOA variants because there are a variety of choices of initial states, phase separating cost Hamiltonians (for many different combinatorial optimization problems), mixer Hamiltonians, and tuning methods for the QAOA angles~\cite{he2023alignment, baertschi2020grover, golden2021threshold, magann2022feedbackbased, bravyi2020obstacles, wurtz2022counterdiabaticity}. The central question of all QAOA variants is how will QAOA scale in terms of obtaining optimal solutions of combinatorial optimization problems as the number of variables increases. This question has different components, including the angle finding problem, how many $p$ rounds need to be applied in order to be competitive with existing classical methods, and how the algorithm performs as problem size increases. It is known that in general, reasonably high $p$ (e.g. more than $p=1$ or $p=2$) will need to be applied in order for QAOA to be perform well at solving combinatorial optimization problems~\cite{farhi2020quantum1, farhi2020quantum2, farhi2022qaoa, basso2022quantum, golden2023numerical}. 
For this reason, the task of increasing $p$ on larger problem sizes is of particular interest, and this is the primary question that is studied in this paper, using state of the art quantum computing hardware. 

Using the whole-chip heavy-hex tailored QAOA circuits that are targeting hardware-compatible Ising models proposed in Refs.~\cite{pelofske2023qavsqaoa,pelofske2023short}, we investigate the task of scaling these QAOA circuits to higher rounds, and to larger heavy-hex chip quantum processors. Notably, this class of random spin glasses contain higher order terms, which increases the problem difficulty, and can be natively addressed by QAOA. The primary challenge with implementing these extremely large QAOA circuits up to higher $p$ is the angle finding task. Refs.~\cite{pelofske2023qavsqaoa,pelofske2023short}, utilized the brute-force approach of full angle gridsearches on the quantum hardware in order to compute good angles for $p=1$ and $p=2$. Unfortunately, this approach scales exponentially with $p$ if the grid resolution is held constant and in practice on-device angle gridsearch learning with $p=3$ is already computationally prohibitive. Refs.~\cite{pelofske2023qavsqaoa,pelofske2023short} observed that problem instances of the same sizes, but with different random coefficient choices, had nearly-identical low-round QAOA energy landscapes.
Our approach in this study to overcoming these angle finding challenges is to make use of \emph{parameter concentration} in QAOA angles in order to transfer high-quality fixed angles from small ($16$ qubit instances) to larger instances. Parameter concentration has been observed analytically and numerically for a number of different QAOA problem types~\cite{brandao2018fixed, farhi2022qaoa, wurtz2021fixedangle, akshay2021parameter, galda2021transferability, lee2021parameters, shaydulin2023evidence, shaydulin2023parameter, galda2023similaritybased, pelofske2023qavsqaoa, pelofske2023short}. We show that training on only a single problem instance provides good angles that can be used for much larger instances. This makes the computation of these parameters very efficient, but previous studies have also used a more computationally intensive approach of training on ensembles of problem instances to obtain good average case parameters. 

The goal of this study is to investigate the \emph{ideal} scaling of QAOA on current quantum computing hardware (with respect to increasing $p$ and increasing the number of variables), using the largest problem sizes that can be feasibly programmed on the hardware. 
This study uses two critical components: 

\begin{enumerate}[noitemsep]
    \item The angle-finding procedure is not performed in a variational outer loop classical optimization procedure, but we rather rely on heuristically computed good QAOA angles found on smaller problem instances and then apply parameter transfer. The angle-finding technique with quantum hardware in the inner loop has been studied before on NISQ hardware~\cite{weidenfeller2022scaling, sack2023largescale}, but there are a number of limitations with making this technique feasible -- including the computational overhead of the angle learning due to challenges such as local minima, and the noise in the computation making the learning task more difficult. Ideally, good QAOA angles would be able to be computed off-chip (classically), and then be used on large-scale quantum hardware. This is what the parameter transfer has enabled us to do for qubit system sizes that cannot be addressed using brute-force computation. 
   \item Because of the relatively high error rates on the current quantum computers, implementing optimization problems whose structure matches the underlying hardware graph reduces the overhead of gate-depth and gate-count. In particular, on quantum processors that have a sparse hardware graph, implementing long range interactions can be quite costly in terms of SWAP gates. Therefore, defining the combinatorial optimization problems that we sample to be compatible with the IBM Quantum processor heavy-hex graph~\cite{pelofske2023qavsqaoa,pelofske2023short} allows the QAOA circuits to be extremely short depth. 
\end{enumerate}

We briefly describe our methods and approach in Section~\ref{section:methods} by giving a description of the higher-order Ising (minimization) optimization problems (Subsection~\ref{section:methods_Higher_order_Ising_models}), the QAOA circuits to sample these optimization problems (Subsection~\ref{section:methods_QAOA_circuits}); we describe the angle finding and parameter transfer methods (Subsection~\ref{section:methods_QAOA_angle_finding}), and give a brief description of the Matrix Product State (MPS) simulation methods (Subsection~\ref{section:methods_MPS_simulations}) and the use of CPLEX to classically find the optimal solutions to the optimization problems (Subsection~\ref{section:methods_Higher_order_Ising_models}). Lastly, the implementation details on IBM quantum computers are given in Subsection~\ref{section:methods_hardware_implementation}.

In Section~\ref{section:results}, the first set of results shows (Subsection~\ref{section:results_transfer_learning}) parameter transfer works very well for these classes of problems up to $p=5$ in a noise-free environment as we show through numerical simulation for up-to $127$ qubits, when trained only on a single 16-qubit instance. In particular, mean expectation values improve consistently with increasing $p$ for all of 100 randomly chosen problem instances at 16, 27, and 127 qubits. 
These results are enabled by classical simulation techniques. As these problem classes grow entanglement relatively slowly grows with increasing $p$, MPS simulations enable us to classically produce the solution distributions that QAOA would achieve on an error-corrected quantum computer for up to $p=5$ and $127$ qubits. Our confidence in the accuracy of these simulations is due to the convergence of the solution values as we increase the MPS bond dimension parameter.
Having established that parameter transfer works in a noise-free computation, we then examine to what extent parameter transfer works on actual NISQ computers. 

In a second set of results (Subsection~\ref{section:results_scaling_p_16_27_127_qubits}), we execute the $100$ problem instances (for each qubit count) on cloud-accessed IBM quantum processors with $16$, $27$, and $127$ qubits using the numerically obtained fixed angles from a single $16$ qubit instance. These results are some of the largest quantum hardware experimental QAOA results reported to date, and include an evaluation of the effectiveness of a relatively simple dynamical decoupling scheme for circuits that make use of the entire NISQ processor. We find the following: 

\begin{enumerate}[noitemsep]
    \item Performance varies significantly among different processors even if they are from the same hardware generation. 
    \item The digital dynamical decoupling sequences we evaluated (pairs of Pauli X gates) improved the performance of three out of four 127 qubit devices, two out of six 27-qubit devices, and the single 16 qubit device. 
    \item Averaged over 100 instances, the best 127 qubit processors improve until $p=2$ and start degrading at higher values of $p$. For 27 qubits, the best processors improve up to $p=3$. Thus, noise appears to effect the higher qubit count devices slightly more than lower qubit count devices despite equal CNOT depth at the same~$p$. 
\end{enumerate}

Overall, our second set of results shows that QAOA parameter transfer works for this class of hardware-compatible optimization problems on current NISQ superconducting qubit processors albeit we can only verify up to $p=3$ as the devices succumb to noise at larger $p$. We thus revisit the question of parameter transferability on quantum hardware in a more systematic fashion in a third set of results limited to $p=1$. We find that parameter concentration remains stable for $p=1$ energy landscapes, run on actual quantum hardware. We show mean energy QAOA angle landscapes for the two parameters at $p=1$ for four different 27 qubit and one 414 qubit systems (Subsection~\ref{section:results_p1_gridsearch}) that are nearly identical. For the 127-qubit backends, we show that best solution distributions are of similar shape on different backends but shifted linearly to account for better average expectation values (Subsection~\ref{section:results_p1_gridsearch}).

\section{Methods}
\label{section:methods}
First we outline the hardware-compatible combinatorial optimization problems in Subsection~\ref{section:methods_Higher_order_Ising_models}. The QAOA algorithm is described in Subsection~\ref{section:methods_QAOA_circuits}; Subsection~\ref{section:methods_QAOA_angle_finding} describes the optimized angle-finding and parameter transfer procedure that allows high-quality angles to be computed for $127$-qubit QAOA circuits, and Subsection~\ref{section:methods_MPS_simulations} describes the MPS simulations. Lastly, Subsection~\ref{section:methods_hardware_implementation} describes the hardware implementation.

\begin{figure}[p!]
    \centering
    \includegraphics[width=0.59\textwidth]{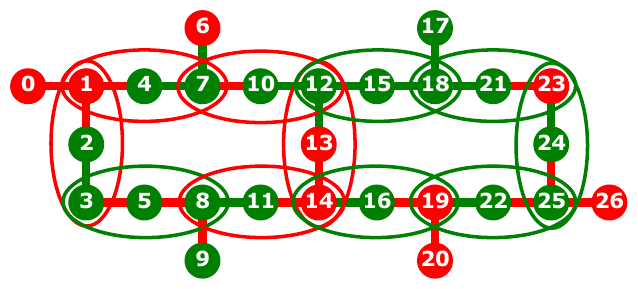}\hfill%
    \includegraphics[width=0.39\textwidth]{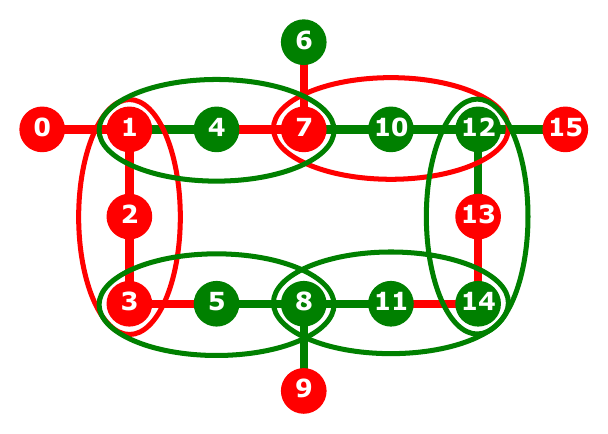}%
    \vspace{-1ex}
    \caption{Examples of higher order Ising models that are hardware-compatible with a heavy-hex hardware graph: 
    \textbf{(left)} A $27$-qubit device: Nodes correspond to linear terms, edges to quadratic terms, and hyperedges encircling three neighboring nodes to cubic terms. Ising coefficients of $-1$ and $+1$ are depicted in red and green, respectively. 
    \textbf{(right)} A $16$-qubit device: Illustrating the terminology of Equation~\eqref{equation:problem_instance}, we have $W = \{2,4,5,10,11,13\}$, with the remaining nodes in $V_2$ being $V_2 \setminus W = \{0,6,9,15\}$. For node $4 \in W$, we have neighbors $\{n_1(l), n_2(l)\} = \{1,7\} \subset V_3$.
    \textbf{(bottom)} A $127$-qubit device: Higher order Ising model comprised of $127$ linear, $144$ quadratic and $71$ cubic terms.
    }
    \vspace{4ex}
    \includegraphics[width=0.95\textwidth]{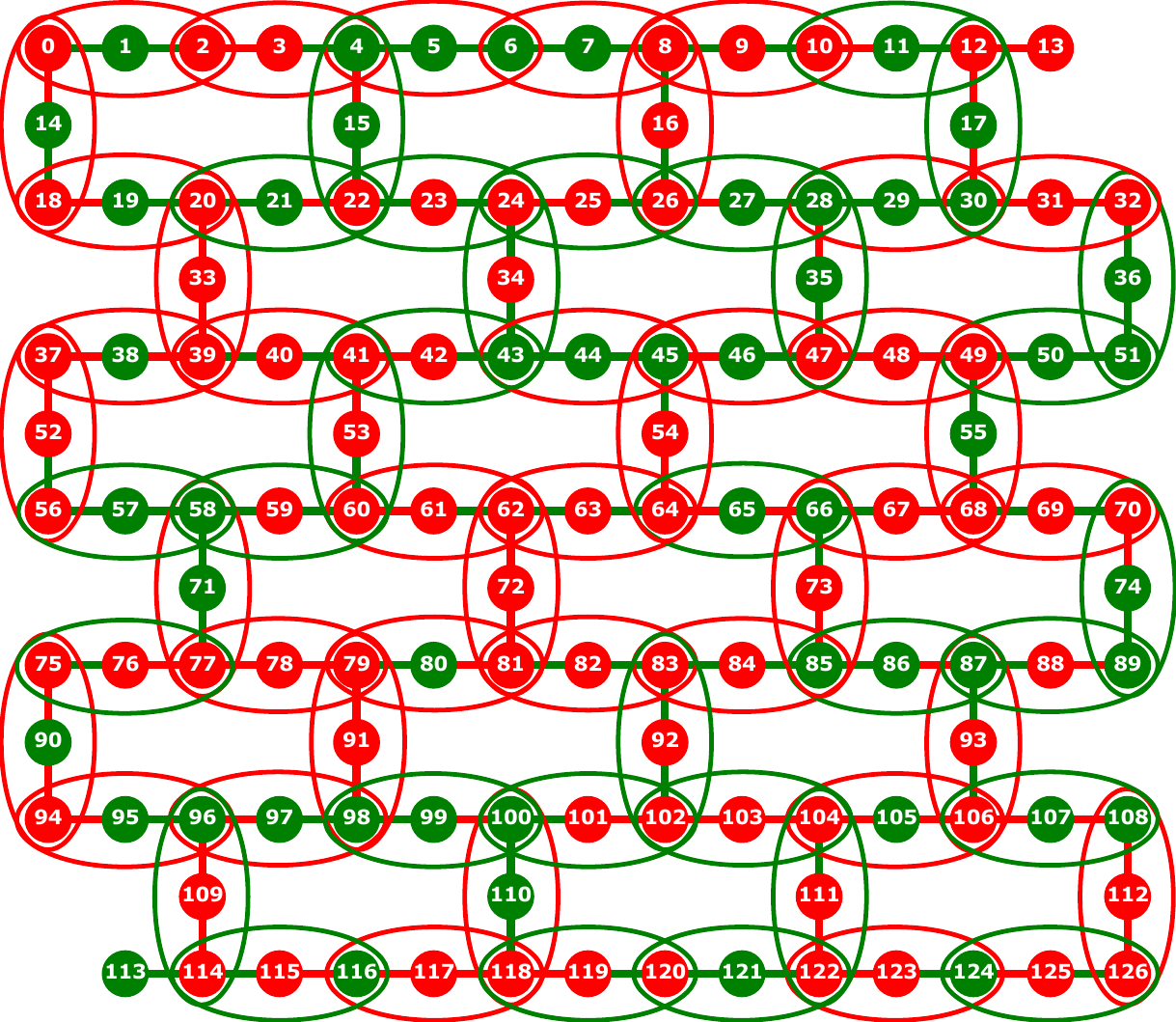}%
    \label{fig:problem_instances}
\end{figure}

\subsection{Heavy-Hex Compatible Ising Models}
\label{section:methods_Higher_order_Ising_models}

The class of \emph{minimization} combinatorial optimization problems that we consider are heavy-hex graph native spin glasses, and were introduced and described in Refs.~\cite{pelofske2023qavsqaoa,pelofske2023short}. This class of models was designed specifically to be heavily optimized for a heavy-hex hardware graph~\cite{chamberland2020topological}, and can include higher order terms (specifically geometrically local cubic terms), thus making the optimization problem more difficult. Importantly, although Refs.~\cite{pelofske2023qavsqaoa,pelofske2023short} used these problems for sampling $127$ qubit heavy-hex native problems, this problem type is well defined for any heavy-hex hardware graph size. Here, we consider random instances of these problem types defined on $16$, $27$, $127$, and $414$ qubit IBM Quantum hardware graphs. 

For a heavy-hex graph $G=(V,E)$ and a vector of spins $z = (z_0, \ldots, z_{n-1}) \in \{ +1,-1\}^n$ we define a cost function 
\begin{equation}
    C(z) = \sum_{v \in V} d_v \cdot z_v + \sum_{(i,j) \in E} d_{i,j} \cdot  z_i \cdot z_j + \sum_{l \in W} d_{l,n_1(l),n_2(l)} \cdot z_l \cdot z_{n_1(l)} \cdot z_{n_2(l)},
    \label{equation:problem_instance}
\end{equation}
and a QAOA cost Hamiltonian $H_C$ by replacing spin variables $z_i$ with Pauli operators $\sigma_z^i$.
Equation~\eqref{equation:problem_instance} defines a random spin glass problem with specific cubic terms: Any subgraph of a heavy-hex lattice is a bipartite graph with vertices $V = \{0,\ldots,n-1\}$ is uniquely bipartitioned as $V=V_2 \sqcup V_3$ with $E\subset V_2\times V_3$, where $V_i$ consists of vertices of maximum degree $i$. $W$ is the set of vertices $l \in V_2$ that have degree equal to $2$, with neighbors denoted by $n_1(l)$ and $n_2(l)$, see Figure~\ref{fig:problem_instances}. Thus $d_v$, $d_{i,j}$, and $d_{l,n_1(l),n_2(l)}$ are the linear, quadratic and cubic coefficients, respectively. The coefficients are chosen randomly from $\{+1, -1\}$ with probability $0.5$, see Figure~\ref{fig:problem_instances}. Figure~\ref{fig:problem_instances} (bottom) shows an example problem instance defined on a $127$ qubit heavy-hex graph.

\paragraph{Instance generation and assessment} In Table~\ref{table:hardware_summary}, we give a summary of the studied hardware devices as well as the problem instances generated to run on these QPUs. For each group of QPUs sharing the same hardware graph, we generate 100 random problem instances according to Equation~\eqref{equation:problem_instance}, which are shared across these devices. 

One additional problem type we evaluate on a subset of the hardware experiments is Equation~\eqref{equation:problem_instance} without cubic terms, i.e., random spin glass problems with only linear and quadratic terms. To assess the achieved QAOA performances in context, we additionally compute for each instance the minimum (ground state) energy and the maximum energy. This is done with CPLEX~\cite{cplexv12} after pre-processing order reduction which introduces auxiliary variables, as outlined in Ref.~\cite{pelofske2023short}. These problems are solved by CPLEX as Mixed Integer Quadratic
Programming (MIQP) problems where the decision variables are all binary.

\begin{table*}[t!]
    \centering
    \begin{tabularx}{\textwidth}{@{\extracolsep{\fill}}llr@{}rl@{\ }lrrr@{}}
        \toprule
        \                           & \multicolumn{5}{c}{Hardware}                                      & \multicolumn{3}{r}{\#Instance Coefficients}    \\
        \cmidrule(lr){2-6}
        \cmidrule(l){7-9}
        Device name                 &Processor & \#Qubits& \#2q-Gates & \multicolumn{2}{l}{Basis gates} & linear    & quadratic & cubic     \\ 
        \midrule[\heavyrulewidth]
        \texttt{ibm\_seattle}       & Osprey r1     & 414   & 475   & ECR,  & ID, RZ, SX, X             & 414       & 475       & 232       \\ 
        \midrule
        \texttt{ibm\_washington}    & Eagle r1      & 127   & 142   & CX,   & ID, RZ, SX, X             & 127       & 142       & 69        \\ 
        \midrule
        \texttt{ibm\_sherbrooke}    & Eagle r3      & 127   & 144   & ECR,  & ID, RZ, SX, X             & 127       & 144       & 71        \\ 
        \texttt{ibm\_brisbane}      & Eagle r3      & 127   & 144   & ECR,  & ID, RZ, SX, X             & 127       & 144       & 71        \\ 
        \texttt{ibm\_cusco}         & Eagle r3      & 127   & 144   & ECR,  & ID, RZ, SX, X             & 127       & 144       & 71        \\ 
        \texttt{ibm\_nazca}         & Eagle r3      & 127   & 144   & ECR,  & ID, RZ, SX, X             & 127       & 144       & 71        \\ 
        \midrule 
        \texttt{ibm\_geneva}        & Falcon r8     & 27    & 28    & CX,   & ID, RZ, SX, X             & 27        & 28        & 11        \\ 
        \texttt{ibm\_auckland}      & Falcon r5.11  & 27    & 28    & CX,   & ID, RZ, SX, X             & 27        & 28        & 11        \\ 
        \texttt{ibm\_algiers}       & Falcon r5.11  & 27    & 28    & CX,   & ID, RZ, SX, X             & 27        & 28        & 11        \\ 
        \texttt{ibmq\_kolkata}      & Falcon r5.11  & 27    & 28    & CX,   & ID, RZ, SX, X             & 27        & 28        & 11        \\ 
        \texttt{ibmq\_mumbai}       & Falcon r5.10  & 27    & 28    & CX,   & ID, RZ, SX, X             & 27        & 28        & 11        \\ 
        \texttt{ibm\_cairo}         & Falcon r5.11  & 27    & 28    & CX,   & ID, RZ, SX, X             & 27        & 28        & 11        \\ 
        \texttt{ibm\_hanoi}         & Falcon r5.11  & 27    & 28    & CX,   & ID, RZ, SX, X             & 27        & 28        & 11        \\ 
        \midrule
        \texttt{ibmq\_guadalupe}    & Falcon r4P    & 16    & 16    & CX,   & ID, RZ, SX, X             & 16        & 16        & 6         \\ 
        \bottomrule
    \end{tabularx}
    \caption{QPU and Instance summary. Numbers of available qubits and 2-qubit gates are accurate at the time in which the experiments of this study were executed.
    Devices mainly differ in their native 2-qubit gates (ECR~vs.~CX), and in the ratios between the numbers of linear, quadratic and cubic instance coefficients that can be accommodated.}
    \label{table:hardware_summary}
\end{table*}

\subsection{Whole Chip QAOA Circuit Description}
\label{section:methods_QAOA_circuits}

The Quantum Alternating Operator Ansatz consists of preparing the initial state $\ket{\psi}$, then for $p$ rounds applying alternatingly the phase separating Hamiltonian $H_C$ parameterized by the real number $\gamma_i$ and the mixing Hamiltonian $H_M$ parameterized by the real number $\beta_i$:
\begin{equation} \label{eq:QAOA_state}
    \ket{\vec{\gamma},\vec{\beta}} = \underbrace{e^{-i\beta_p H_M} e^{-i\gamma_p H_C}}_{\text{round }p}\cdots \underbrace{e^{-i\beta_1 H_M} e^{-i\gamma_1 H_C}}_{\text{round }1} \ket{\psi}
\end{equation}

In each round, $H_C$ is first applied which separates out the basis states of the state vector by phases $e^{-i\gamma C(z)}$. Next, $H_M$ gives parameterized interference between solutions with different cost values. After $p$ rounds, the state $\ket{\vec{\gamma},\vec{\beta}}$ is measured in the computational basis and thus finds a sample $z$ of cost value $C(z)$ with probability $|\braket{z|\vec{\gamma},\vec{\beta}}|^2$. Notably, the QAOA cost Hamiltonian can include higher order polynomial terms~\cite{campbell2022qaoa, basso2022performance}, without requiring ancilla qubit overhead. We make use of this property of QAOA in order to sample higher-order Ising models that are heavy-hex hardware-compatible, introduced in Refs.~\cite{pelofske2023qavsqaoa,pelofske2023short}. 

\begin{figure}[t!]
    \centering
    \includegraphics[width=\textwidth]{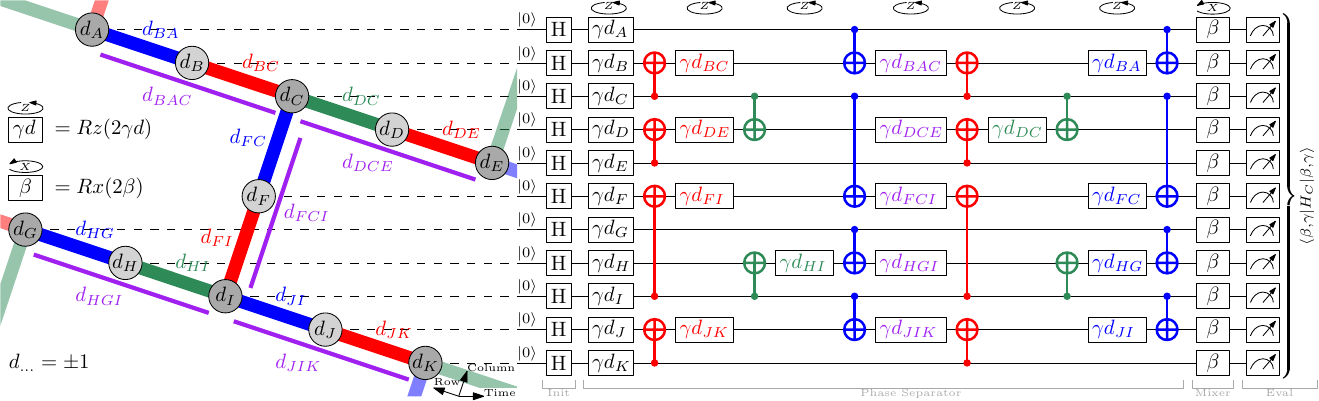}
    \caption{\textbf{From Refs.}~\cite{pelofske2023qavsqaoa,pelofske2023short}: QAOA circuit description for heavy-hex graph compatible higher order Ising models of arbitrary size. The graph is bipartite and has an arbitrary 3-edge-coloring given by K\H{o}nig's line coloring theorem. 
    \textbf{(left)} 3-edge-coloring and bipartite grey-shading of the nodes. Adjacent purple lines denote the cubic terms terms. 
    \textbf{(right)} Any quadratic term (colored edge) gives rise to a combination of two CNOTs and a Rz-rotation in the phase separator, giving a CNOT depth of 6 due to the degree-3 nodes. When targeting the degree-2 nodes with the CNOT gates, these constructions can be nested to implement the cubic terms with just one additional Rz-rotation. }
    \label{fig:QAOA_circuit_description}
\end{figure}

Figure~\ref{fig:QAOA_circuit_description} shows the QAOA circuit construction algorithm used in this study for one layer of the algorithm ($p=1$) which is the same for all layers, specifically targeting the Ising model type defined in Equation~\eqref{equation:problem_instance}. The transverse field mixer QAOA implementation is used in all circuits. A greedy Breadth-first search (BFS) 3-edge-coloring is computed each time a circuit is constructed, and that same edge coloring is then used for all $p$ layers in that circuit. 

The operators $e^{-i\beta H_M}$ and $e^{-i\gamma H_C}$ are $2\pi$-periodic, hence we can restrict the QAOA angle search space to $\beta_i,\gamma_i \in [0,2\pi)$ for each round $1\leq i \leq p$. However, careful consideration of the parity of solution values as well as symmetries when starting in the state $\ket{\psi} = \ket{+^n}$ and measuring in the computational basis allows us to further restrict the search space to
$\beta_1, \ldots, \beta_{p-1} \in [0,\pi)$, $\beta_p \in [0,\tfrac{\pi}{2})$, and $\gamma_1, \ldots, \gamma_p \in [0,\pi)$, see Ref.~\cite{pelofske2023short}.

\subsection{QAOA Angle Finding with JuliQAOA and Parameter Transfer}
\label{section:methods_QAOA_angle_finding}

Arguably the most difficult aspect of implementing most variants of QAOA is determining good angles. Specifically, it is known that in general QAOA needs to be applied for a reasonably high number of rounds ($p$)~\cite{farhi2020quantum1, farhi2020quantum2} in order to get to high quality solutions of combinatorial optimization problems. However, this requires high quality angles (since in almost all cases there are no analytical solutions for optimal QAOA angles) for each $p$, and there are a total of $2p$ parameters that need to be optimized. The standard variational hybrid quantum-classical approach to this is to repeatedly evaluate the expectation value of the cost Hamiltonian $H_C$ for different sets of angles on a quantum computer, using a classical algorithm to guide exploration in angle space. This approach is quite costly however with respect to total compute time. Moreover, because current quantum computers are quite noisy, learning good angles in this manner is in general hard, and in particular are infeasible for a $127$ qubit instance. A promising approach to mitigate some of these problems is to find good angles on smaller, more tractable instances and then use those same angles on larger instances. This technique, often referred to as \emph{parameter transfer} or \emph{parameter concentration}, has been shown to be effective, both analytically and numerically, for a number of different problem types~\cite{brandao2018fixed, farhi2022qaoa, wurtz2021fixedangle, akshay2021parameter, galda2021transferability, lee2021parameters, shaydulin2023evidence, shaydulin2023parameter, galda2023similaritybased, pelofske2023qavsqaoa, pelofske2023short, streif2020training}. Motivated by the existing evidence for parameter transfer working for different problem sizes, and by the experimental evidence for parameter concentration across different random heavy-hex native Ising models observed in~\cite{pelofske2023qavsqaoa,pelofske2023short}, we utilize parameter transfer in order to obtain good fixed angles for this class of random Ising models with higher order terms. Specifically, we obtain good angles for $p=1, 2, 3, 4, 5$ on a single random $16$ qubit instance, and then validate that those angles transfer to other random $16$ qubit instances, and random $27$ qubit, as well as random $127$ qubit instances. These angles are not optimal QAOA angles, rather they are high-quality heuristic angles, in particular meaning that we obtain reliable improvements in the mean energy as a function of $p$ increasing. 

The method we use to compute these good angles is the high-performance, QAOA-specific quantum simulator \texttt{JuliQAOA}~\cite{golden2023juliQAOA} with $1000$ basin hopping iterations and angle extrapolation (fixing the angles found at previous $p$-steps and initializing at those angles when proceeding to the next $p$). This allows us to find very high quality angles, in this case running $O(10^5)$ exact statevector simulations on one $16$ qubit instance (derived from the \texttt{ibmq\_guadalupe} architecture) with higher order terms for $p=1,\ldots,5$. The QAOA angles are computed on one arbitrary $16$ qubit instance so as to determine how well the parameters transfer for just one instance - and moreover, performing this computation once is much more efficient than repeating this for an entire ensemble of problems (although, this is a likely more robust approach that could be investigated in future study). Having trained on only one problem instance also allows us to evaluate how well the parameters transfer to the other $99$ random $16$ qubit instances. \texttt{JuliQAOA} has been used in several previous QAOA publications, with the goal of computing very high quality QAOA angles on general types of combinatorial optimization problems~\cite{golden2023qaoasat, pelofske2023high, golden2023numerical, golden2021threshold}. The fixed angles used for the experiments shown in Subsection~\ref{section:results_scaling_p_16_27_127_qubits} are given explicitly below.

The trained QAOA angles up to $p=5$ on a single $16$ qubit problem instance (with cubic terms) using \texttt{JuliQAOA}~\cite{golden2023juliQAOA}, which were used for the parameter transfer onto much larger problem instances, are:

\begin{itemize}[noitemsep]
    \item $p_1 : \beta = [0.38919], \gamma = [6.04302]$
    \item $p_2 : \beta = [0.48912, 0.27367], \gamma = [6.09758, 5.95396]$
    \item $p_3 : \beta = [0.50502, 0.35713, 0.19264], \gamma = [6.14054, 6.01729, 5.94123] $
    \item $p_4 : \beta = [0.54321, 0.41806, 0.28615, 0.16041], \gamma = [6.16242, 6.05959, 5.98417, 5.9299] $
    \item $p_5 : \beta = [0.53822, 0.44776, 0.32923, 0.23056, 0.12587], \gamma = [6.16555, 6.08373, 6.01445, 5.9616, 5.93736] $
\end{itemize}

These exact QAOA angles for $p=1$ and $p=2$ can be directly compared to the angles that were computed using high resolution grid-searches for this same class of optimization problems in refs. \cite{pelofske2023qavsqaoa, pelofske2023short}; this comparison shows that the $p=1,2$ angles agree reasonably well, but are not exactly the same (note that there are angle symmetries that must be accounted for in order to compare these angles).

\subsection{MPS simulations}
\label{section:methods_MPS_simulations}

We use MPS formalism to compute approximations to $\ket{\vec{\gamma},\vec{\beta}}$ in Equation~\eqref{eq:QAOA_state}. Specifically, a version of time-evolving block decimation~\cite{vidal2003efficient} has been used to simulate the action of $e^{-i\gamma_k H_C}$ and $e^{-i\beta_k H_M}$ for $k=1,\ldots,p$. MPS tensors are ordered in the same way as the qubits are labeled in Figure~\ref{fig:problem_instances}. The accuracy of MPS simulations is determined by bond dimension, denoted by $\chi$ here. In general, the accuracy is improved with increasing $\chi$. Significant portion of the terms in $e^{-i\gamma_k H_C}$ are non-local and accurate simulation with MPS requires the bond dimension to grow quickly.

The Hamiltonian $H_C$ is a sum of $\sigma^z_i$, $\sigma^z_i \sigma^z_j$ and $\sigma^z_i \sigma^z_j \sigma^z_k$ interactions. All those terms commute and $\sigma^z_i \sigma^z_j \sigma^z_k$ terms span the entire graph, so $e^{-i\gamma_k H_C}$ can be written as $\prod_{\vec{\alpha}} e^{-ih_{\alpha_1,\alpha_2,\alpha_3}}$, where $\vec{\alpha} = (\alpha_1,\alpha_2,\alpha_3)$ and each $h_{\alpha_1,\alpha_2,\alpha_3}$ is a three-body term acting on qubits $(\alpha_1,\alpha_2,\alpha_3)$. Importantly, each $e^{-ih_{\alpha_1,\alpha_2,\alpha_3}}$ can be written as a Matrix Product Operator with bond dimension 2. This is achieved by standard tensor network methods~\cite{orus2014practical} that include series of tensor reshapes and SVDs of the original $8 \times 8$ matrix constructed from $e^{-ih_{\alpha_1,\alpha_2,\alpha_3}}$.

Further, three-body interactions are divided into groups such that gates $e^{-ih_{\alpha_1,\alpha_2,\alpha_3}}$ and $e^{-ih_{\alpha'_1,\alpha'_2,\alpha'_3}}$ belong to the same group if and only if the sets $\{ \alpha_1,\alpha_1+1,\ldots,\alpha_3 \}$ and $\{ \alpha'_1,\alpha'_1+1,\ldots,\alpha'_3 \}$ are disjoint. This step is needed, so that the exact simulation of all gates $e^{-ih_{\alpha_1,\alpha_2,\alpha_3}}$ in a given group increases the MPS bond dimension by at most factor of 2, for some MPS tensors. After all the gates in a given group are applied, MPS is compressed, so the bond dimension does not increase beyond the predefined maximal value. The above manipulations of $e^{-i\gamma_k H_C}$ are performed to decrease the cost of MPS simulations and in turn, to increase accuracy of the simulation. As a result, simulations with small $p$ can be performed exactly and accuracy is expected to gradually deteriorate as $p$ is increased.

\begin{figure}
    \centering
    \includegraphics[width=\columnwidth]{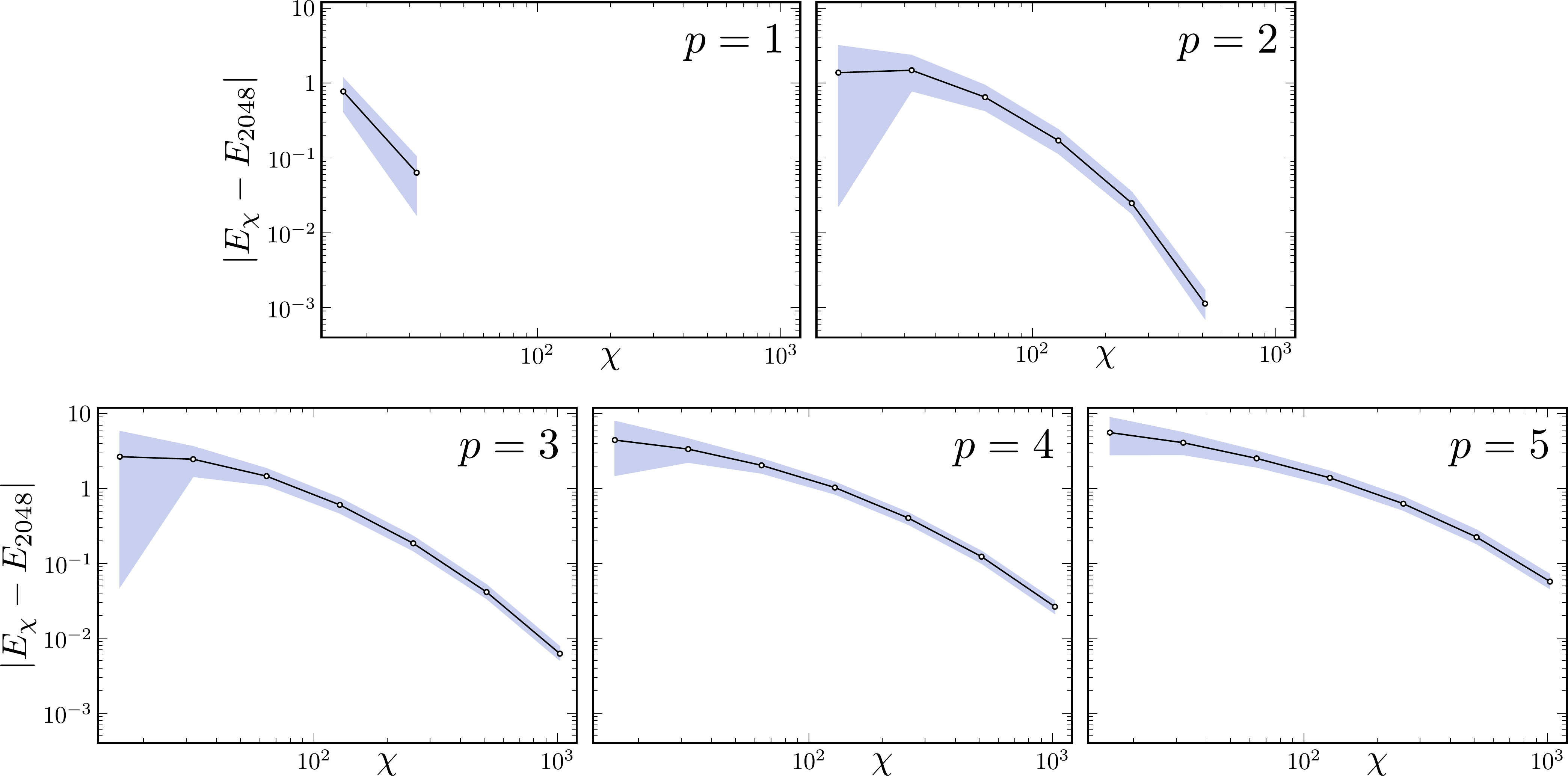}
    \caption{Error analysis performed for MPS simulations. Here we show error in the energy $|E_{\chi} - E_{2048}|$ as a function of bond dimension $\chi$ for different values of $p$. Solid black lines represent the error in the energy averaged over one hundred instances of $H_C$. All computed errors, for all instances, are contained within gray areas around black lines. The gray areas are relatively small, especially at large $\chi$. This indicates that different instances of $H_C$ result in similar errors. Errors are small; they are all below $0.1$, which is observed in the hardest case of $p=5$. Simulations with $p<5$ incur smaller errors.
    }
    \label{fig:MPS_errors}
\end{figure}

We perform MPS simulations with $\chi = 2^m$, for $m = 4,\ldots,11$ to estimate the impact of errors imposed by finite bond dimension. The summary of our results is presented in Figure~\ref{fig:MPS_errors}. All the panels show the error in the energy $\Delta E$ as a function of bond dimension. It is measured as $\Delta E = |E_{\chi} - E_{2048}|$, where $E_{\chi} = \bra{\psi^\mathrm{MPS}_\chi}H_C \ket{\psi^\mathrm{MPS}_\chi}$. Here, $\ket{\psi^\mathrm{MPS}_\chi}$ is an MPS approximation to $\ket{\vec{\gamma},\vec{\beta}}$ in Equation~\eqref{eq:QAOA_state} obtained by a simulation with maximum bond dimension of $\chi$. Our most accurate simulations are performed with $\chi=2048$, and hence we treat $E_{2048}$ as the best approximation to the exact energy. Solid, black lines in Figure~\ref{fig:MPS_errors} represent $\Delta E$ averaged over one hundred instances of $H_C$. All simulation errors, for all instances of $H_C$, are within gray areas shown in Figure~\ref{fig:MPS_errors}.

MPS simulations become exact for $p=1$ and $p=2$ at $\chi=64$ and $\chi=1024$ respectively. As pointed out above, the proper treatment of $e^{-i\gamma_k H_C}$ allows us to perform exact simulations at relatively small values of $\chi$. The error $\Delta E$ drops to zero in those cases. Those values are not shown in $p=1$ and $p=2$ panels of Figure~\ref{fig:MPS_errors}. Simulations with $p>2$ are no longer exact but the errors are small and do not exceed $0.1$ for $p=5$. Note that $\Delta E$ is an absolute error in the energy. In relative terms, the error is below $10^{-3}$, given $E_{2048} \approx -150$ on average. It is important to note that all simulation errors, for all instances of $H_C$ are similar to each other, especially in the large $\chi$ limit. This is indicated by very thin gray error areas around the mean values of the error in all panels. Our error analysis strongly suggests that our MPS simulation is dependable and sufficiently accurate (for considered values of $p$) to represent results that would have been obtained on a quantum computer in the limit of vanishingly small noise.

Since MPS is a unitary tensor network, one can draw bitstrings $z$ from the probability distribution $P(z) = |\braket{z|\psi^\mathrm{MPS}_\chi}|^2$~\cite{ferris2012perfect}. That is, one can approximately calculate samples that would have been drawn from $\ket{\vec{\gamma},\vec{\beta}}$ in Equation~\eqref{eq:QAOA_state}, assuming that the quantum computer has been executing operations noiselessly. We use that fact to generate samples shown in Figure~\ref{fig:MPS_samples}.

On average, computing $\ket{\psi^\mathrm{MPS}_\chi}$ for $p = 5$ and $\chi = 2048$ took less than $1.5$ hours on a 48-core computational node.

\begin{figure}[t!]
    \centering
    \includegraphics[height=0.27\textwidth,trim={00 0 0 0},clip]{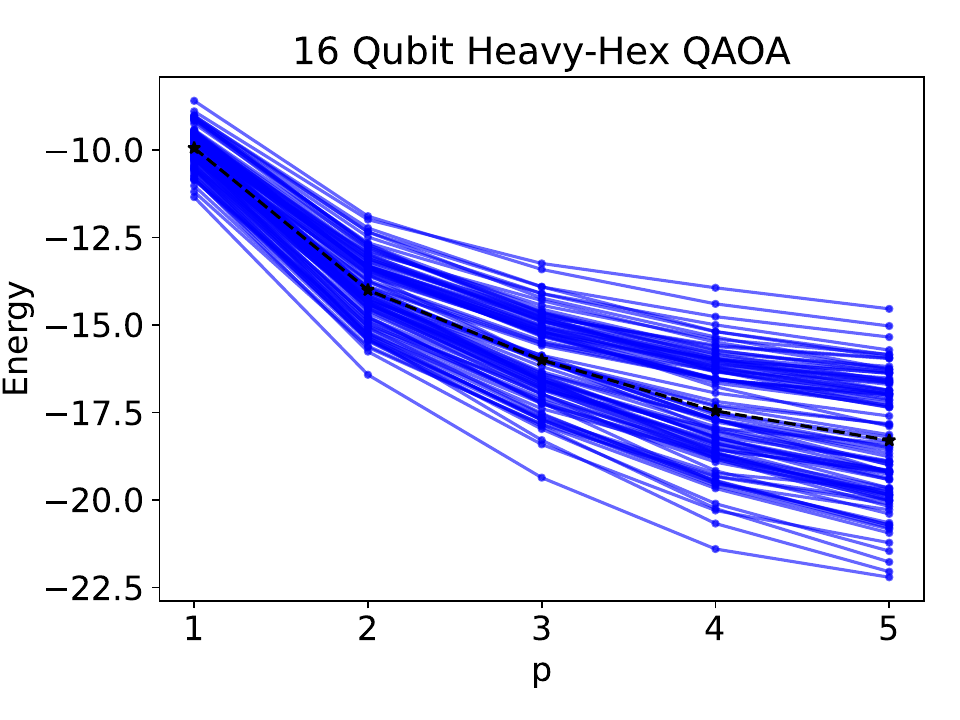}\hfill%
    \includegraphics[height=0.27\textwidth,trim={40 0 0 0},clip]{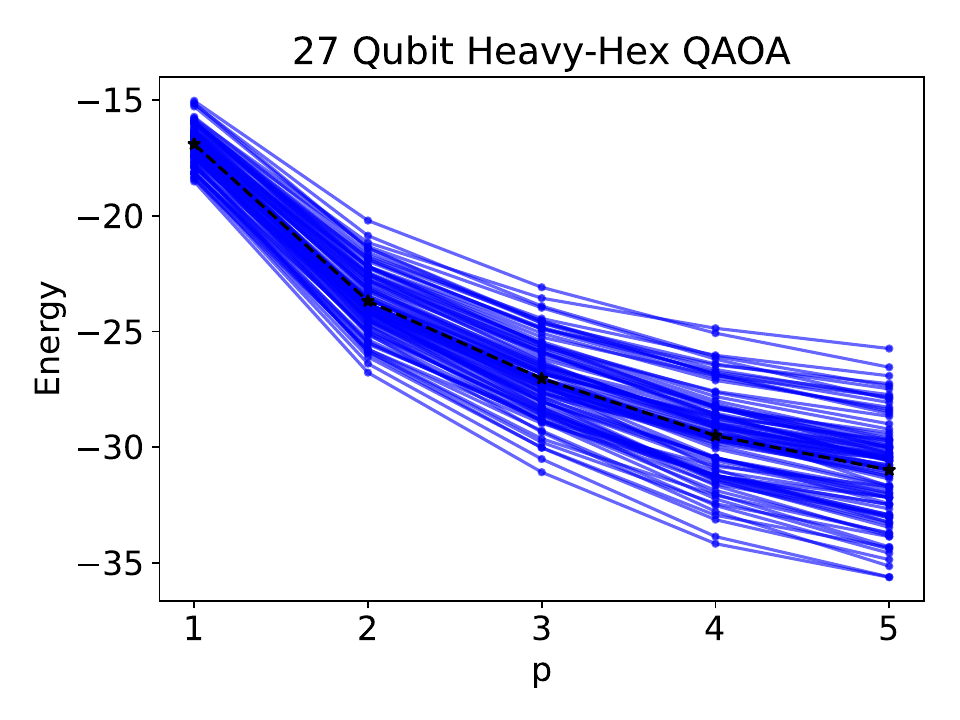}\hfill%
    \includegraphics[height=0.265\textwidth,trim={55 -2 0 0},clip]{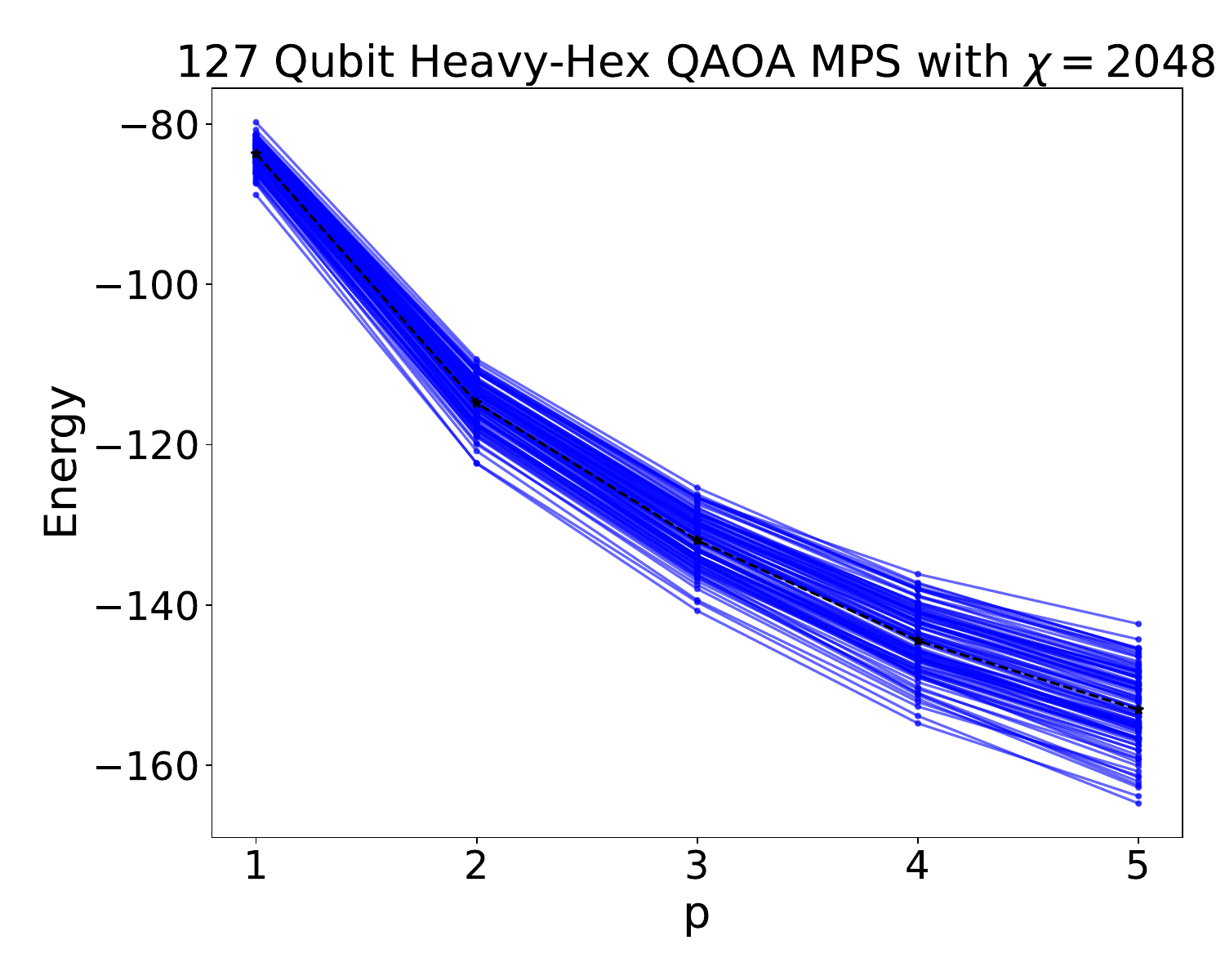}%
    \vspace*{-2ex}
    \caption{Classical simulations of mean energies demonstrating (noiseless) concentration of QAOA parameters.
    We simulate 100 random instances for each circuit size using fixed QAOA angles (trained on a single 16 qubit instance):\linebreak
    \textbf{(left)} The angles for $1\leq p \leq5$ are used to execute QAOA on 100 random 16-qubit higher-order heavy-hex instances,\linebreak
    \textbf{(center)} The same angles are used for 100 random $27$-qubit instances,
    \textbf{(right)} MPS simulation with bond dimension $\chi=2048$ is used for 100 random $127$-qubit instances.
    For growing circuit size $16,27,127$, \emph{for every random higher-order Ising model}, as $p$ increases \emph{the mean energy strictly improves}, 
    showing that parameter transfer succeeds in a noiseless setting.
    In each plot, also the mean energy across the instance ensemble is plotted as a dashed black line.  
    }
    \vspace*{1ex}
    \label{fig:parameter_transfer_scaling_plots}
\end{figure}

\begin{figure}[t!]
    \centering
    \includegraphics[width=0.24\textwidth]{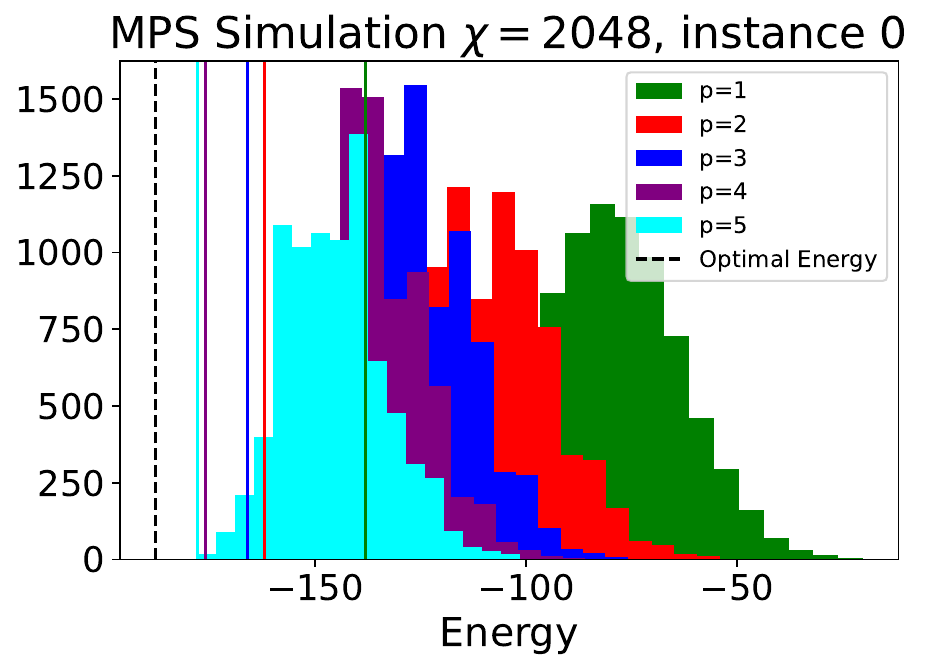}\hfill%
    \includegraphics[width=0.24\textwidth]{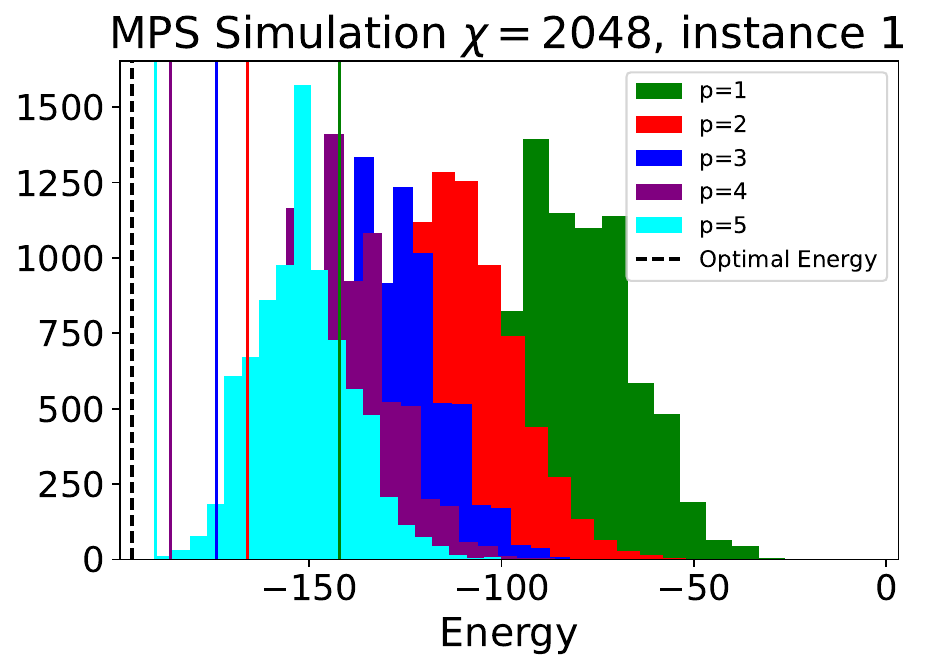}\hfill%
    \includegraphics[width=0.24\textwidth]{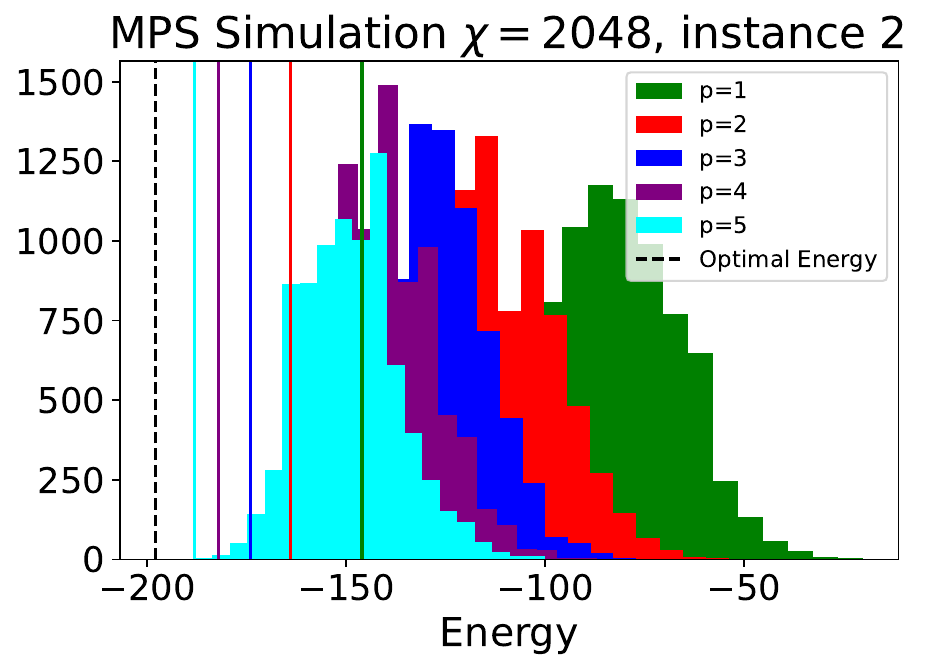}\hfill%
    \includegraphics[width=0.24\textwidth]{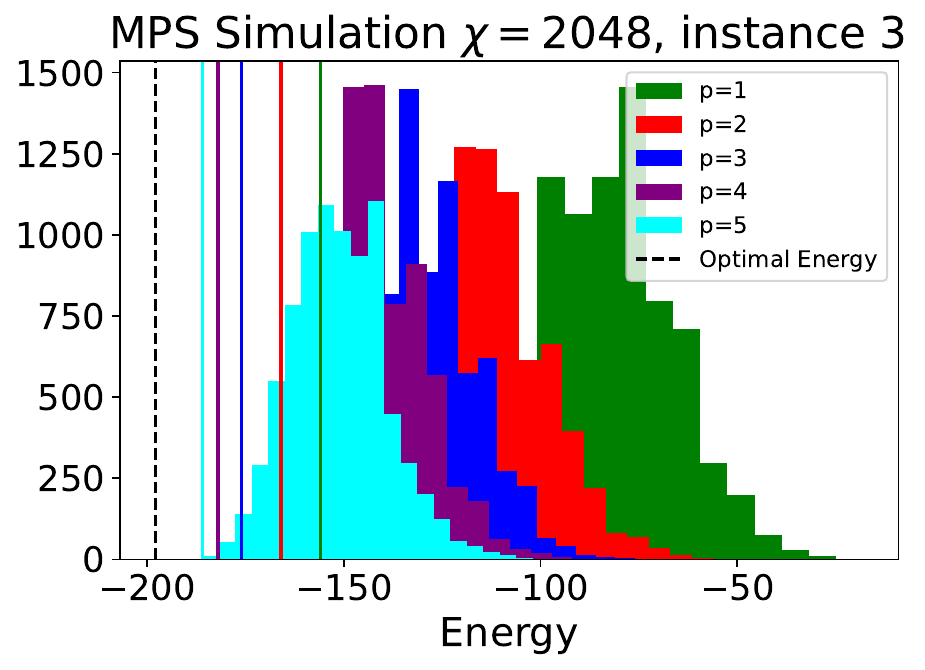}\hfill%
    \includegraphics[width=0.24\textwidth]{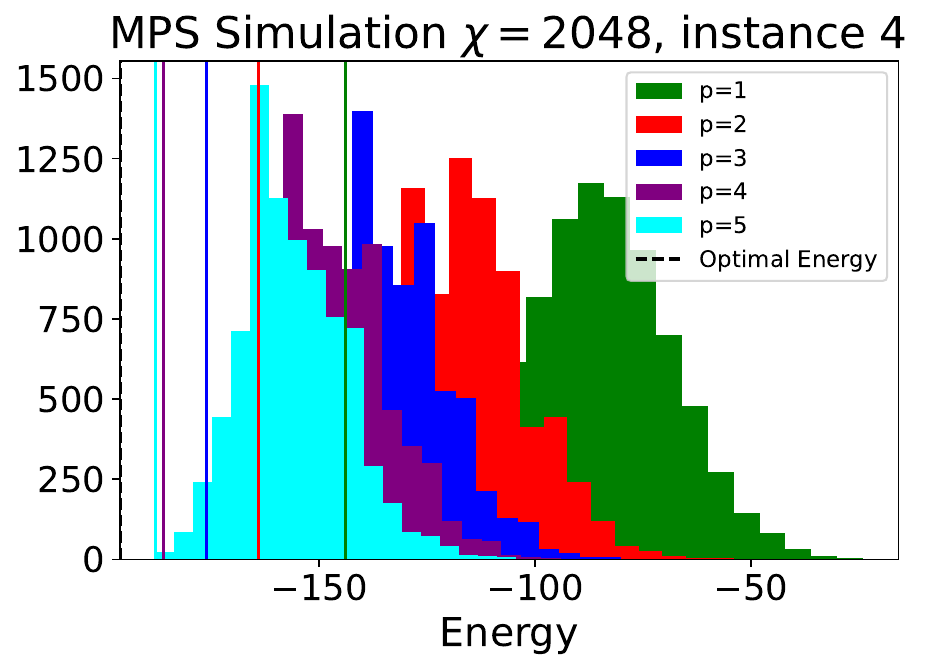}\hfill%
    \includegraphics[width=0.24\textwidth]{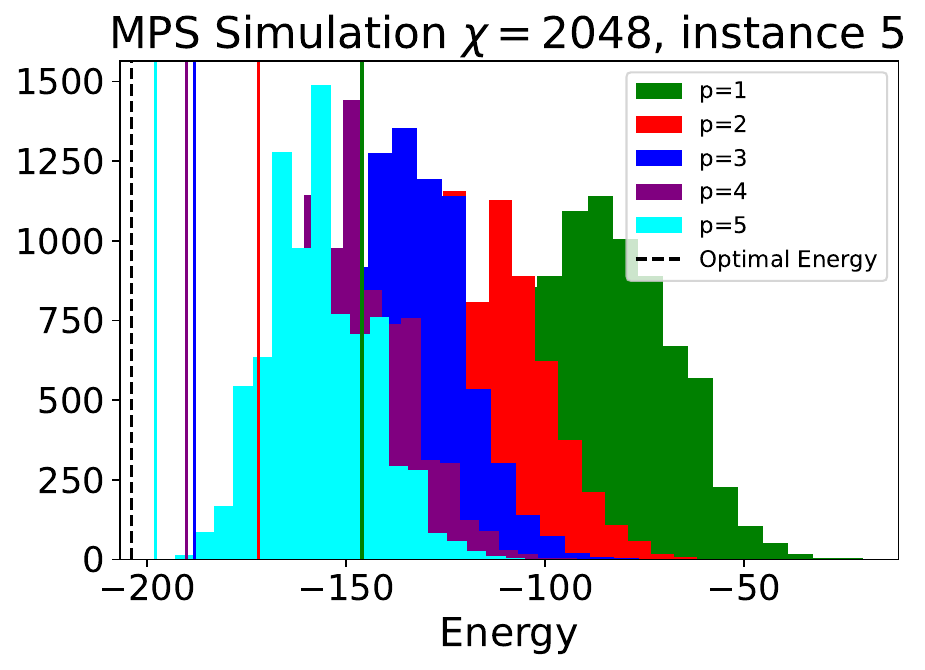}\hfill%
    \includegraphics[width=0.24\textwidth]{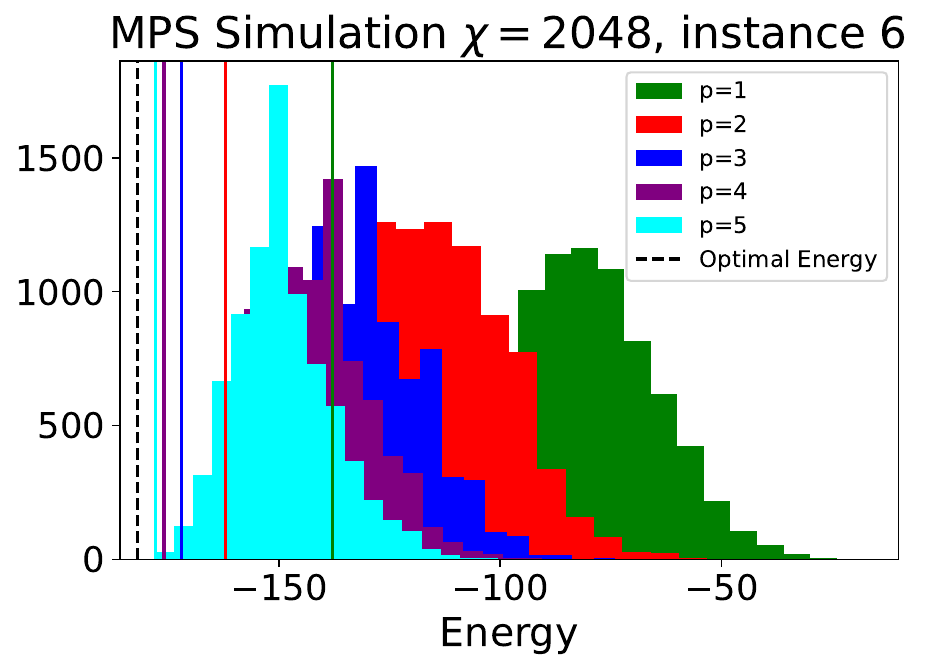}\hfill%
    \includegraphics[width=0.24\textwidth]{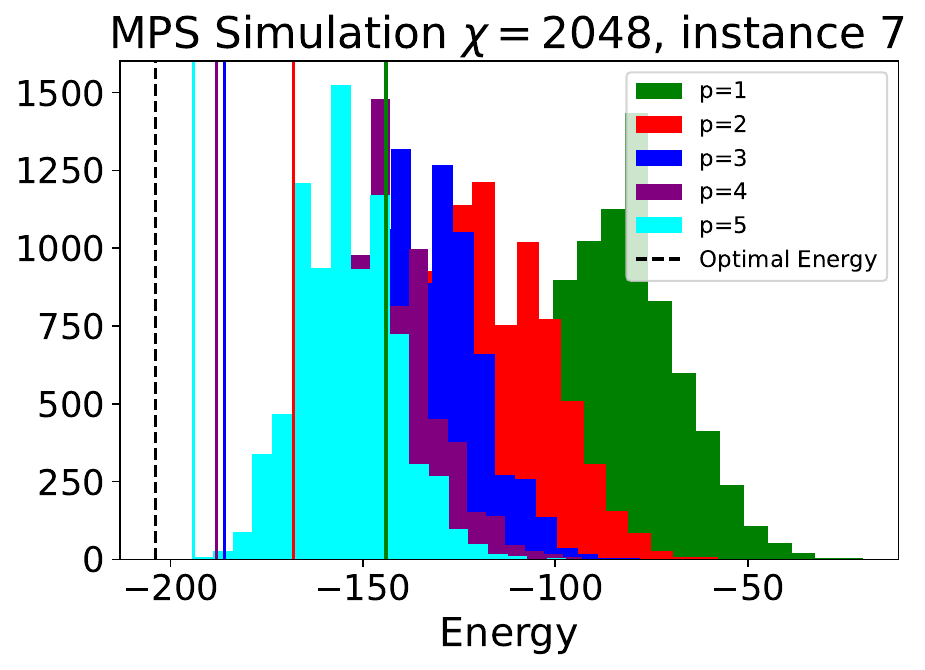}%
    \vspace*{-1ex}
    \caption{MPS simulation sample distributions, using a bond dimension of $\chi=2048$ for $127$-qubit QAOA circuits sampling eight of the higher-order Ising model problem instances. For each $p$, a total of $8192$ samples are computed. The ground state energy is marked in all plots with a dashed vertical black line, and the minimum energy found within each $p$ energy distribution is marked with a vertical solid line. In addition to the means of the distributions improving as a function of $p$ (shown for all $100$ problem instances in Figure~\ref{fig:parameter_transfer_scaling_plots}), we see here that the minimum energy sampled also improves as $p$ increases. Notably, none of the distributions sampled the optimal energy, although the minimum energies from $p=4$ and $p=5$ are close to the optimal energy. These distributions show the ideal QAOA sampling capabilities, using the classical simulation method of MPS, if the quantum computation was noiseless. }
    \label{fig:MPS_samples}
\end{figure}

\subsection{IBM Quantum Hardware Implementation Details}
\label{section:methods_hardware_implementation}
The quantum circuits are passed through the Qiskit~\cite{Qiskit} transpiler in order to adapt the circuits to the hardware native gateset, such as adapting the circuits to use the two qubit unidirectional \texttt{echoed cross-resonance} ECR~\cite{sheldon2016procedure} gate. The QAOA circuits are heavily optimized for the heavy-hex hardware graph, so the compilation uses the fixed hardware graph and the compiler optimization is not able to reduce the two qubit gate count.

We also evaluate a relatively simple, and hardware gate-native, digital dynamical decoupling scheme of pulses of pairs of Pauli X gates, scheduled both As Soon As Possible (ASAP) and As Late As Possible (ALAP). This is implemented in Qiskit using the digital dynamical decoupling pass~\cite{PadDynamicalDecoupling}. Dynamical decoupling is an open loop quantum control error suppression technique for mitigating decoherence on idle qubits~\cite{viola1998dynamical, suter2016colloquium, viola1999dynamical, ahmed2013robustness, larose2022mitiq}, which can be approximated using digital sequences of single qubit gates that are mathematically equivalent to applying the identity gate. Section~\ref{section:appendix_QAOA_circuit_drawings} contains detailed compiled circuit renderings for $p=1$ whole-chip $127$ qubit QAOA circuits. Dynamical decoupling is used for these QAOA circuits specifically because it can mitigate errors on idle qubits encountered in the processor under some noise conditions, but importantly does not introduce compute overhead of additional samples or circuit executions since it is a compilation procedure that adds gates during idle periods of time for qubits. 

The hardware results are reported in terms of the mean approximation ratio, which for random Ising models is defined over the full range of unconstrained energy values for a specific problem instance denoted as $\text{Min}$ and $\text{Max}$, which for a specific energy sample $e$ is defined as:

\begin{equation}
\text{Approximation Ratio} = \frac{\text{Max} - e}{\text{Max} - \text{Min}}
\end{equation}

The goal is to get the approximation ratio of the samples to be as close to $1$ as the combinatorial optimization solver can get - an approximation ratio of $1$ means that the sampled solutions are optimal. 

Note that this definition of approximation ratio is consistent with the standard usage of approximation ratio - but it also means that random samples can on average have an approximation ratio of $0.5$. Typically we will report the approximation ratio as the \emph{mean} approximation ratio over a large distribution of samples -- this is computed by taking the mean approximation ratio for each individual sample and then taking the mean over all of the approximation ratios for all of the samples.

\subsection{Compiled Whole-Chip QAOA Circuit Diagram}
\label{section:appendix_QAOA_circuit_drawings}

\begin{figure}[t!]
    \centering
    \includegraphics[width=0.24\textwidth]{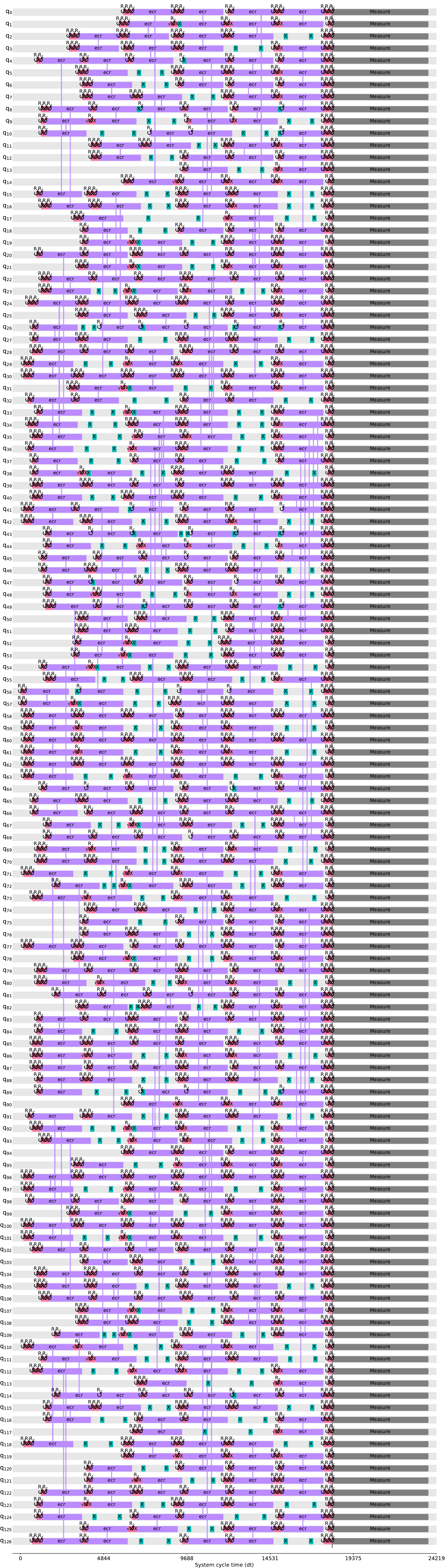}\ %
    \includegraphics[width=0.24\textwidth]{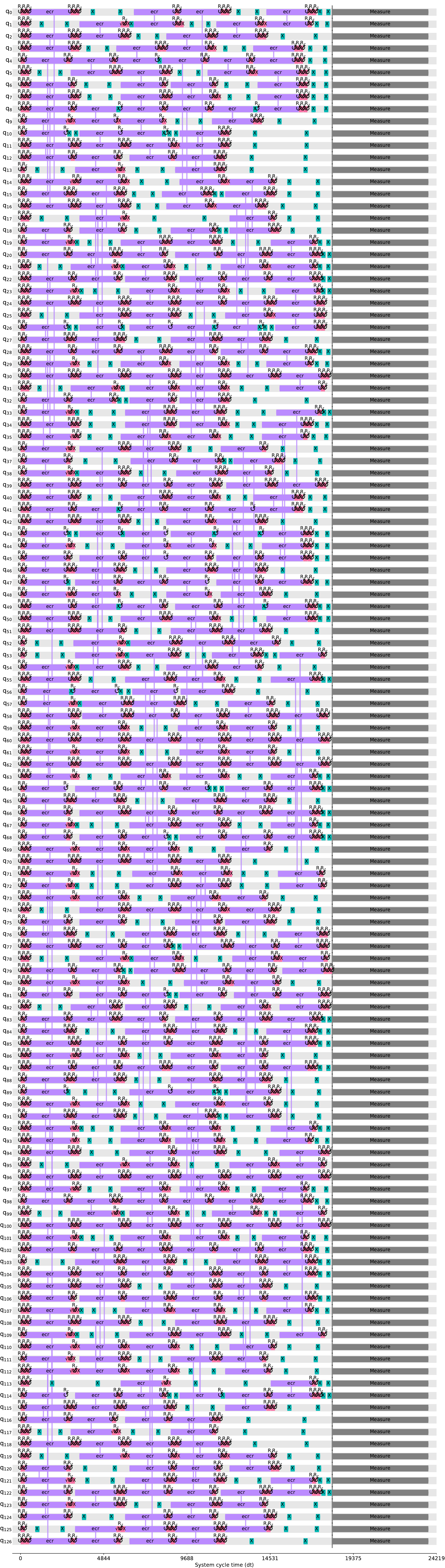}\hfill\hfill%
    \includegraphics[width=0.24\textwidth]{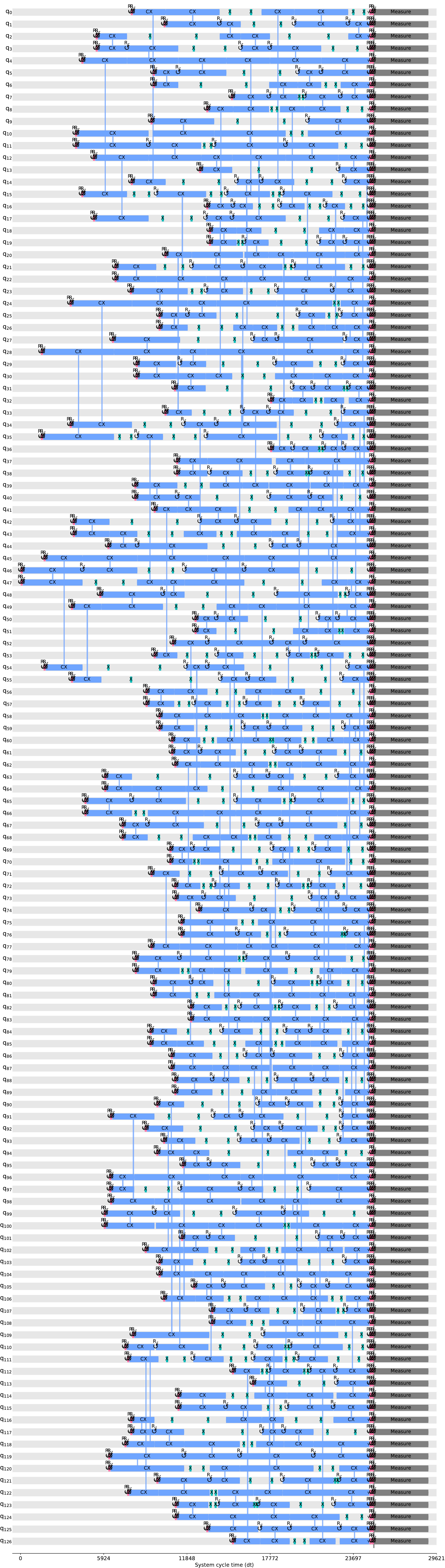}\ %
    \includegraphics[width=0.24\textwidth]{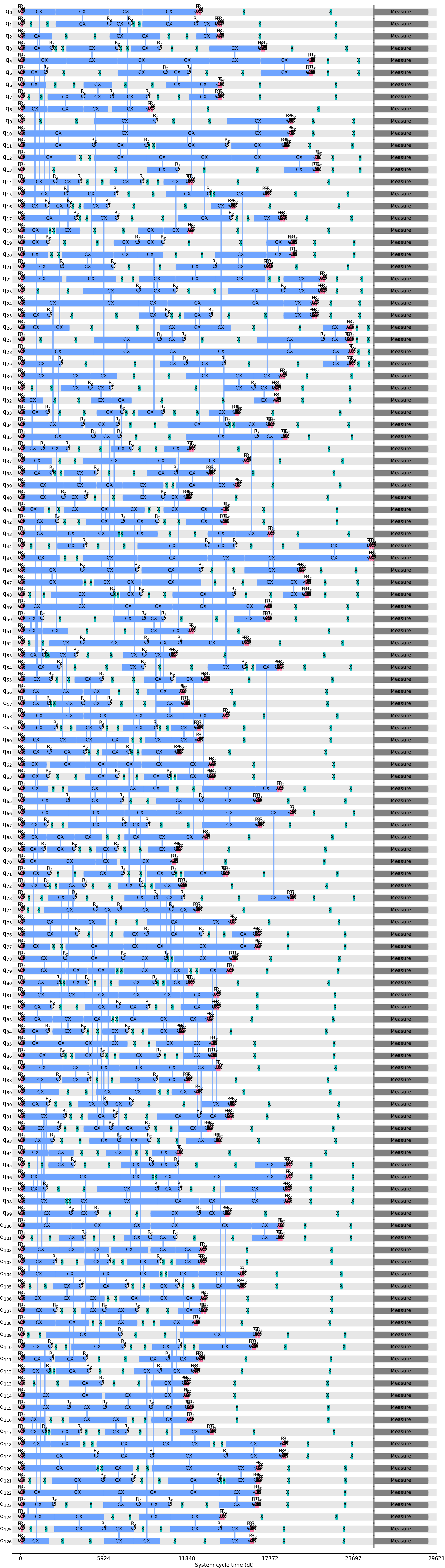}\\%
    \hspace*{0.5cm}\texttt{ibm\_sherbrooke ALAP}\hfill%
    \texttt{ibm\_sherbrooke ASAP}\hfill\hfill%
    \texttt{ibm\_washington ALAP}\hfill%
    \texttt{ibm\_washington ASAP}\hspace*{0.5cm}%
    \caption{$p=1$ timeline QAOA circuit for sampling Ising models that contain higher order terms. Compiled to 
    \textbf{(both left)} \texttt{ibm\_sherbrooke} with native 2-qubit gate~\texttt{ECR} and
    \textbf{(both right)} \texttt{ibm\_washington} with native gate~\texttt{CX}.\newline
    For both devices, we compile with Pauli \texttt{X} gate pair dynamical decoupling passes inserted and scheduled 
    \texttt{ALAP}~\textit{(sub~left)} or \texttt{ASAP}~\textit{(sub~right)}.
    Gate times of \texttt{ECR} are more uniform, resulting in a denser gate scheduling, compared to the \texttt{CX} gate times, whose heterogeneity was not considered in the layered circuit design of Figure~\ref{fig:QAOA_circuit_description}. The ASAP scheduled circuits contain more overall idle qubit time, after a qubit has had at least one gate applied, resulting in more dynamical decoupling sequences being inserted compared to ALAP scheduled circuits. }
    \label{fig:medium_p1_DD_timeline_compiled}
\end{figure}

Figure~\ref{fig:medium_p1_DD_timeline_compiled} shows the compiled and scheduled $127$ qubit whole chip QAOA circuits ($p=1$) with dynamical decoupling sequences inserted, drawn using Qiskit~\cite{Qiskit}. The \texttt{rz} gates are virtual gates~\cite{mckay2017efficient}, meaning they have no error rate, and the \texttt{rz} gates are represented as black circular arrow markers. The \texttt{x} gates are represented as vertical green lines, the \texttt{sx} gates are represented as vertical red lines, and the \texttt{cx} gates (e.g. CNOT gates) are represented by vertical blue lines that connect two qubit lines, and the \texttt{ECR} gates are represented by vertical purple lines that connect two qubit lines. The width of the gate instructions represent the time duration of the gates. The state of all qubits are measured at the end of the circuit, represented by dark grey blocks. The ASAP scheduling inserts more pairs of Pauli X gates compared to the ALAP scheduling. 

The timeline circuit diagrams proceed as a function of time on the x-axis, and this representation of the circuits shows that the \texttt{ibm\_washington} compiled circuits in Figure~\ref{fig:medium_p1_DD_timeline_compiled} used more time per circuit to execute a single circuit compared to the \texttt{ibm\_sherbrooke} compiled circuits.

Note that the \texttt{ibm\_washington} compiled circuits in Figure~\ref{fig:medium_p1_DD_timeline_compiled} correspond to QAOA circuits that sample \texttt{ibm\_washington} topology native Ising models (see Refs.~\cite{pelofske2023qavsqaoa,pelofske2023short}), which means that there are two missing CNOTs in the hardware graph compared to the \texttt{ibm\_sherbrooke} hardware graph.

\begin{figure}[t!]
    \centering
    \includegraphics[width=0.49\textwidth]{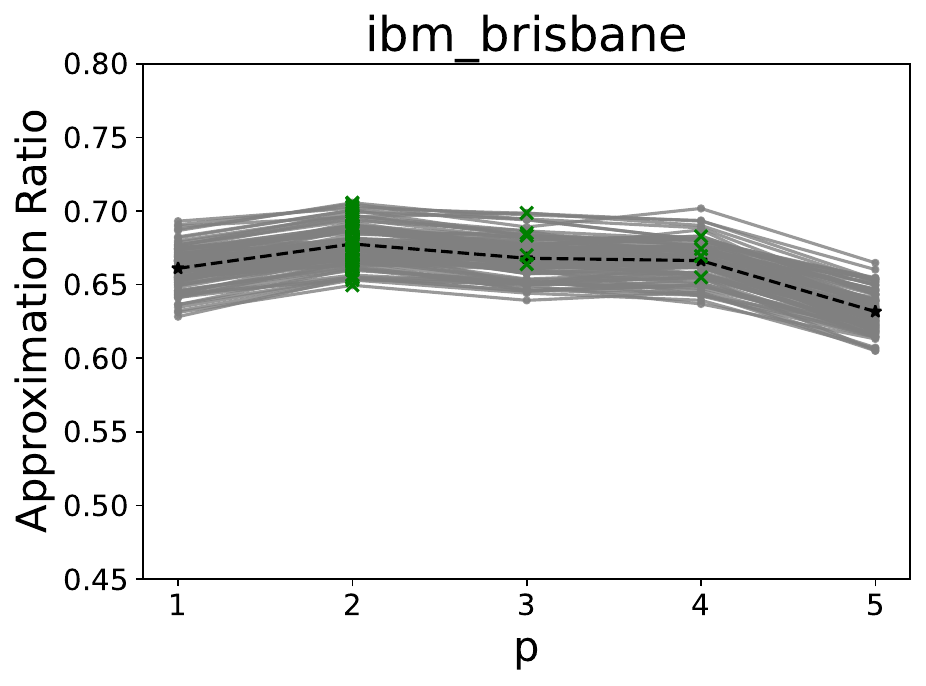}\hfill%
    \includegraphics[width=0.49\textwidth]{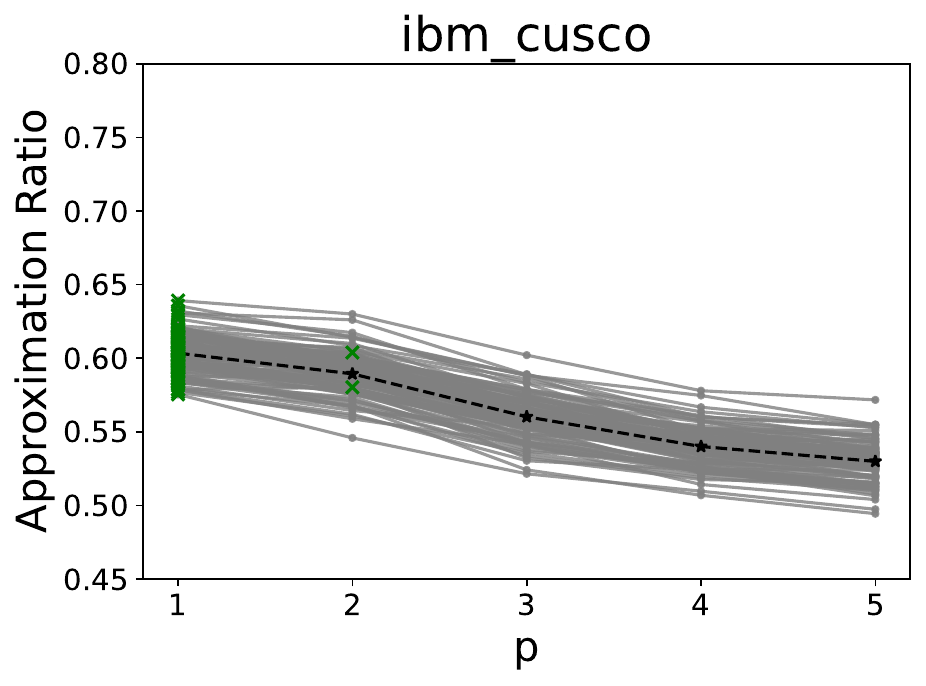}\\[-4ex]%
    \includegraphics[width=0.49\textwidth]{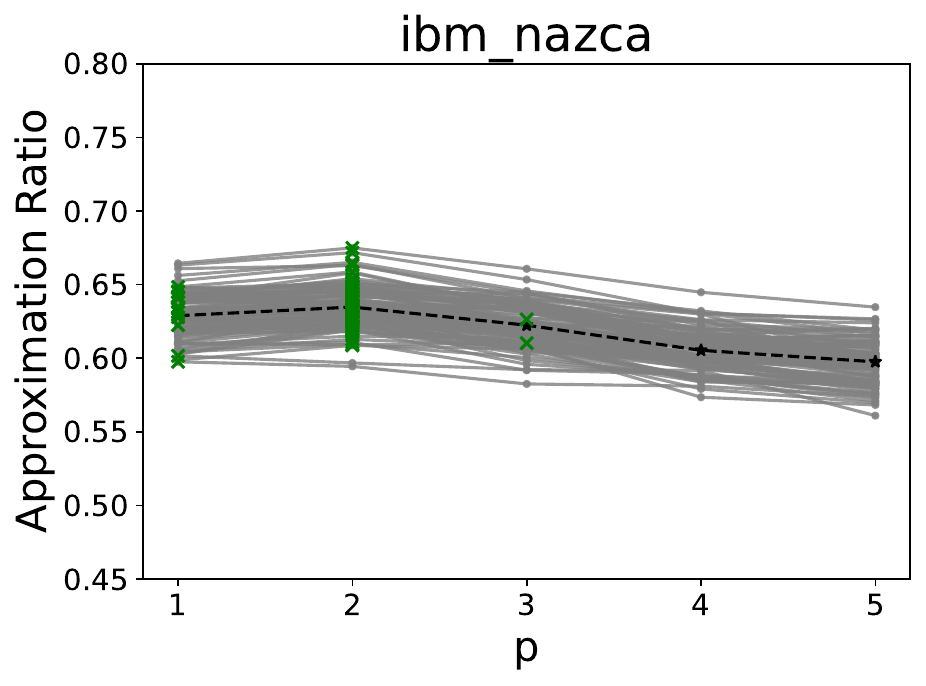}\hfill%
    \includegraphics[width=0.49\textwidth]{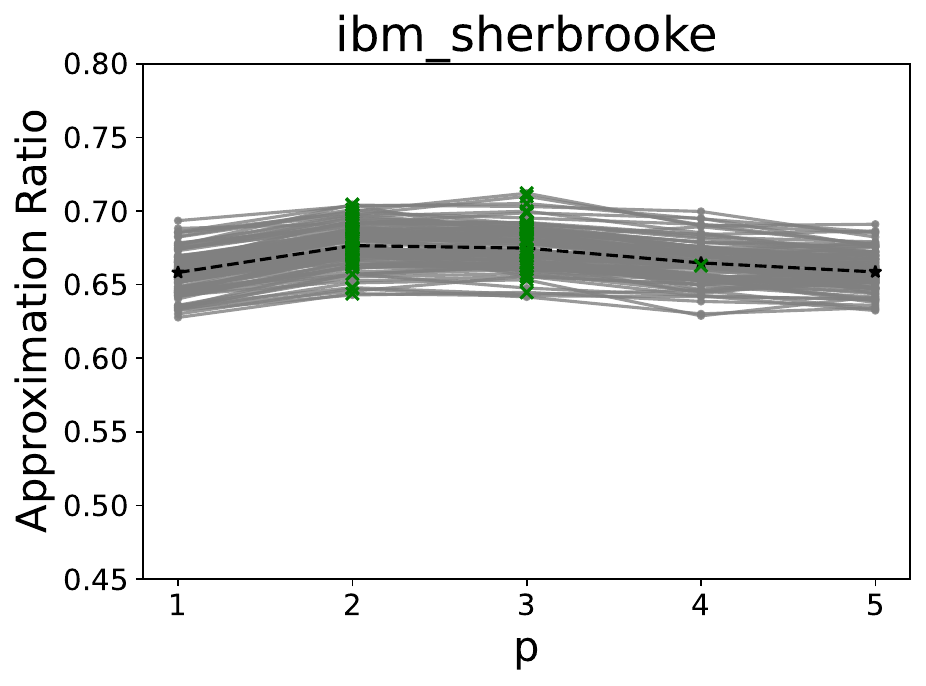}%
    \vspace{-2ex}
    \caption{Mean experimental QAOA approximation ratios of 100 hardware-compatible instances \emph{without dynamical decoupling}:
    Each \textbf{grey} line is an instance shared across devices with experiments run for $p=1,2,3,4,5$. 
    Each \textbf{green} `$\times$' marker gives the~$p$ that achieves lowest mean energy for the corresponding instance.
    The \textbf{black} dashed line shows the average mean QAOA energy across all 100 random spin glass instances with cubic terms.
    Each data point is computed from 20,000 shots on a 127-qubit device. For each $p$, angles are fixed across all devices and instances.
    }
    \label{fig:parameter_transfer_mean_energy_increasing_p_127_qubits}
\end{figure}

\section{Results}
\label{section:results}

Subsection~\ref{section:results_transfer_learning} presents numerical simulations showing that under noiseless conditions QAOA parameter transfer works well and can be applied to significantly larger problem sizes than what was trained on. Subsection~\ref{section:results_scaling_p_16_27_127_qubits} then uses these fixed angles to execute QAOA circuits on a variety of IBM Quantum hardware. Subsection~\ref{section:results_p1_gridsearch} presents a low $p$ comparison between $127$ qubit quantum processors, showing a clear improvement on newer generations of IBM quantum computers. Subsection~\ref{section:results_p1_gridsearch} show $p=1$ QAOA energy landscapes, on whole-chip higher order Ising models, computed on various IBM quantum computers with qubit counts ranging from $27$ qubits up to $414$ qubits showing consistent parameter transfer as the problem sizes increase but the energy landscapes remain relatively unchanged. Subsection~\ref{section:results_CPLEX_compute_time} reports the classical compute time required for CPLEX to optimally solve (minimize) the given combinatorial optimization problem instances.

\begin{figure}[t!]
    \centering
    \includegraphics[width=0.49\textwidth]{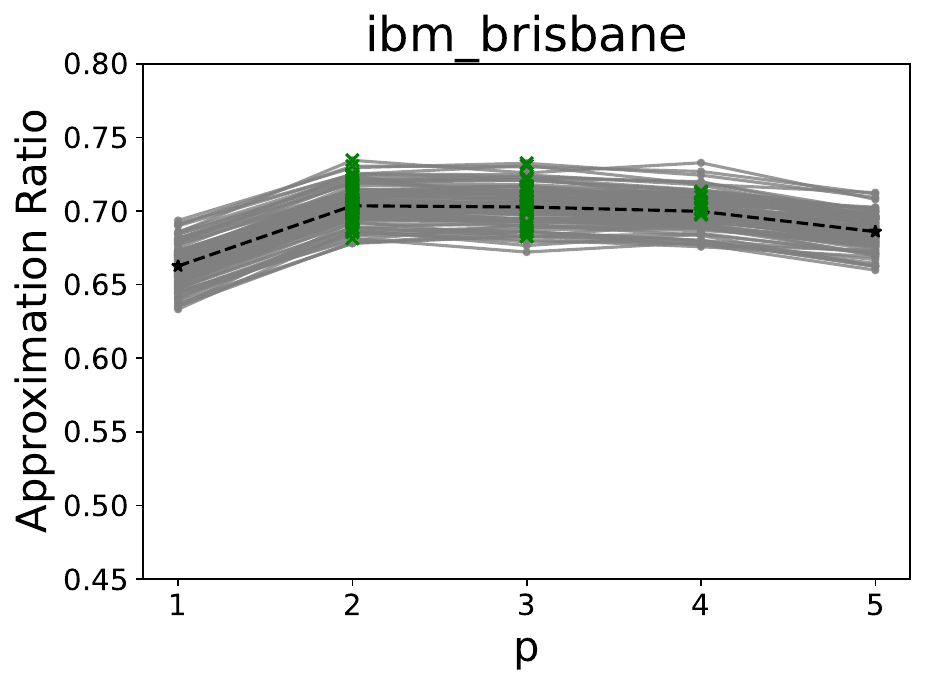}\hfill%
    \includegraphics[width=0.49\textwidth]{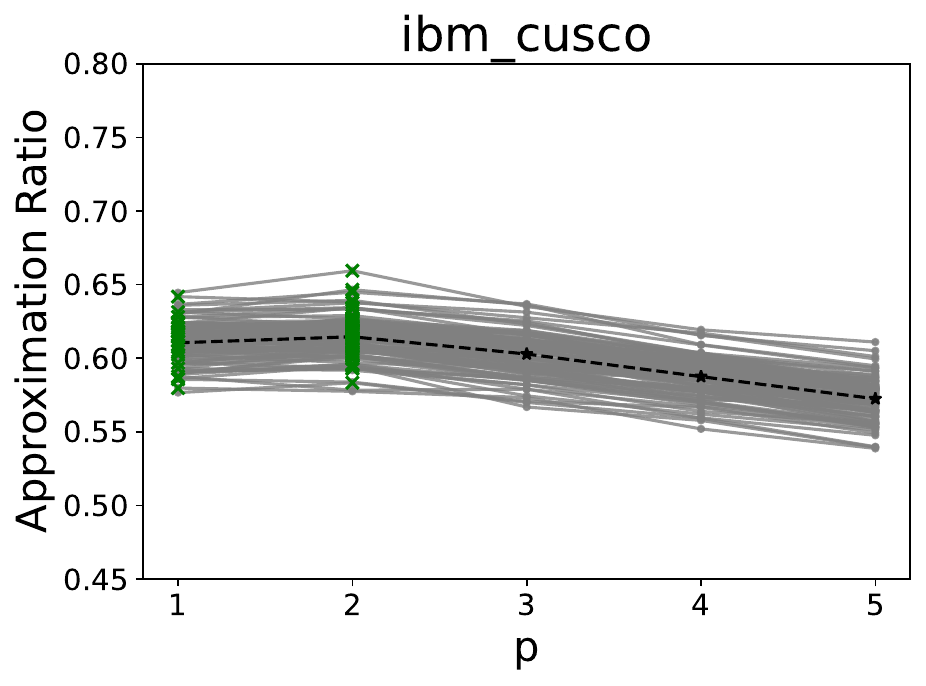}\\[-4ex]%
    \includegraphics[width=0.49\textwidth]{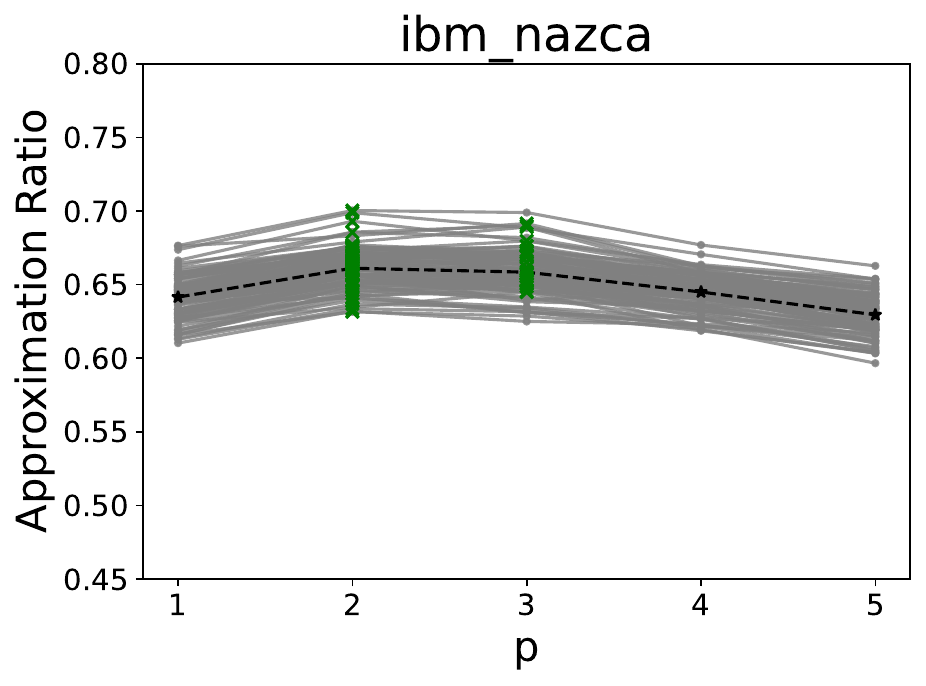}\hfill%
    \includegraphics[width=0.49\textwidth]{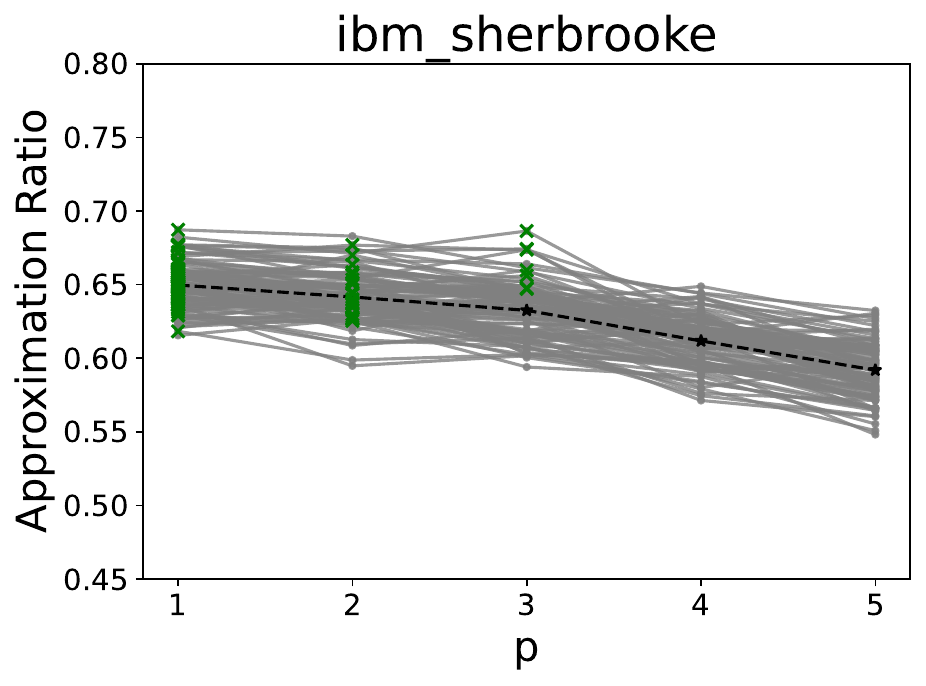}%
    \vspace{-2ex}
    \caption{Mean experimental QAOA approximation ratios of 100 random instances \emph{with ALAP-scheduled dynamical decoupling}:
    Each \textbf{grey} line is an instance shared across devices; 
    each \textbf{green} `$\times$' marker gives its corresponding~$p$ that achieves lowest mean energy.
    The \textbf{black} dashed line shows the average mean QAOA energy across all instances. Cf.~Figure~\ref{fig:parameter_transfer_mean_energy_increasing_p_127_qubits}.
    }
    \label{fig:ALAP_DD_parameter_transfer_mean_energy_increasing_p_127_qubits}
\end{figure}

\subsection{Numerical Simulations of Parameter Transfer of QAOA Angles}
\label{section:results_transfer_learning}

As introduced in Subsection~\ref{section:methods_QAOA_angle_finding}, parameter concentration is the following property of a QAOA problem: QAOA parameter (angle) values that are optimized for an instance $I$ of a particular combinatorial optimization problem (such as random spin glasses or Maximum Cut) are transferable to other instances of similar structure, but potentially of significantly different size from the original $I$. Parameters from an instance $I$ are transferable to an instance $I'$ if the quality of the solutions found by fixed QAOA angles are similar for both $I$ and $I'$. While more formal definitions of transferability are possible, we pragmatically define that parameters transfer from $I$ to $I'$ up to a maximum number of rounds $p_{\max}$ if the mean solution quality for both $I$ and $I'$ improves with increasing number of rounds $p$ up to and including $p_{\max}$. 
Figure~\ref{fig:parameter_transfer_scaling_plots} presents the numerical simulation results for the scaling of increasing $p$ QAOA using the fixed parameter transfer angles on $100$ random ensembles for $16$, $27$, and $127$ qubit instances, using the methods described in Subsection~\ref{section:methods_QAOA_angle_finding}. Specifically, the same $\vec{\beta}, \vec{\gamma}$ (for each $p$) are used for all numerical simulations in these plots (Section~\ref{section:methods_QAOA_angle_finding} explicitly gives what these fixed angles are). The $16$ and $27$ qubit data is the mean energy taken from $10,000$ samples per circuit with no noise model, simulated classically using Qiskit~\cite{Qiskit}. Simulations of 127-qubit system are performed with MPS. Here we quote expectation values of $H_C$ computed by direct tensor contraction. That computation is equivalent to the limit of an infinite number of shots. Figure~\ref{fig:parameter_transfer_scaling_plots} shows that the parameter transfer succeeded, and in particular allows us to obtain good angles for up to $p=5$, verified by classical MPS simulations. Figure~\ref{fig:MPS_errors} studies the errors in MPS simulations for all $100$ random $127$ qubit hardware-compatible instances, as a function of $\chi$, including the largest QAOA circuit depth we tested (which is $p=5$). Figure~\ref{fig:MPS_samples} shows distributions of samples for the QAOA circuits, computed using the MPS simulation method (with a bond dimension of $\chi=2048$), which shows what the expected performance of QAOA is under noiseless conditions, for a subset of the $127$ variable problem instances.

\begin{figure}[t!]
    \centering
    \includegraphics[width=0.49\textwidth]{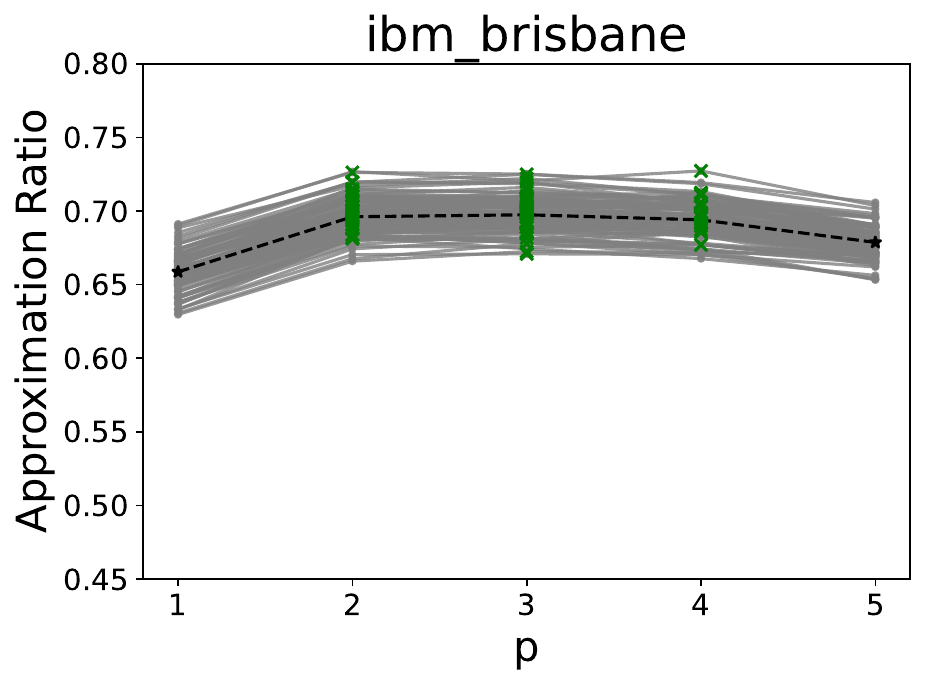}
    \includegraphics[width=0.49\textwidth]{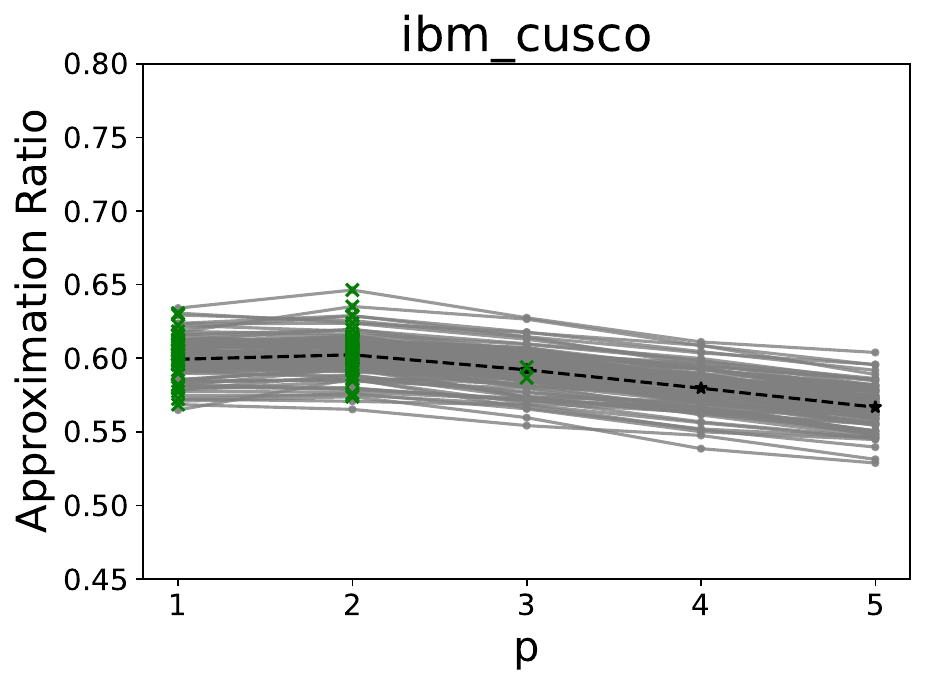}\\
    \includegraphics[width=0.49\textwidth]{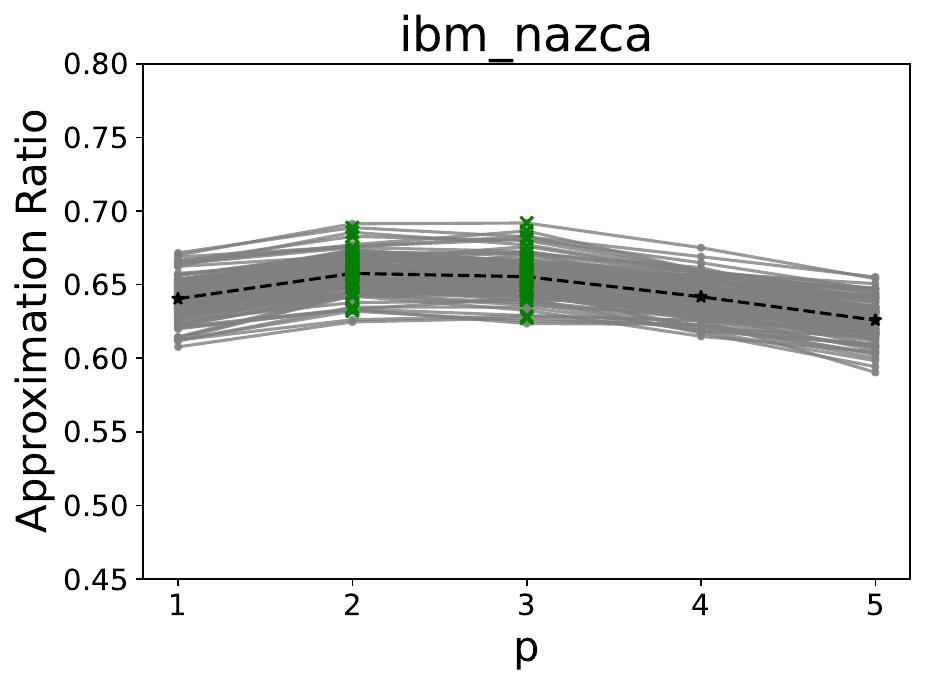}\hfill%
    \includegraphics[width=0.49\textwidth]{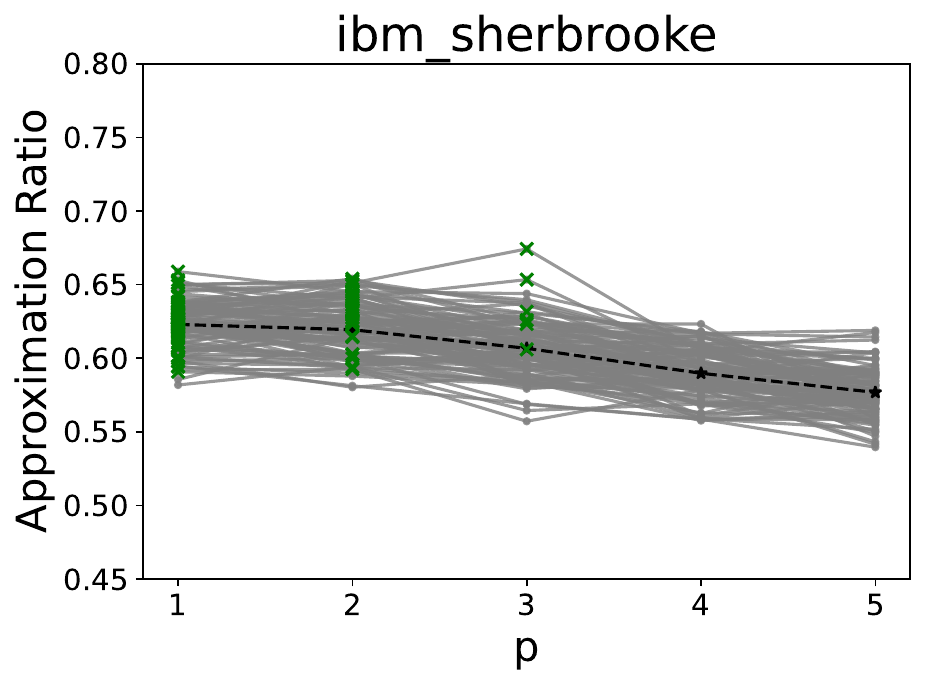}%
    \vspace{-2ex}
    \caption{Mean experimental QAOA energies of 100 random instances \emph{with ASAP-scheduled dynamical decoupling}:
    Each \textbf{grey} line is an instance shared across devices; 
    each \textbf{green} `$\times$' marker gives its corresponding~$p$ that achieves lowest mean energy.
    The \textbf{black} dashed line shows the average mean QAOA energy across all instances. Cf.~Figure~\ref{fig:parameter_transfer_mean_energy_increasing_p_127_qubits}.
    }
    \label{fig:ASAP_DD_parameter_transfer_mean_energy_increasing_p_127_qubits}
\end{figure}

\subsection[Scaling p on 16, 27, and 127 qubit IBM Quantum Processor Hardware]{Scaling $p$ on 16, 27, and 127 qubit IBM Quantum Processor Hardware}
\label{section:results_scaling_p_16_27_127_qubits}

The results presented in this section are reported as the mean energy of the samples of the problem Ising models, from a total of $20,000$ shots per parameter and device. 
The plots in this section use the angles learned from a $16$ qubit instance, giving good approximation ratios as $p$ increases for the ideal computation. Figure~\ref{fig:parameter_transfer_scaling_plots} in Subsection~\ref{section:results_transfer_learning} shows the scaling in $p$ under noiseless conditions obtained with these angles. In particular, these numerical simulations show that in the noiseless setting we would get improving energy for each step of $p$. In this section, we execute the whole-chip QAOA circuits on various IBM Quantum computers, specifically using the fixed angles discussed in Subsection~\ref{section:results_transfer_learning} for $p=1$ up to $p=5$. This is therefore an evaluation of how well the transfer-learned angles perform on the heavy-hex graph hardware. 

\begin{figure}[t!]
    \centering
    \includegraphics[width=0.31\textwidth,trim={0 45 0 0},clip]{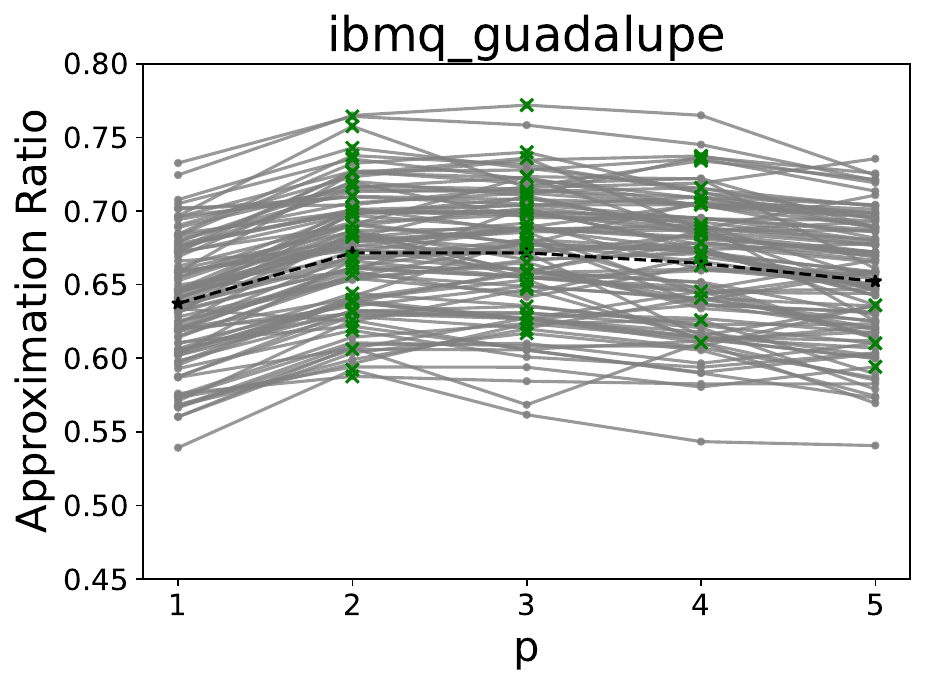}\\%
    \includegraphics[width=0.32\textwidth,trim={0 45 0 0},clip]{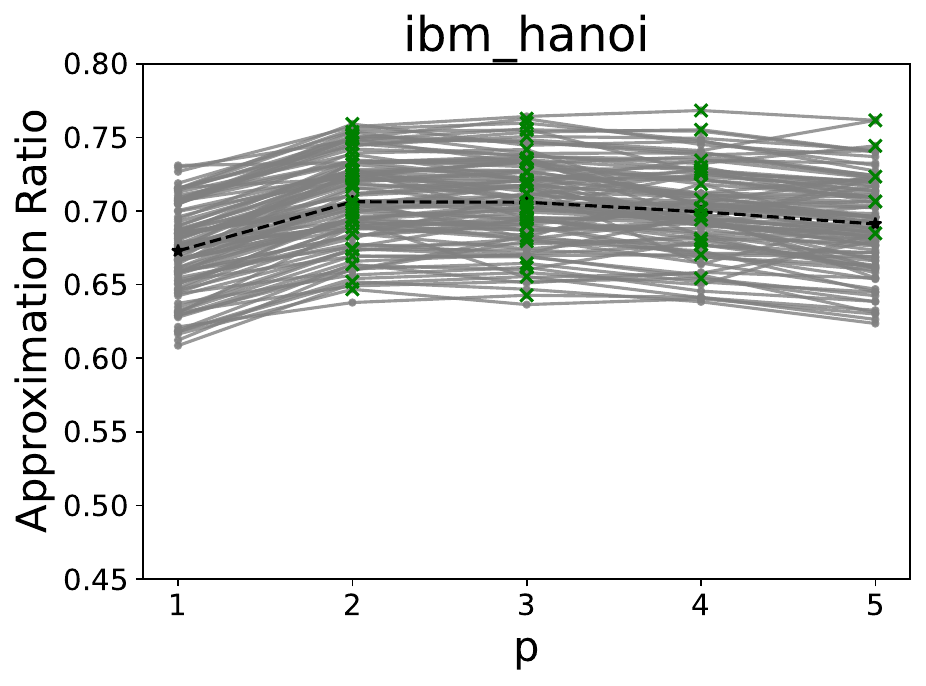}\hfill%
    \includegraphics[width=0.32\textwidth,trim={0 45 0 0},clip]{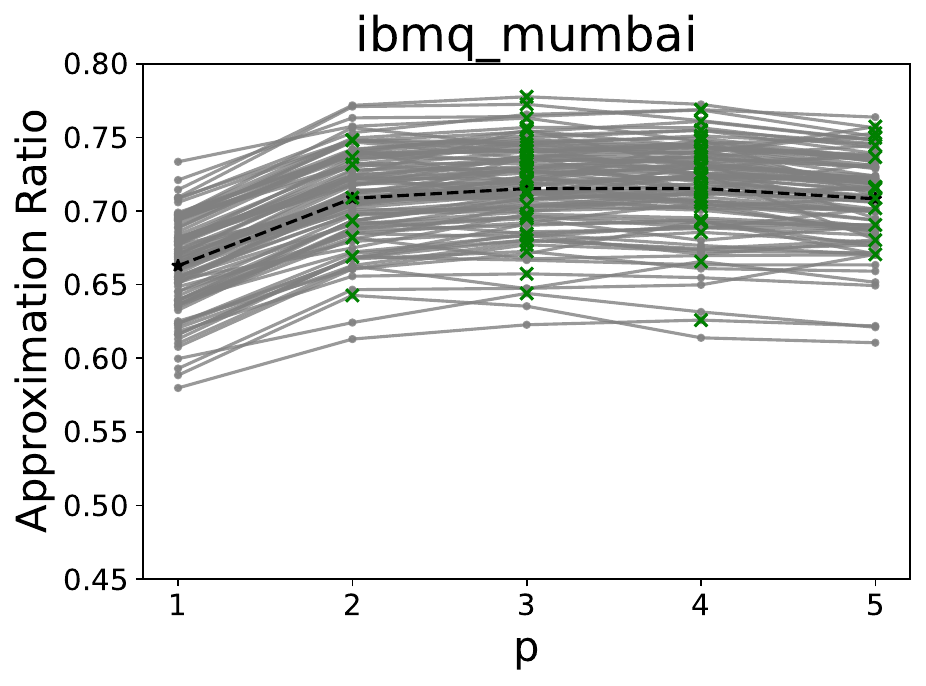}\hfill%
    \includegraphics[width=0.32\textwidth,trim={0 45 0 0},clip]{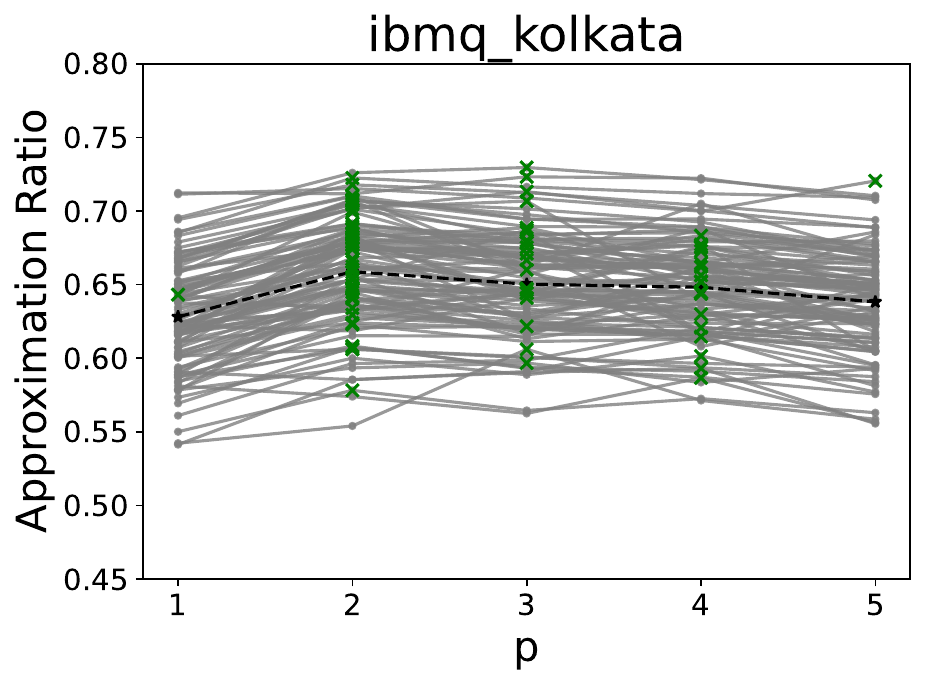}\\%
    \includegraphics[width=0.32\textwidth]{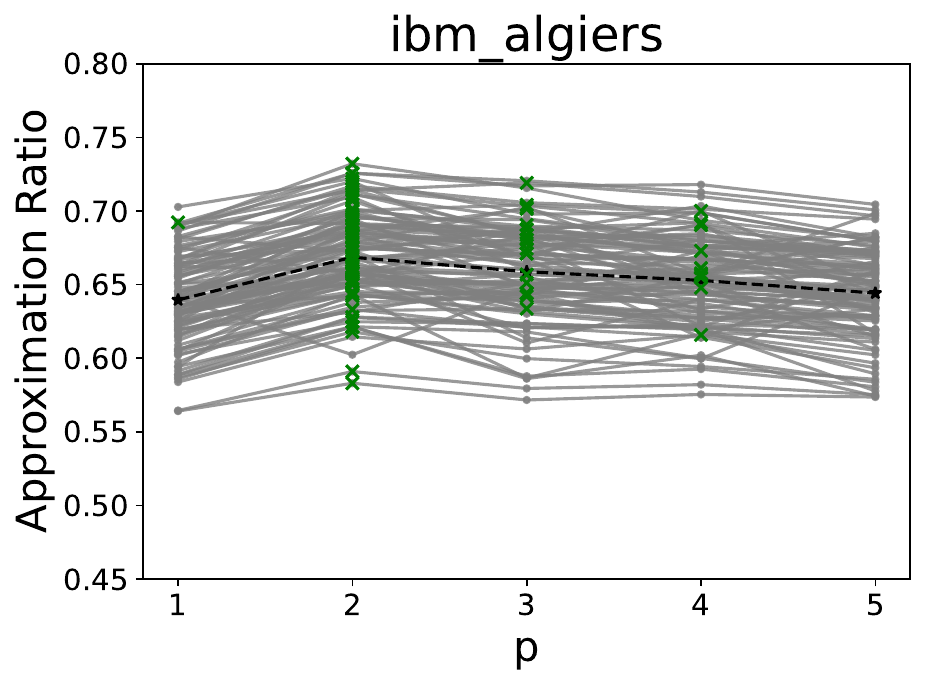}\hfill%
    \includegraphics[width=0.32\textwidth]{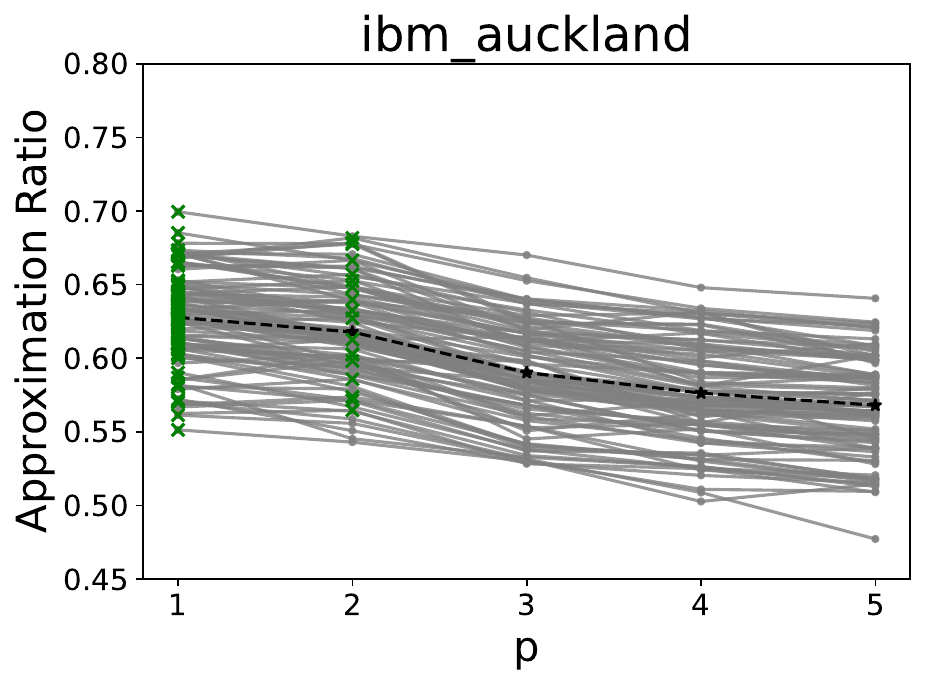}\hfill%
    \includegraphics[width=0.32\textwidth]{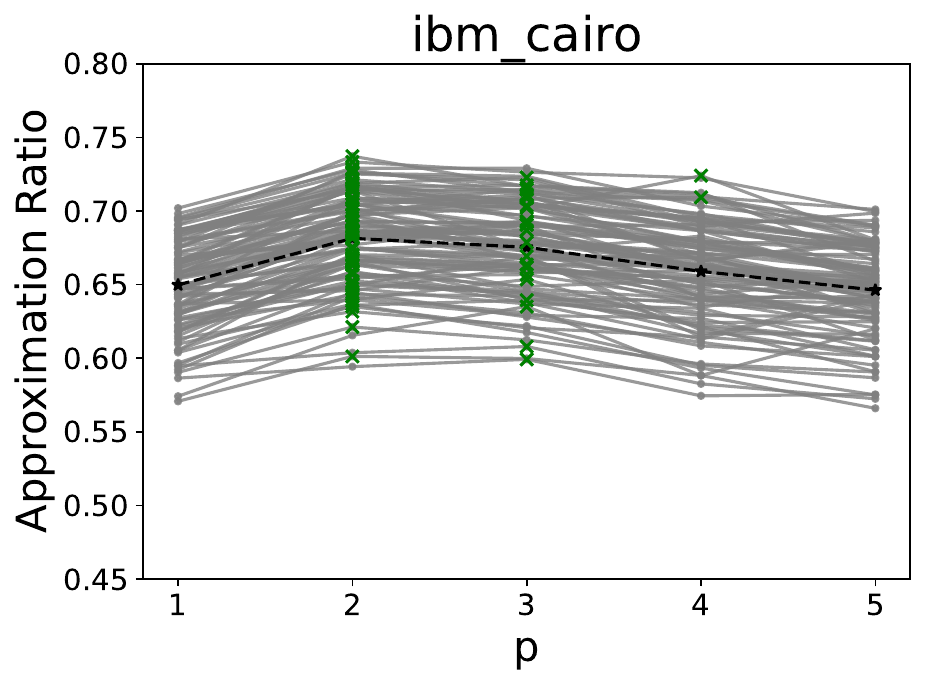}%
    \vspace*{-2ex} 
    \caption{Mean approximation ratio of the samples measured from the whole-chip QAOA circuits, \emph{without dynamical decoupling}:
    Each \textbf{grey} line is an instance run for $1\leq p \leq 5$, with its \textbf{green} `$\times$' marker the~$p$ that achieves lowest mean energy.
    The \textbf{black} dashed line shows the average mean QAOA energy taken across all 100 random higher-order instances.
    Each data point is computed from 20,000 shots. The same $100$ problem instances were executed on a total of six $27$-qubit devices, and $100$ $16$-qubit instances were executed on \texttt{ibmq\_guadalupe}; with angles shared for any given~$p$.
    }
    \label{fig:parameter_transfer_mean_energy_increasing_p_16_27_qubits}
\end{figure}

\begin{figure}[t!]
    \centering
    \includegraphics[width=0.31\textwidth,trim={0 45 0 0},clip]{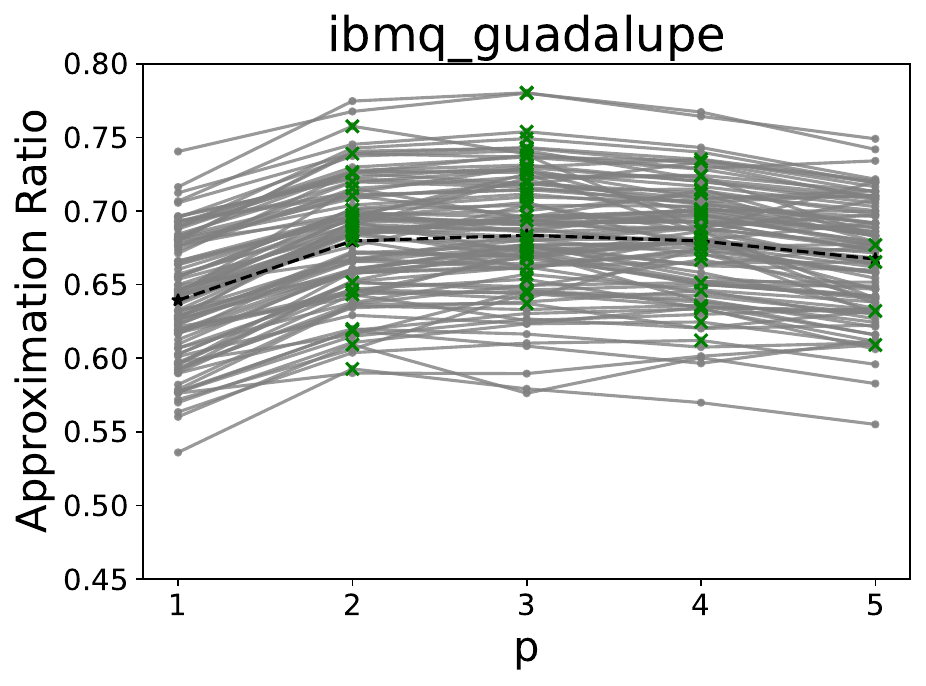}\\%
    \includegraphics[width=0.32\textwidth,trim={0 45 0 0},clip]{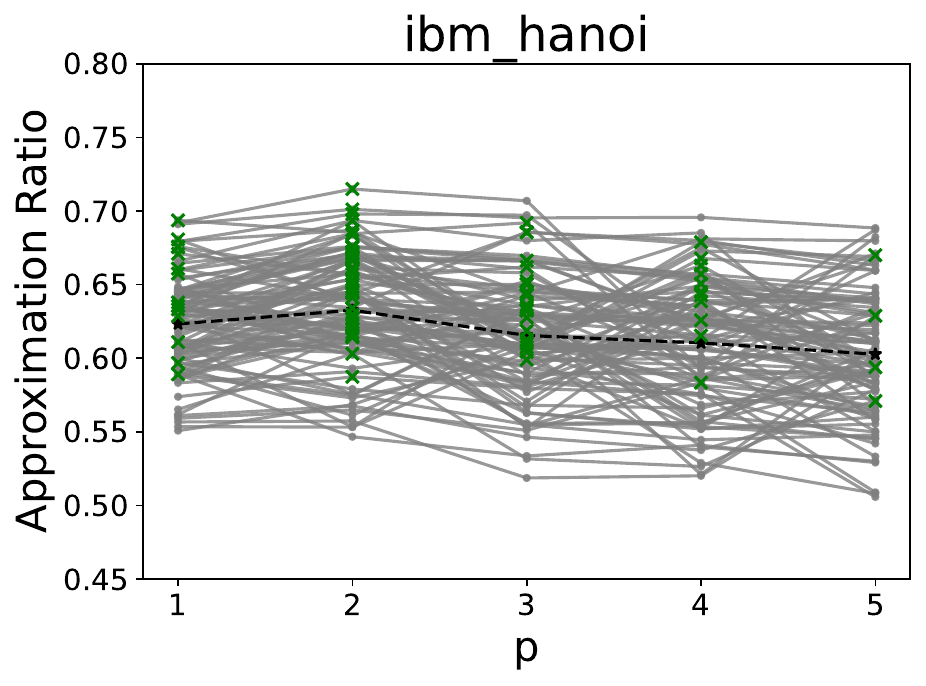}\hfill%
    \includegraphics[width=0.32\textwidth,trim={0 45 0 0},clip]{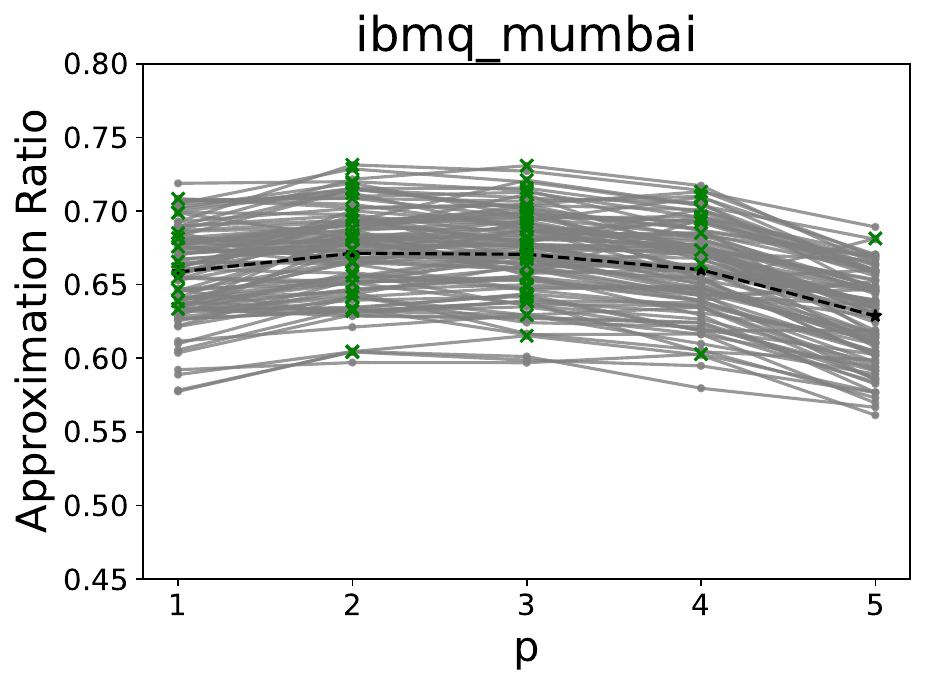}\hfill%
    \includegraphics[width=0.32\textwidth,trim={0 45 0 0},clip]{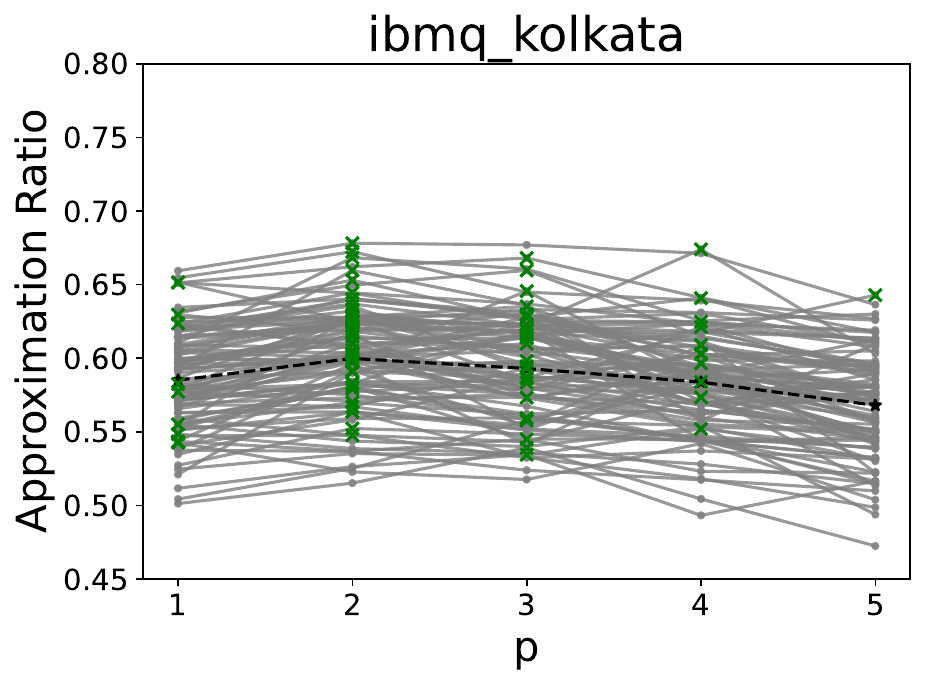}\\%
    \includegraphics[width=0.32\textwidth]{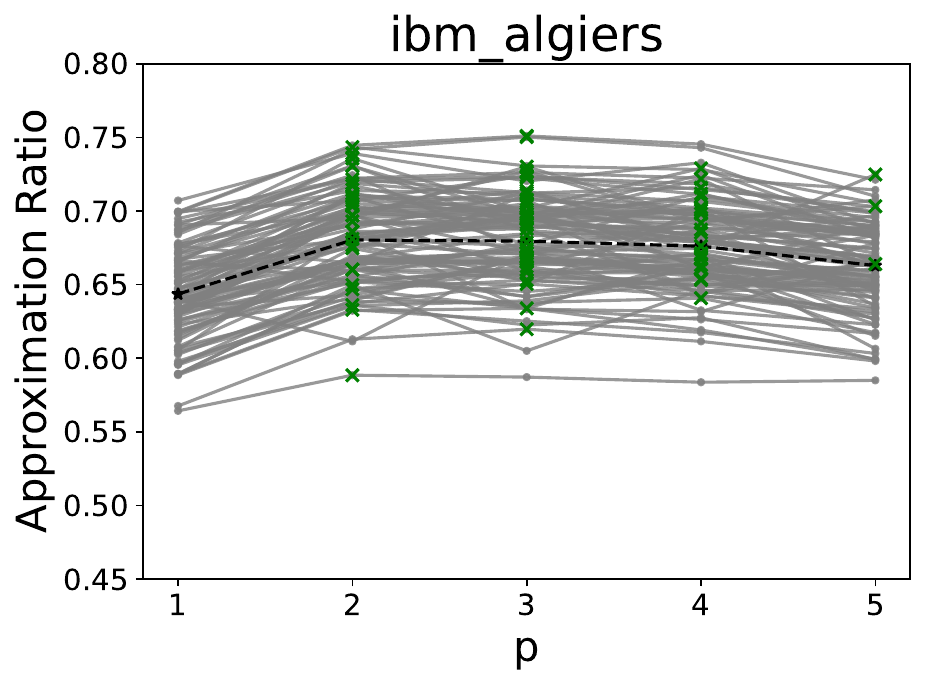}\hfill%
    \includegraphics[width=0.32\textwidth]{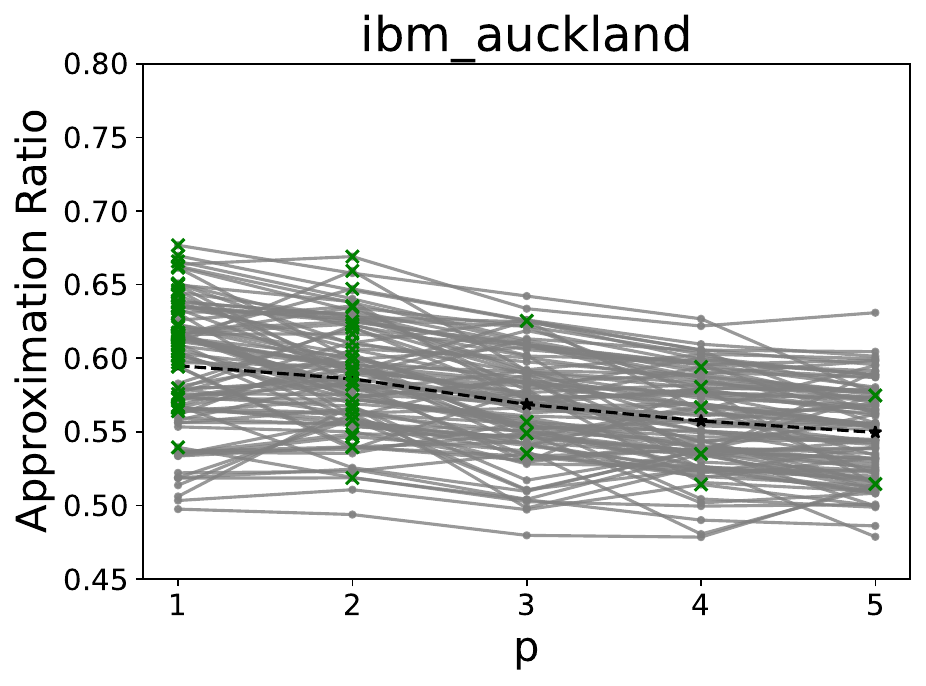}\hfill%
    \includegraphics[width=0.32\textwidth]{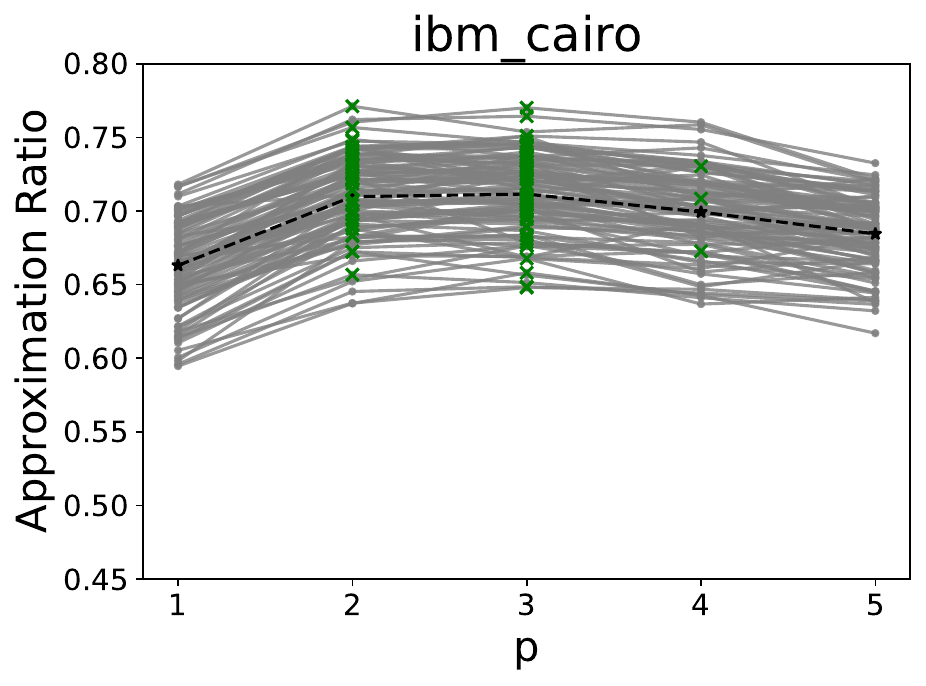}%
    \vspace*{-2ex}
    \caption{Mean approximation ratio of the samples from the QAOA circuits run with \emph{ALAP-scheduled dynamical decoupling}:
    Each \textbf{grey} line is an instance run for $1\leq p \leq 5$, with its \textbf{green} `$\times$' marker the~$p$ that achieves lowest mean energy.
    The \textbf{black} dashed line shows the average mean QAOA energy taken across all 100 random instances. Cf.~Figure~\ref{fig:parameter_transfer_mean_energy_increasing_p_16_27_qubits}.
    }
    \label{fig:ALAP_DD_parameter_transfer_mean_energy_increasing_p_16_27_qubits}
\end{figure}

\begin{figure}[t!]
    \centering
    \includegraphics[width=0.31\textwidth,trim={0 45 0 0},clip]{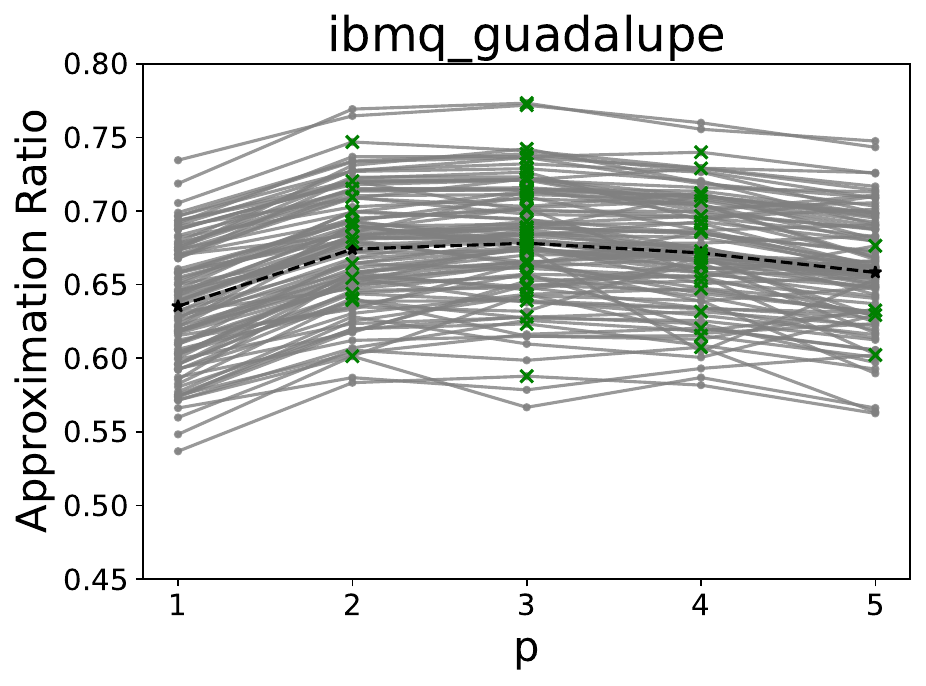}\\%
    \includegraphics[width=0.32\textwidth,trim={0 45 0 0},clip]{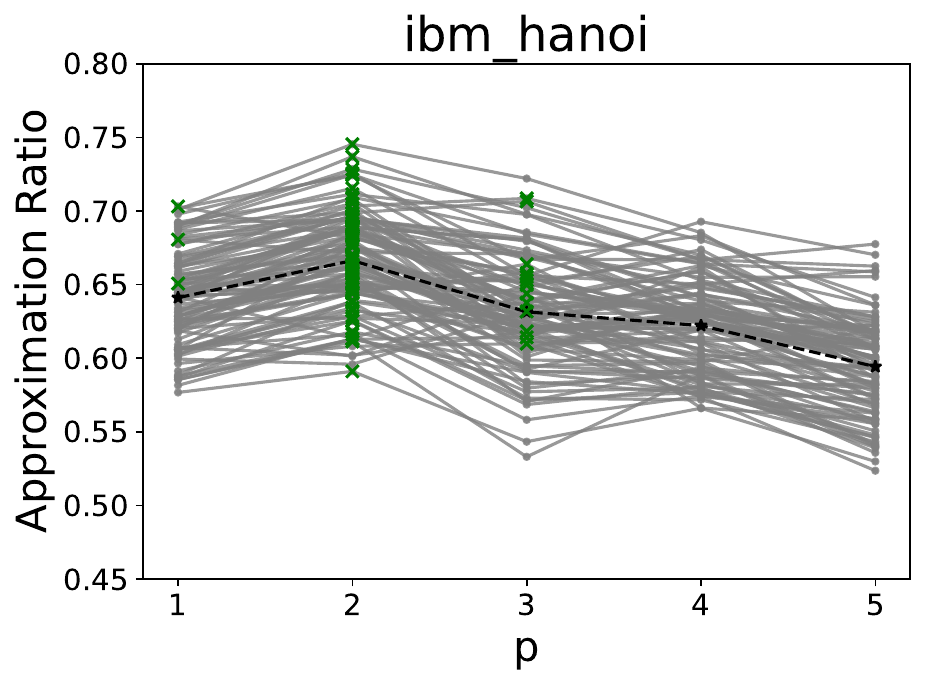}\hfill%
    \includegraphics[width=0.32\textwidth,trim={0 45 0 0},clip]{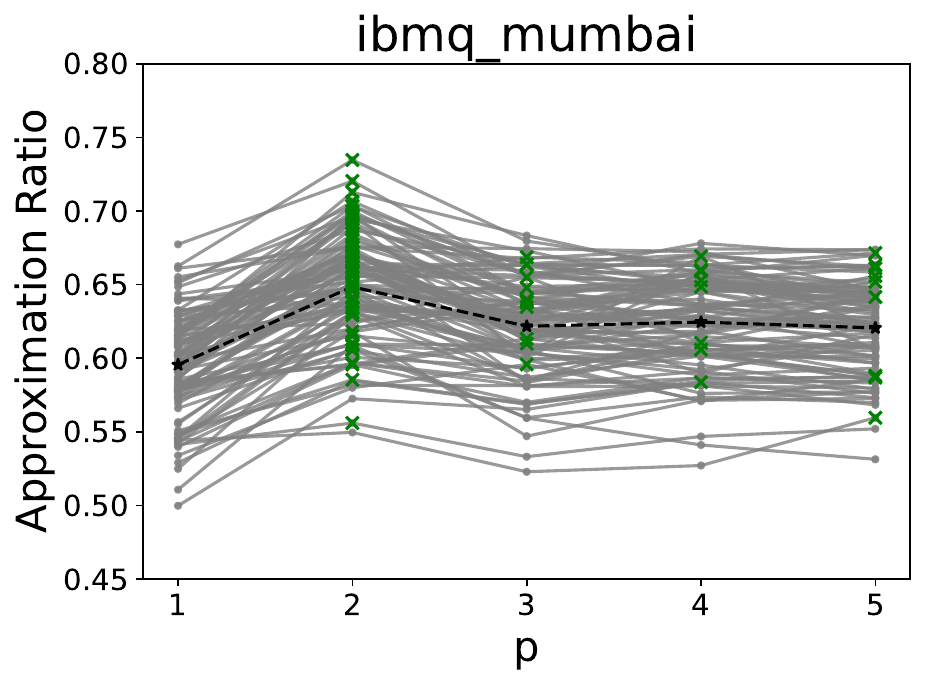}\hfill%
    \includegraphics[width=0.32\textwidth,trim={0 45 0 0},clip]{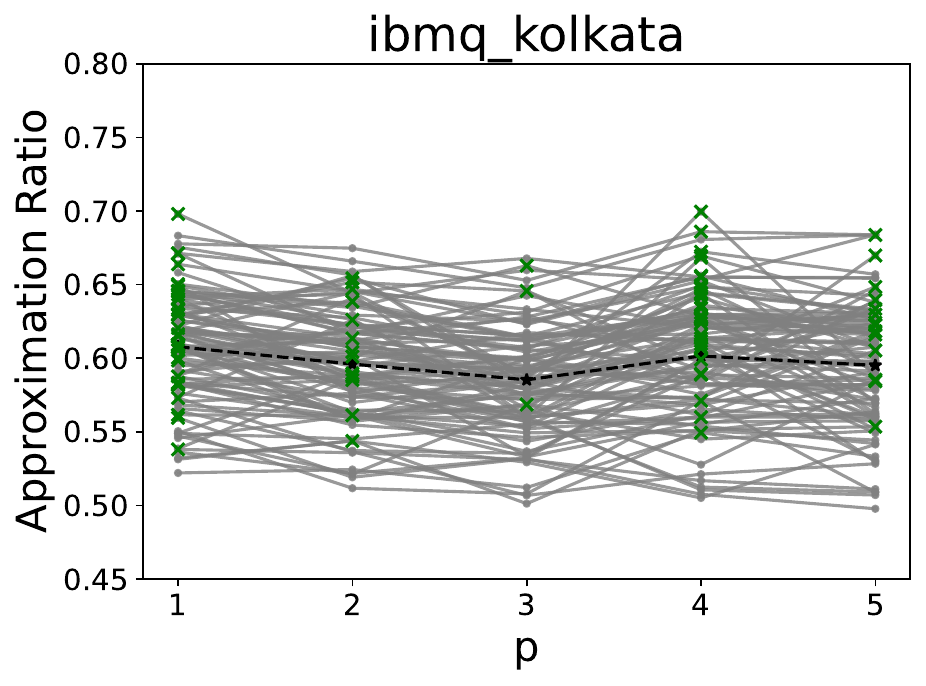}\\%
    \includegraphics[width=0.32\textwidth]{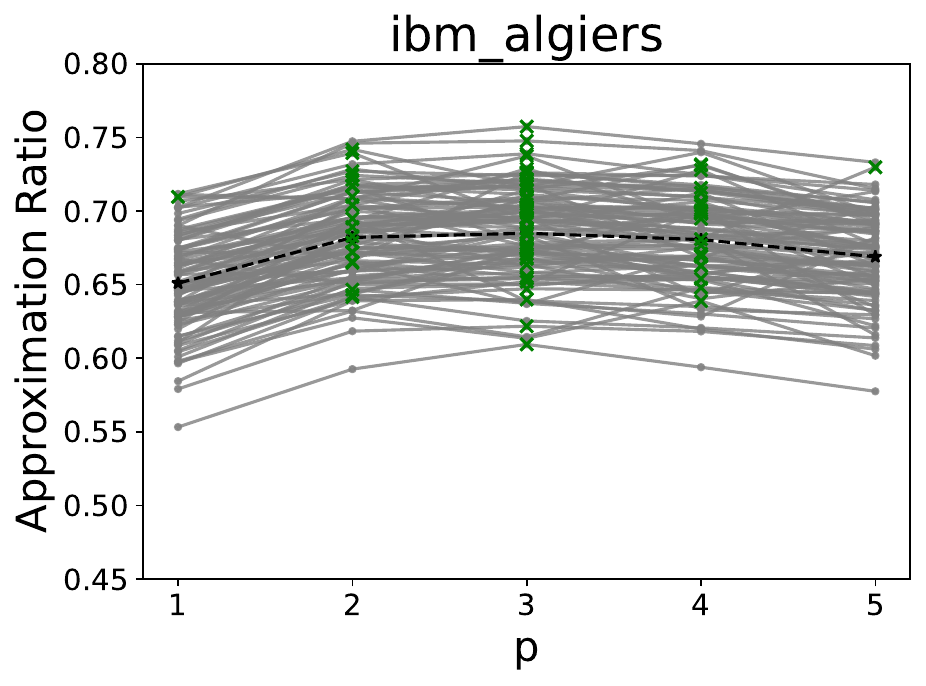}\hfill%
    \includegraphics[width=0.32\textwidth]{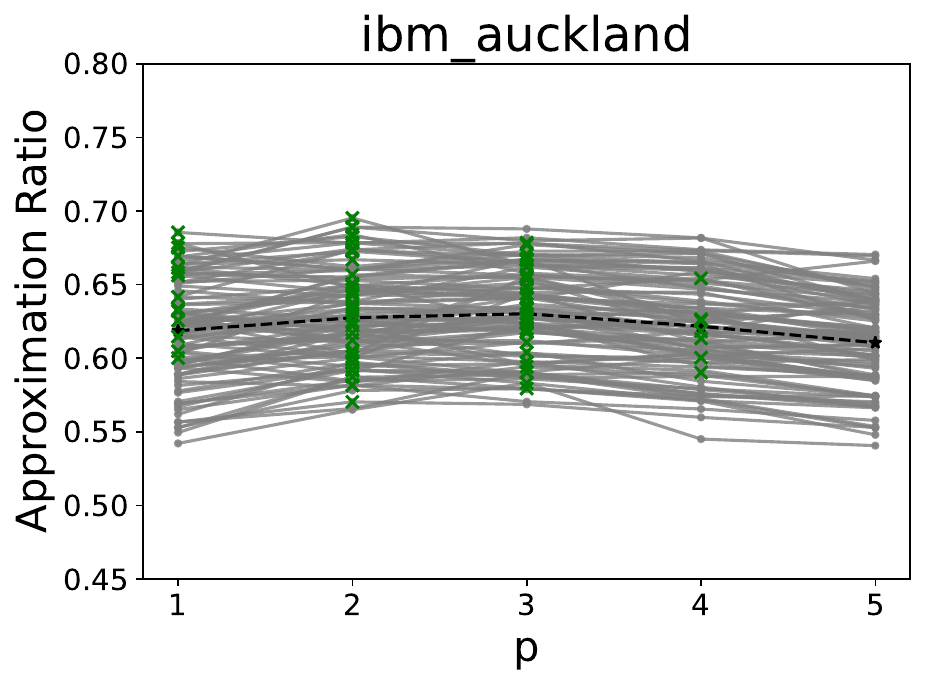}\hfill%
    \includegraphics[width=0.32\textwidth]{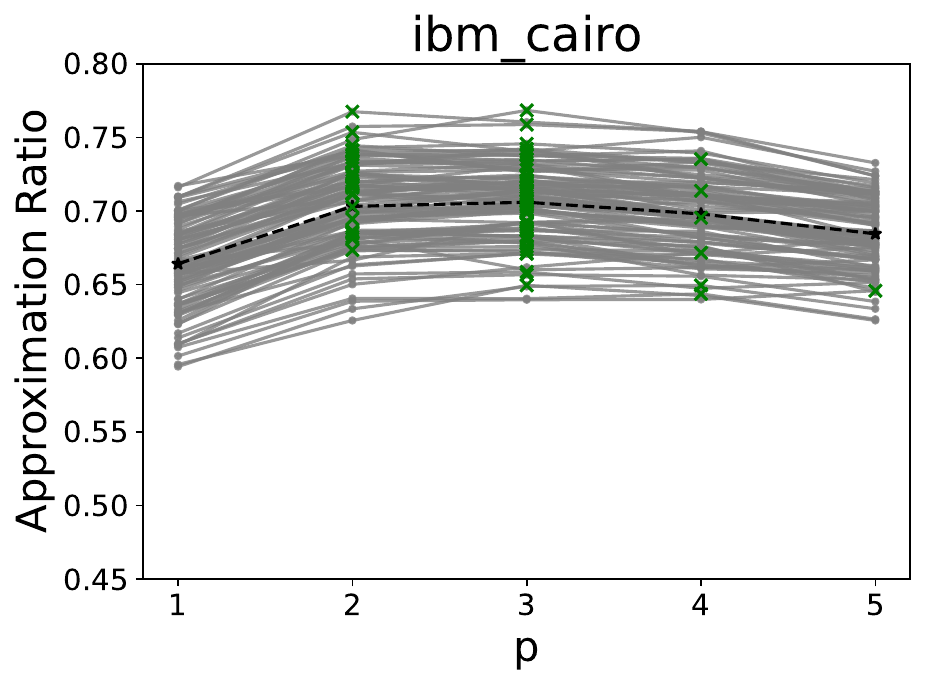}%
    \vspace*{-2ex}
    \caption{Mean approximation ratio of the samples from the QAOA circuits run with \emph{ASAP-scheduled dynamical decoupling}:
    Each \textbf{grey} line is an instance run for $1\leq p \leq 5$, with its \textbf{green} `$\times$' marker the~$p$ that achieves lowest mean energy.
    The \textbf{black} dashed line shows the average mean QAOA energy taken across all 100 random instances. Cf.~Figure~\ref{fig:parameter_transfer_mean_energy_increasing_p_16_27_qubits}.
    }
    \label{fig:ASAP_DD_parameter_transfer_mean_energy_increasing_p_16_27_qubits}
\end{figure}

The bare QAOA circuits results are plotted in Figure~\ref{fig:parameter_transfer_mean_energy_increasing_p_127_qubits} for four 127 qubit backends and Figure~\ref{fig:parameter_transfer_mean_energy_increasing_p_16_27_qubits} for six 27 qubit devices and a single 16 qubit device (\texttt{ibmq\_guadalupe}). Recall that without noise, these figures (if represented in terms of Hamiltonian energy, instead of approximation ratio) would look identical to the corresponding energy plots from Figure~\ref{fig:parameter_transfer_scaling_plots}. Figures~\ref{fig:ALAP_DD_parameter_transfer_mean_energy_increasing_p_127_qubits} and~\ref{fig:ALAP_DD_parameter_transfer_mean_energy_increasing_p_16_27_qubits} show the hardware-executed mean energy for the QAOA circuits using ALAP-scheduled dynamical decoupling QAOA circuits for the 127 and 27 qubit systems respectively. Figures~\ref{fig:ASAP_DD_parameter_transfer_mean_energy_increasing_p_127_qubits} and~\ref{fig:ASAP_DD_parameter_transfer_mean_energy_increasing_p_16_27_qubits} show the same, but with ASAP-scheduled digital dynamical decoupling sequences.

For the 127 qubit device from Figure~\ref{fig:parameter_transfer_mean_energy_increasing_p_127_qubits}, we see that NISQ reality does indeed look different: 
As a first observation, three out of four quantum processors (\texttt{ibm\_brisbane},\texttt{ibm\_nazca},\texttt{ibm\_sherbrooke})at least improve the mean approximation ratio as averaged over the 100 instances until $p=2$, as indicated by the black dashed line, but fail to improve for higher $p$, due to noise. The green crosses indicate for each instances the $p$ at which the minimum energy (maximum approximation ratio) was achieved; we see that some instances are sampled best at $p=3,4$, or also $p=1$ for a few of the instances. The remaining backend \texttt{ibm\_cusco} performs best at $p=1$. 
Secondly, despite all backends featuring an Eagle r3 QPU, performance differences are significant with \texttt{ibm\_brisbane} achieving best average approximation ratios of almost $0.7$ and \texttt{ibm\_cusco} only achieving about $0.6$. Overall, these differences are consistent with the reported two-qubit gate fidelities for these devices. 

In a fourth observation, we look at the two corresponding 127 qubit plots with digital dynamical decoupling sequences, i.e., Figure~\ref{fig:ALAP_DD_parameter_transfer_mean_energy_increasing_p_127_qubits} for ALAP, and Figure~\ref{fig:ASAP_DD_parameter_transfer_mean_energy_increasing_p_127_qubits} for ASAP. Overall, the two different scheduling schemes seem to perform similar. However, both ALAP and ASAP have a positive effect on the performance of three of the quantum processors: \texttt{ibm\_brisbane} improves to approximately $0.72$ values and almost achieves a maximum approximation ratio at $p=3$ instead of at $p=2$, but not quite. The two lower performing devices \texttt{ibm\_cusco} and \texttt{ibm\_nazca} also see significant improvements. Strikingly, however, \texttt{ibm\_sherbrooke}'s performance takes a significant hit with both ALAP and ASAP scheduled dynamical decoupling. 

Our observations are similar for the 16 and 27 qubit systems from Figure~\ref{fig:parameter_transfer_mean_energy_increasing_p_16_27_qubits}: While most backends still have their average minimum performance at $p=2$, in most cases many instances find their minimum energy (maximum approximation ratio) at $p=3$. \texttt{ibmq\_mumbai} is a notable exception as it reaches the minimum average at $p=3$ and in fact remains nearly flat even to $p=4$. Secondly, we again see performance difference among the 27 qubit systems ranging from a mean across the ensemble of $100$ instances of $0.70$ for \texttt{ibmq\_mumbai} to a $0.60$ mean value for \texttt{ibm\_auckland}, which actually achieves its maximum at $p=1$. 

Our fourth observation with respect to dynamical decoupling for the 27 qubit backends (see Figure~\ref{fig:ASAP_DD_parameter_transfer_mean_energy_increasing_p_16_27_qubits} and Figure ~\ref{fig:ALAP_DD_parameter_transfer_mean_energy_increasing_p_16_27_qubits}) is less optimistic than for the 127 qubit count: dynamical decoupling only helps two out of six backends, namely \texttt{ibm\_algiers} and \texttt{ibm\_cairo}, which actually matches \texttt{ibmq\_mumbai}'s performance without dynamical decoupling. 
ASAP-scheduled digital dynamical decoupling shows an average increase of the mean energy up to $p=3$ for \texttt{ibm\_auckland} albeit at relatively poor performance. The 16 qubit backend \texttt{ibmq\_guadalupe} profits from dynamical decoupling with a minimum mean energy up to $p=3$. In summary, the particular dynamical decoupling scheme that we applied did not uniformly improve these NISQ QAOA computations, but in some cases it did clearly improve the computation.

\begin{figure}[h!]
    \centering
    \includegraphics[width=0.49\textwidth]{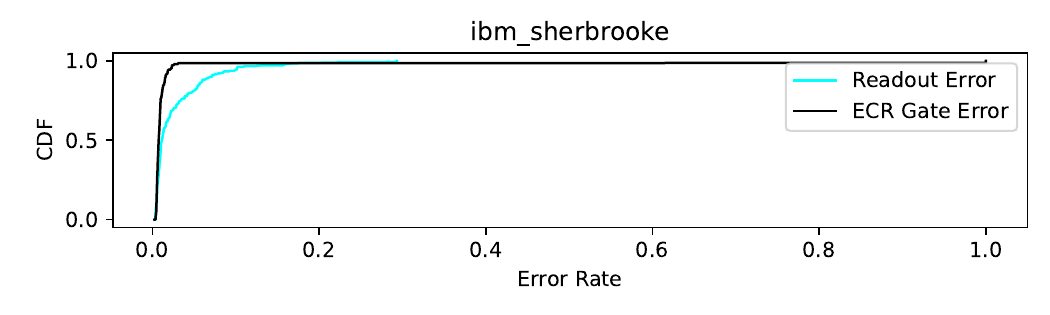}
    \includegraphics[width=0.49\textwidth]{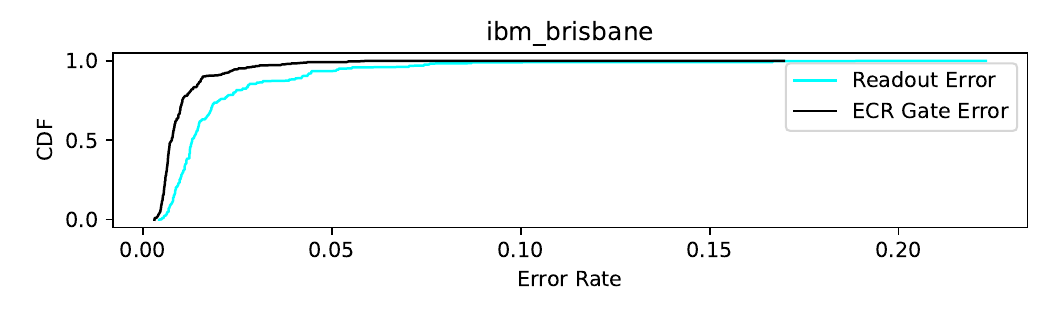}
    \includegraphics[width=0.49\textwidth]{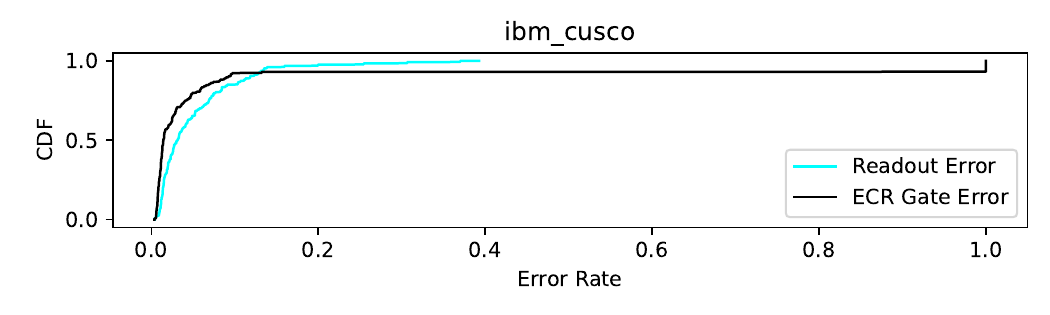}
    \includegraphics[width=0.49\textwidth]{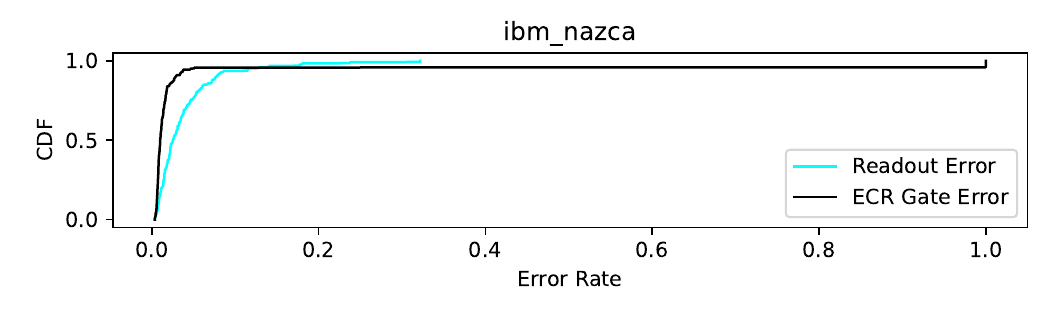}
    \includegraphics[width=0.49\textwidth]{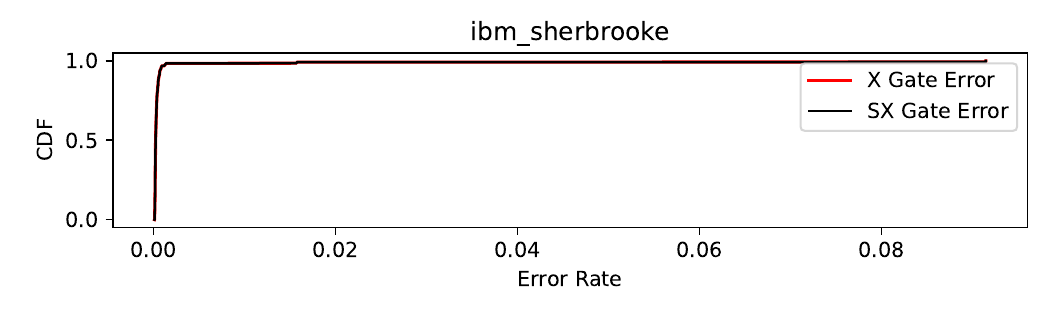}
    \includegraphics[width=0.49\textwidth]{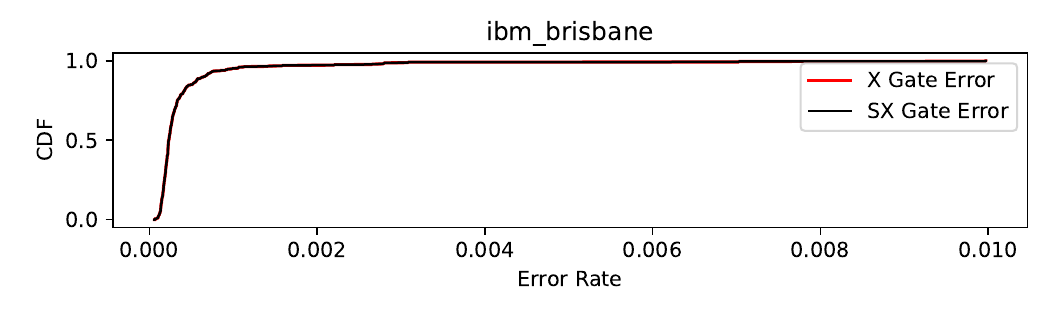}
    \includegraphics[width=0.49\textwidth]{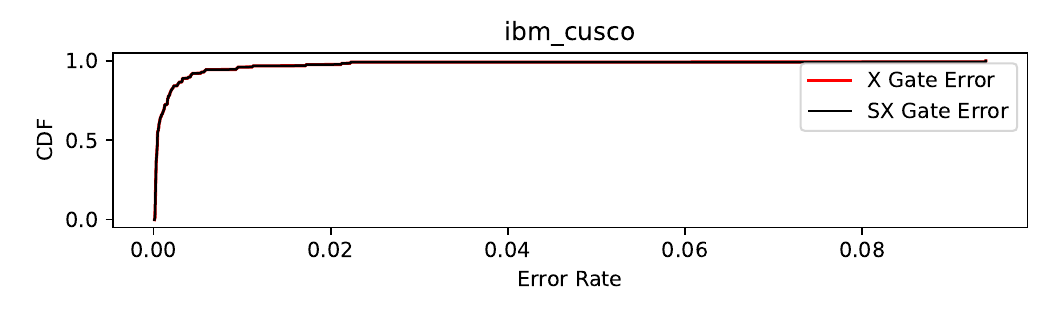}
    \includegraphics[width=0.49\textwidth]{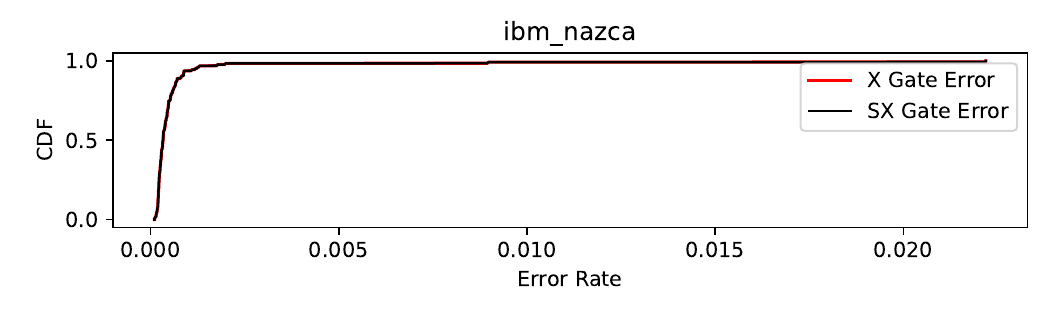}
    \caption{Gate level error rates for the four $127$ qubit IBM processors, as measured by the vendor at the time the circuits were run. These measures are aggregated from all of the executed circuits and all gate operations for each device (including all qubits and two qubit gate operations), and presented as CDFs. Note that error rates of 1 are in the ECR gates are not actually calibrated error rates of 1, but instead placeholder values from the backend denoting that the connection has not been calibrated. }
    \label{fig:hardware_error_rates}
\end{figure}

Appendix~\ref{section:appendix_optimal_solution_tables} contains tables (Table~\ref{table:QAOA_minimums_127_variable} and Table~\ref{table:QAOA_minimums_16_27_variable}) showing the exact optimal energy for all $300$ fixed higher-order problem instances studied in this section, along with the minimum energies sampled across the QAOA circuits when executed on hardware. The tables also include the maximum energies of the problem instances, which gives a quantification of the range of the energy spectrum of these higher order Ising models. Notably, these tables show that the IBM Quantum processors were able to find the optimal solution with at least one sample for the $27$ and $16$ variable problem instances, but were never able to find the optimal solution to the $127$ variable problem instances. Figure \ref{fig:hardware_error_rates} reports CDF distributions for the gate level calibrated error rates on the four $127$ qubit IBM processors, reported by the vendor, at the time these circuits were executed. These gate error rate distributions show that some device clearly have higher gate error rates than other devices, and the QAOA result quality can be compared to these gate level error rates - where we see that the lower error rate device generally perform better.

\subsection[p=1 QAOA Hardware Angle Gridsearch Results]{$p=1$ QAOA Hardware Angle Gridsearch Results}
\label{section:results_p1_gridsearch}

\begin{figure}[p!]
    \centering
    \includegraphics[width=0.49\textwidth]{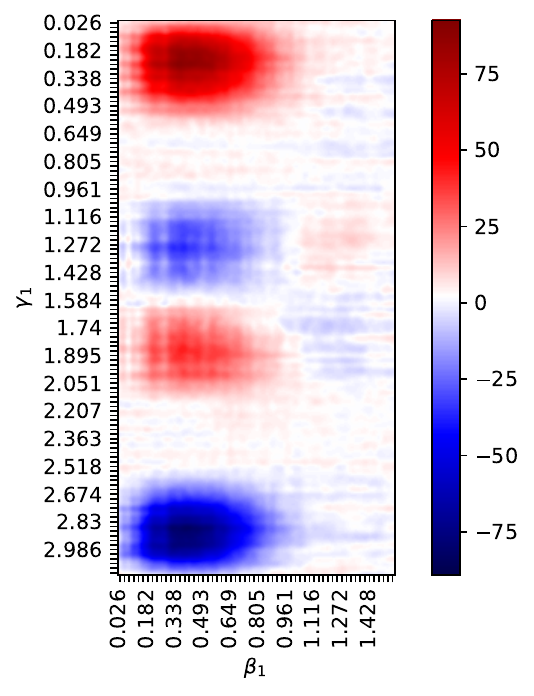}\hfill%
    \includegraphics[width=0.49\textwidth]{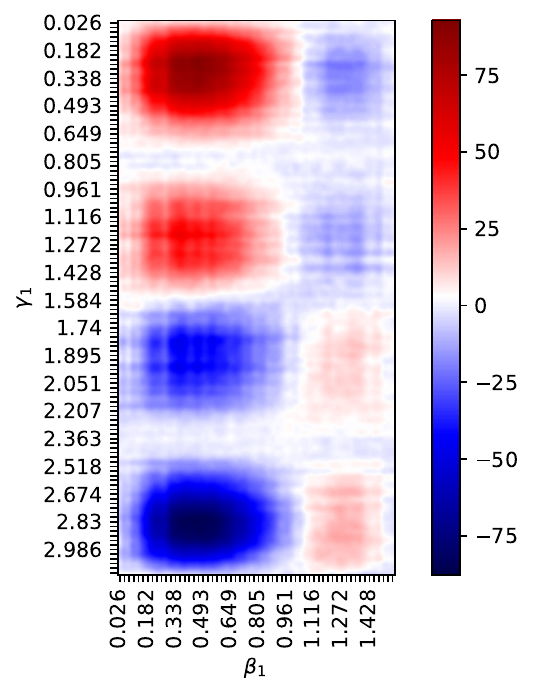}%
    \vspace*{-2ex}
    \caption{Experimental mean energy landscapes for a $p=1$ QAOA on \texttt{ibm\_seattle} for two 414 qubit instances:
    \textbf{(left)} A higher-order Ising model with up to cubic terms. 
    \textbf{(right)} An Ising model with linear and quadratic terms.
    }
    \vspace*{1ex}
    \label{fig:ibm_seattle_p1_heatmap}
\end{figure}

\paragraph{$414$ qubit $p=1$ QAOA on \texttt{ibm\_seattle}:}%
Figure~\ref{fig:ibm_seattle_p1_heatmap} shows $p=1$ angle gridsearch on \texttt{ibm\_seattle}\footnote{\texttt{ibm\_seattle} was decommissioned before more complete whole-chip QAOA experiments could be executed. } for a random Ising model instance with cubic terms and without cubic terms. The angle gridsearch is presented in terms of the mean energy computed from the distribution of $10,000$ samples drawn for each $\beta_1, \gamma_1$ angle. A total of $7,200$ linearly spaced $\beta_1, \gamma_1$ are evaluated, as in Refs.~\cite{pelofske2023qavsqaoa,pelofske2023short}. The higher order Ising model is comprised of $475$ quadratic terms, $414$ linear terms, and $232$ ZZZ terms (e.g. hyperedges). The Ising model with no higher order terms is comprised of $475$ quadratic terms and $414$ linear terms. 

\begin{figure}[p!]
    \centering
    \includegraphics[width=0.49\textwidth]{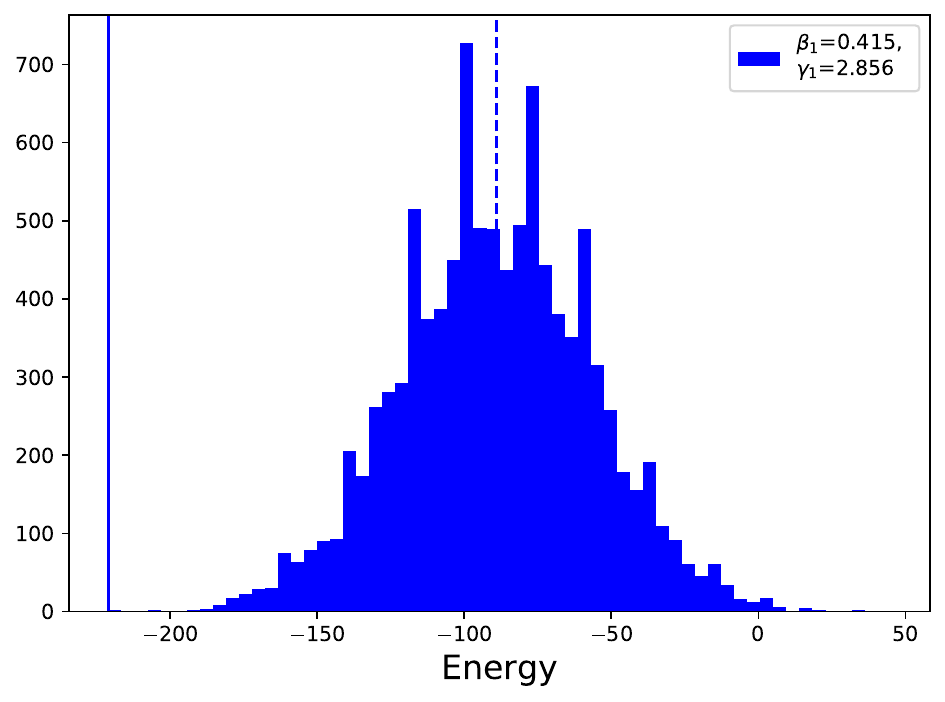}\hfill%
    \includegraphics[width=0.49\textwidth]{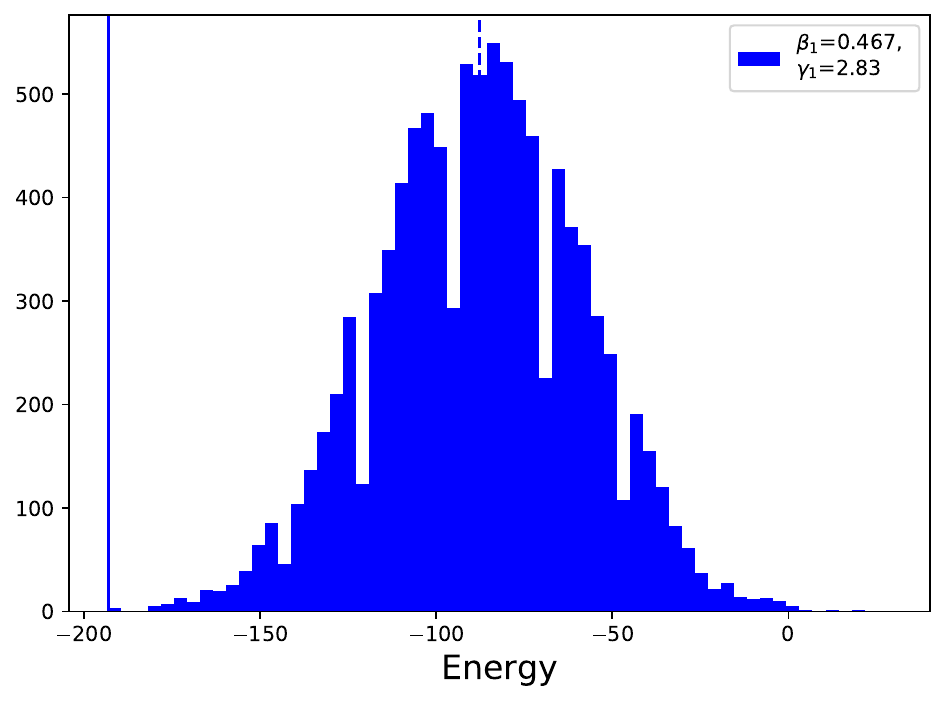}%
    \vspace*{-2ex}
    \caption{Experimental sample energy distribution for a $p=1$ QAOA on \texttt{ibm\_seattle} for two 414 qubit instances,
    for optimal angles found in the hardware angle gridsearch that minimize the mean energy (shown in Figure~\ref{fig:ibm_seattle_p1_heatmap}):
    \textbf{(left)} The higher-order model with cubic terms, with a mean energy of $-89.14$ for angles $\beta=0.415,\ \gamma=2.856$. 
    \textbf{(right)} The model with linear and quadratic terms, with a mean energy of $-87.72$ for angles $\beta = 0.467,\ \gamma=2.83$. 
    The mean energies are marked with vertical dashed blue lines. The vertical solid lines mark the minimum sample energy found among the 10,000 samples at these angles; however, during the whole angle gridsearch, the overall minimum sample energies lie at $-241$ for the higher-order Ising model and $-221$ for the Ising model on the right.\newline
    For context, the energy spectra of the instances range from $-637$ (ground state) to $+623$ (maximum) for the higher-order model on the left
    and from $-567$ (ground-state) to $+565$ (maximum) for the Ising model on the right. 
    }
    \label{fig:ibm_seattle_p1_energy_distribution}
\end{figure}

Notably, the hardware-computed $p=1$ energy landscape on these $414$ qubit instances are very similar to the $p=1$ energy landscapes shown in Ref.~\cite{pelofske2023short}. Figure~\ref{fig:ibm_seattle_p1_energy_distribution} shows the full energy distribution (of $10,000$ samples) for the best $p=1$ angles on the hardware-gridsearch, along with the optimal energy. Note that the minimum energies found from the $p=1$ sampling are far away from the optimal solution energy.


\begin{figure}[th!]
    \centering
    \hspace*{1cm}\texttt{ibm\_geneva}\hfill\texttt{ibm\_auckland}\hfill\texttt{ibm\_cairo}\hfill\texttt{ibmq\_mumbai}\hspace*{1.3cm}\ \\%
    \includegraphics[width=0.24\textwidth]{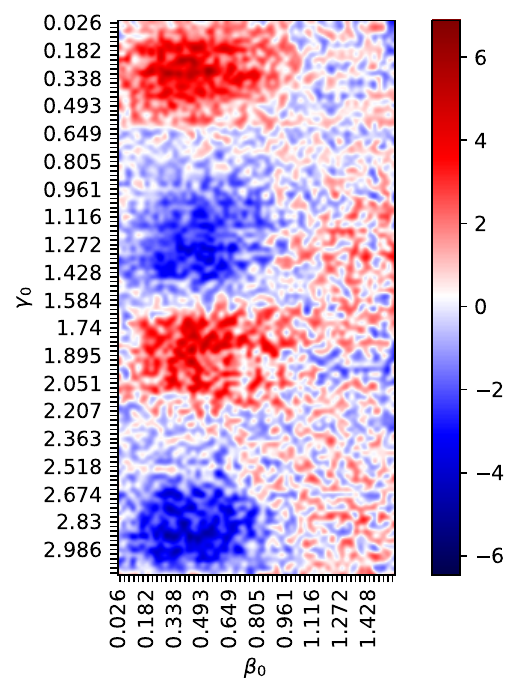}\hfill%
    \includegraphics[width=0.24\textwidth]{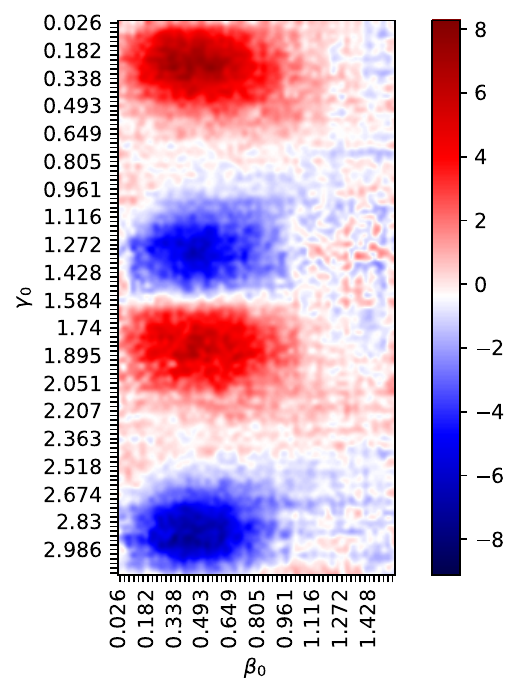}\hfill%
    \includegraphics[width=0.24\textwidth]{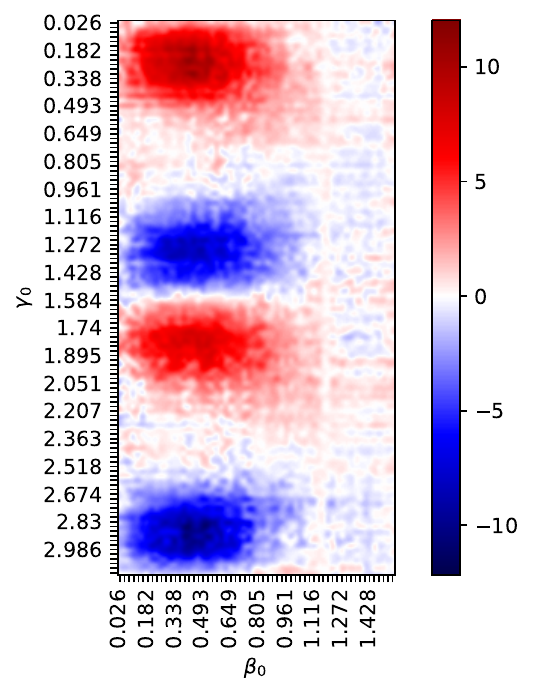}\hfill%
    \includegraphics[width=0.24\textwidth]{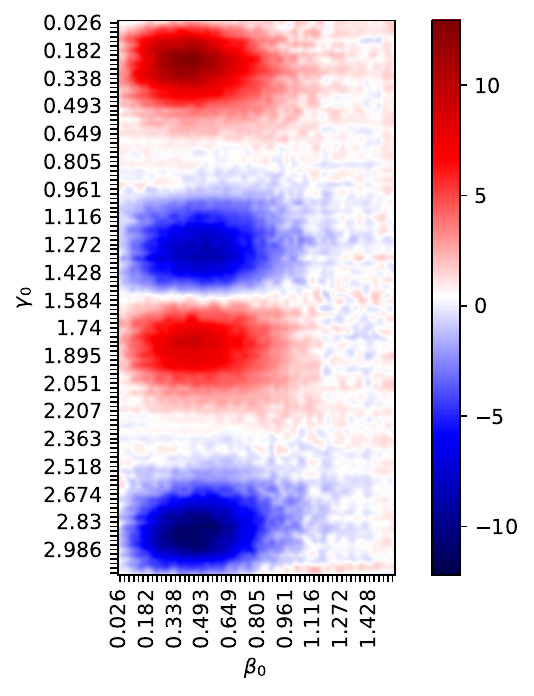}%
    \vspace*{-2ex}
    \caption{Mean energy landscapes for $p=1$ QAOA on a single 27-qubit instance sampled using 4 different IBM quantum processors that have the same heavy-hex hardware graph. The energy landscapes show how the different QAOA angles perform - blue denotes better optimization of the Ising models, since we are solving them as minimization optimization problems. Notably, these search landscapes show the variability of the different quantum processors, where some clearly have noisier search landscapes than others. Each region of the heatmaps are average energies computed over a large distribution of hardware measurements. 
    }
    \vspace*{1ex}
    \label{fig:27_qubit_IBMQ_heatmaps_medium}
\end{figure}

\paragraph{$27$ qubit $p=1$ gridsearch:}%
Figure~\ref{fig:27_qubit_IBMQ_heatmaps_medium} shows hardware $p=1$ angle gridsearch mean energy heatmaps on several IBM Quantum processors. Notably, the energy landscapes are very similar to the $414$ qubit whole-lattice heavy-hex QAOA in Subsection~\ref{section:results_p1_gridsearch}, and the previously reported $127$ qubit whole-lattice heavy-hex QAOA results from Ref.~\cite{pelofske2023short}. Notice that the energy landscape from \texttt{ibm\_geneva} is considerably more noisy compared to the other device energy heatmaps.

\begin{figure}[th!]
    \centering
    \includegraphics[width=0.49\textwidth]{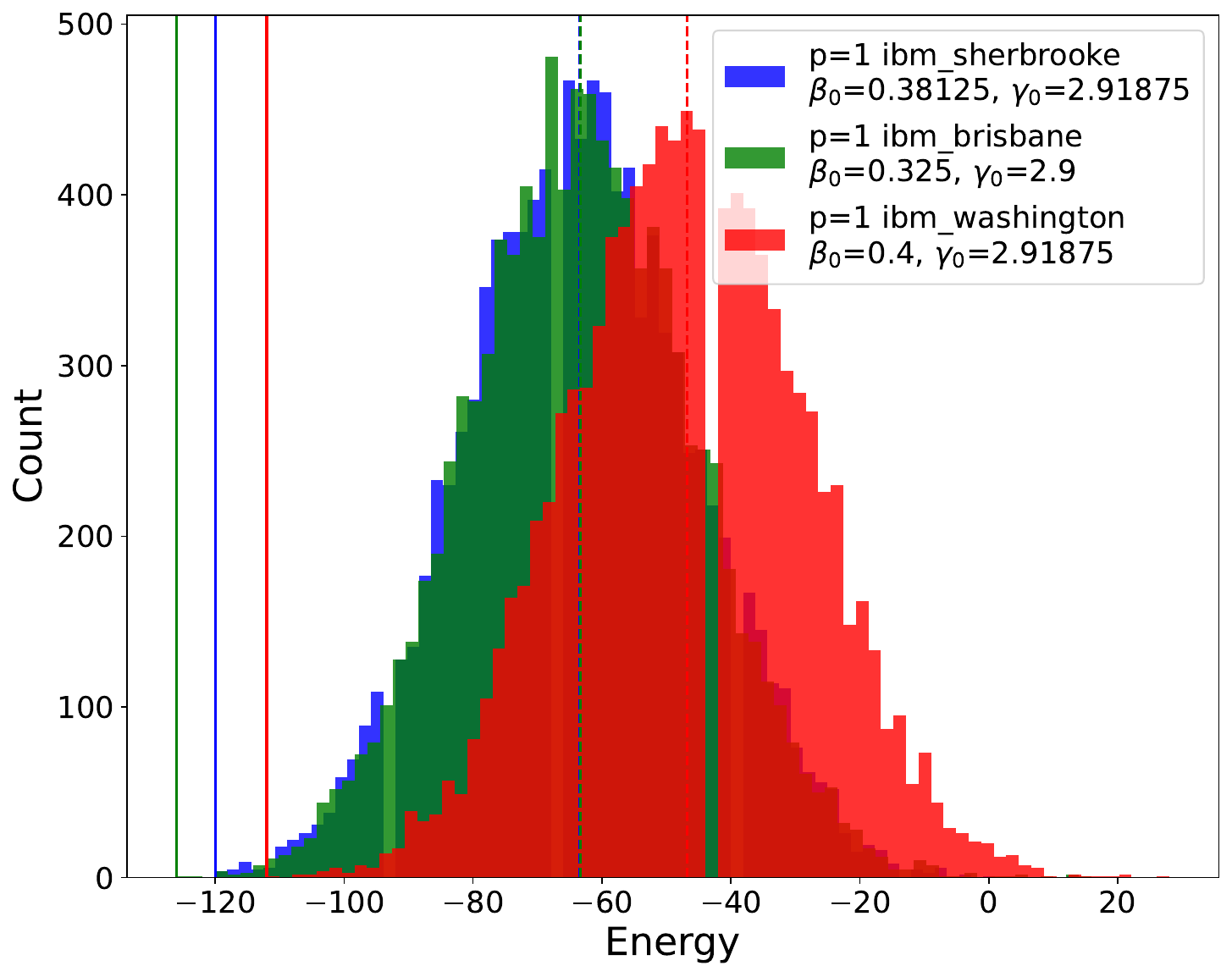}\hfill%
    \includegraphics[width=0.49\textwidth]{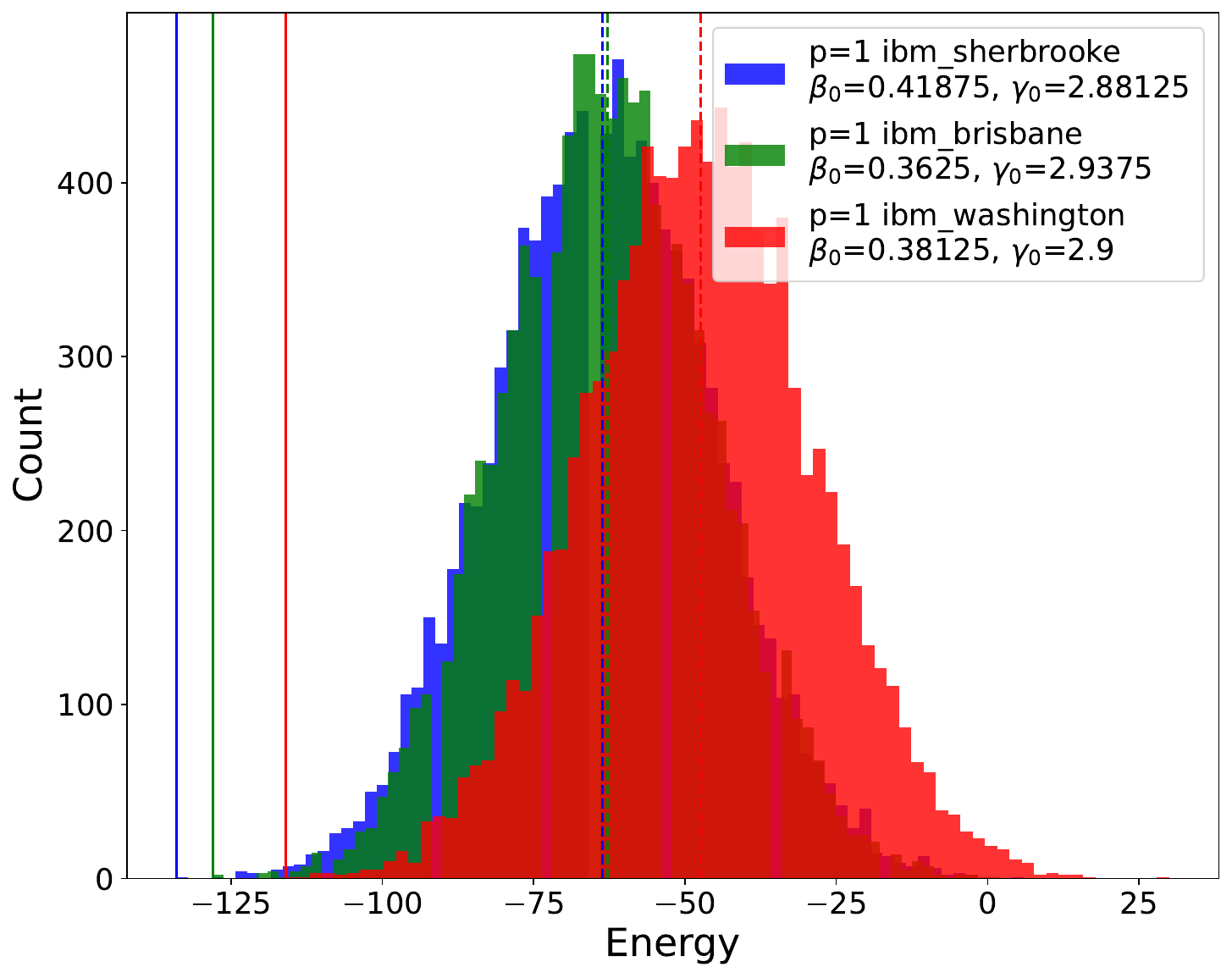}\\%
    \includegraphics[width=0.49\textwidth]{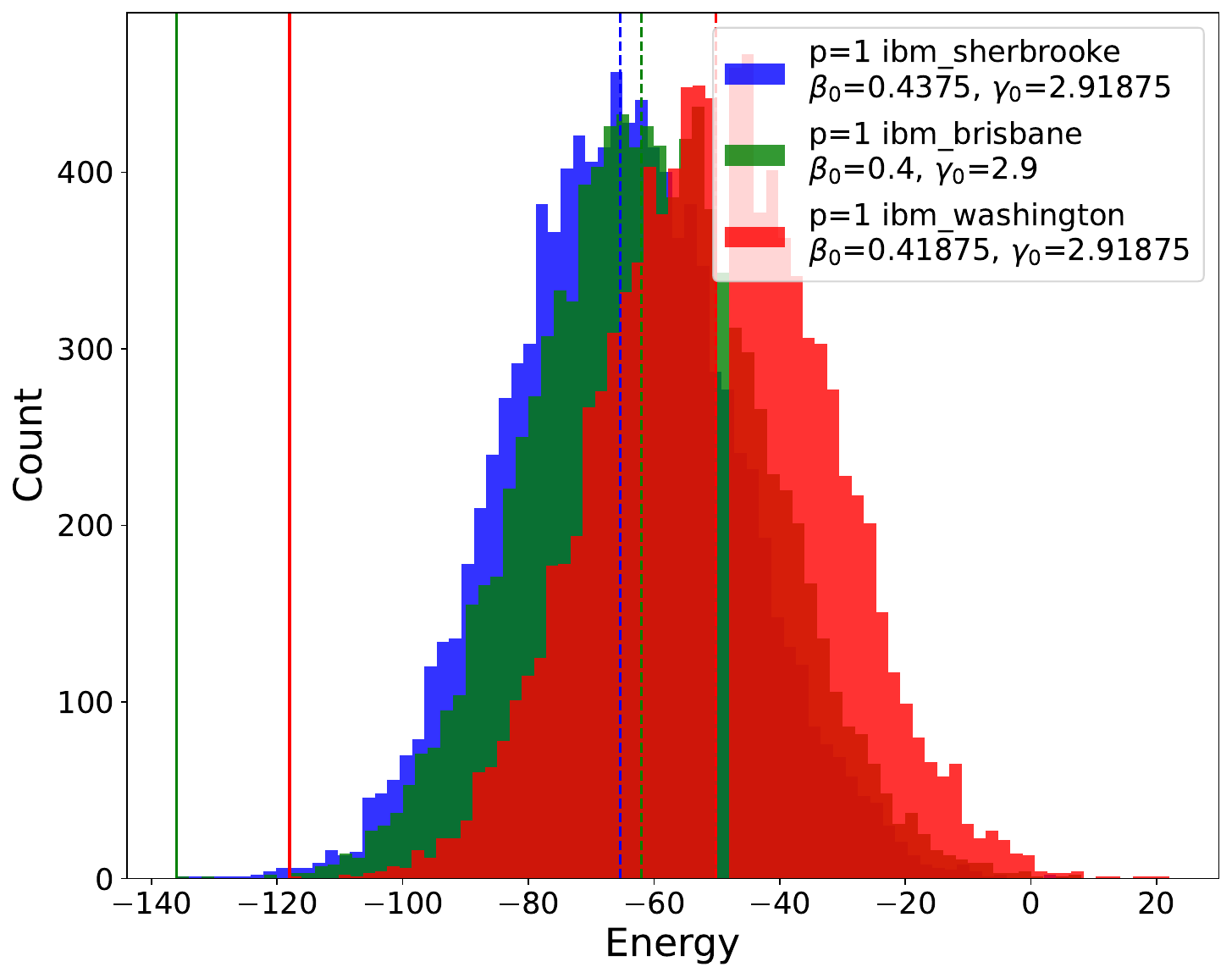}\hfill%
    \includegraphics[width=0.49\textwidth]{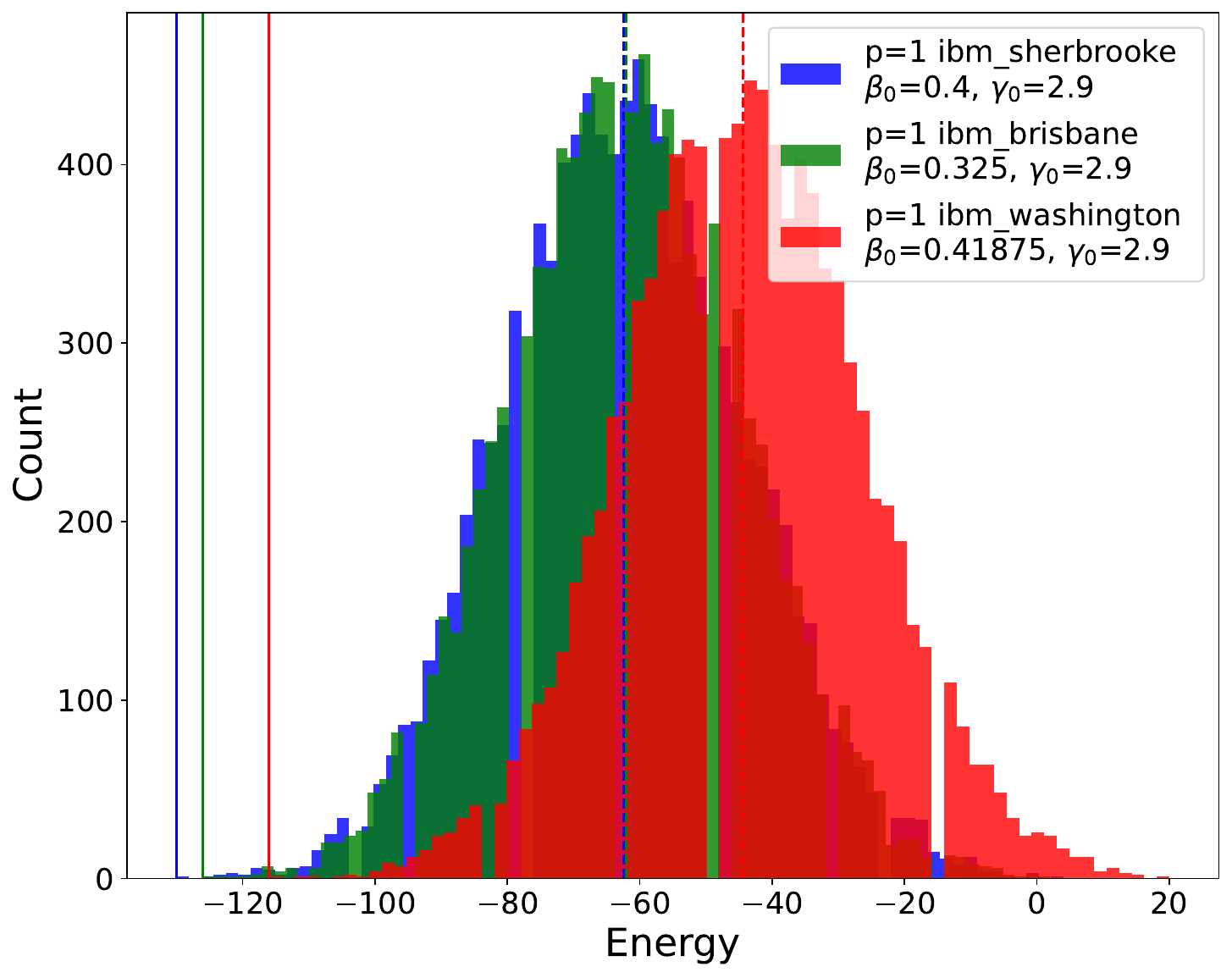}%
    \vspace*{-1ex}
    \caption{Experimental sample energy distributions for a $p=1$ QAOA for 4 random 127-qubit instances executed across 3 IBMQ devices, for the best device-specific angles found by a focused angle gridsearch. The minimum sample energies and mean energies from these distributions are marked with solid and dashed vertical lines, respectively.}
    \label{fig:p1_compare_hardware}
\end{figure}

\paragraph{Comparison of difference 127 qubit IBMQ Processors with Whole-Chip $p=1$ QAOA Circuits:}%
A straightforward question that can be asked using whole-chip circuits is how different processors compare, when executing the same circuit. This offers a clear way to benchmark device performance, using all available hardware components. In this section, we use the short depth QAOA circuits to compare three of the $127$ qubit IBM Quantum superconducting qubit processors; \texttt{ibm\_washington}, \texttt{ibm\_brisbane}, and \texttt{ibm\_sherbrooke}. We do this using a focused QAOA angle gridsearch for $p=1$ QAOA depth, using higher order Ising models that are compatible with all three of these processors - which in particular means hardware compatible with \texttt{ibm\_washington}, as its hardware graph is a subgraph of the other two. The angle gridsearch is performed on-device, using $\beta_1 = 0.4$ and $\gamma_1 = 2.9$ as the center of the grid (based on the observed parameter concentration, especially of the $p=1$ angle gridsearch heatmaps in Ref.~\cite{pelofske2023short}), and a grid of $81$ linearly spaced points $\pm 0.15$. $10,000$ shots are taken for each angle. Figure~\ref{fig:p1_compare_hardware} shows the energy distributions from using these three quantum computers to sample $4$ different random higher order Ising models, where the reported distribution is of the $10,000$ samples with the lowest mean energy among the focused angle gridsearch. This distribution shows that the newer generation of the $127$ qubit processors (see Table~\ref{table:hardware_summary}) performed definitely better than the previous generation \texttt{ibm\_washington} device. Notably, the best angles varied slightly depending on the device, due to the noise in the computation.

\subsection{CPLEX Classical Compute Time}
\label{section:results_CPLEX_compute_time}
Here we report the classical compute time from CPLEX that is required to optimally solve all of the optimization problem instances. This time is reported in seconds from the python CPLEX module; this time does not include the compute time used to perform the order reduction, or datastructure parsing. Note that the order reduction procedure that is used to solve the problem instances using CPLEX introduces auxiliary variables, and therefore inflates the total number of decision variables that must be solved by CPLEX~\cite{pelofske2023short}. 

\begin{figure}[h!]
    \centering
    \includegraphics[width=0.60\textwidth]{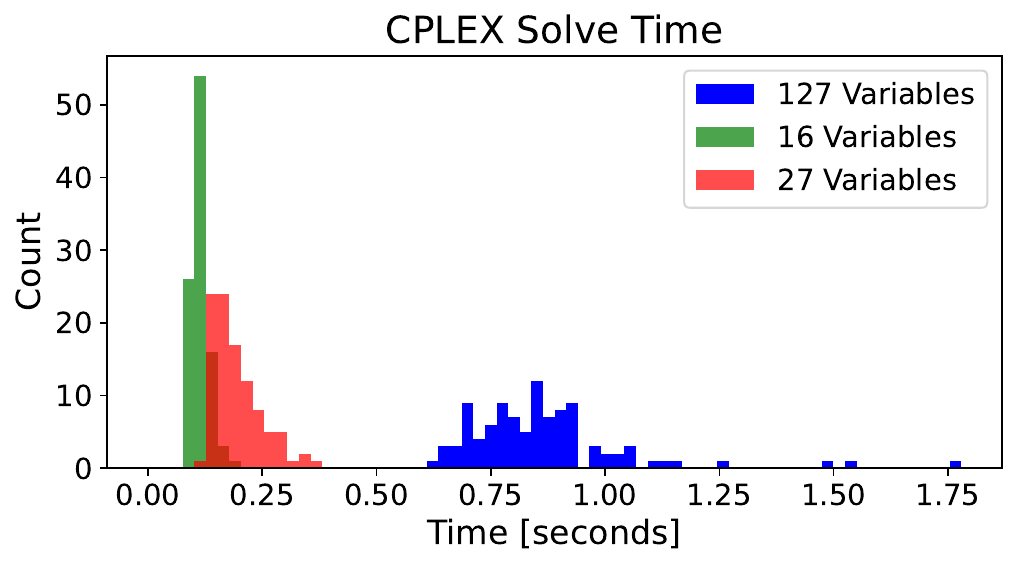}
    \caption{CPLEX compute times required to exactly solve all $300$ heavy-hex compatible problem instances with geometrically local cubic terms. Note that the actual optimization problem being solved by CPLEX is the order reduced version of the original problem, meaning that there are added auxiliary variables compared to the original problem. }
    \label{fig:CPLEX_compute_time}
\end{figure}

Figure \ref{fig:CPLEX_compute_time} shows the distributions of CPLEX solve times for the $100$ problem instances for the $3$ problem sizes. These timing statistics show the level of computation time that hardware runs of QAOA would need to achieve to be competitive with state of the art classical optimization solvers (albeit, specifically for the class of sparse optimization problems used in this study). 

The heavy-hex $414$ qubit problem instance with cubic terms (used in Section \ref{section:results_p1_gridsearch}) was solved exactly with CPLEX in $3.129$ seconds, and without cubic terms was solved exactly in $0.074$ seconds.

\section{Conclusion}
\label{section:discussion}

We have demonstrated that parameter transfer of QAOA angles up to $p=5$ can be successfully applied to large (up to $127$ qubit) systems using training on a single small ($16$ qubit) instance. This provides  evidence -- in addition to what has been presented in the literature on various other optimization problems -- that parameter concentration can be used as an efficient method for computing high-quality (although not necessarily optimal) QAOA angles. We used converged classical MPS simulations with up to a bond dimension of $\chi=2048$ to calculate the noiseless mean expectation values, as well as the sample distributions, of the $127$ qubit QAOA circuits sampling these hardware compatible higher-order Ising model instances using the transfer-learned angles. 

We also demonstrated the scaling of whole-chip QAOA on heavy-hex hardware-native spin glass models, with respect to $p$, on several IBM Quantum superconducting qubit processors. This demonstration comprises large circuits that fully evaluate the current performance of these IBM Quantum processors using highly NISQ-friendly and short depth QAOA circuits. We find that the peak of QAOA performance on hardware is at $p=2,3$ on most of the IBM Quantum processors. This result shows the current state of competition between the error inherent in the computation, and the improving approximation ratios from larger $p$ (and good angles learned at higher $p$). These type of sparse short depth circuits are in contrast to dense circuits, such as quantum volume circuits~\cite{pelofske2022qv, cross2019validating}, but allow probing of usage of an entire hardware graph. We observed that the relatively simple Pauli X pair digital dynamical decoupling sequences improved the mean QAOA computation on some of the IBM Quantum processors, but on other devices it actually made the computation worse. 

While the scale of the number of qubits used in these QAOA simulations far exceeds what can be exactly classically simulated using full state vector simulations, the sparsity of the underlying hardware graph means that simulating the mean expectation value for low QAOA rounds is possible. At high rounds, we expect the classical simulation of such QAOA circuits to also begin to struggle, and an interesting future avenue of study is to determine where this point is. In this work we have used MPS simulations in order to simulate the QAOA circuits up to $p=5$ in order to verify that the parameter transfer procedure was successful, but is unclear how classically simulate-able higher round QAOA is when targeting Ising models defined on heavy-hex graphs. For example, a Hamiltonian dynamics simulation was performed on a $127$ qubit heavy-hex IBM Quantum device~\cite{kim2023evidence}, the experiments for which were then classically simulated efficiently using a number of different approaches~\cite{begusic2023fast, tindall2023efficient, kechedzhi2023effective, liao2023simulation, begusic2023fastconverged, rudolph2023classical, patra2023efficient, shao2023simulating, anand2023classical, tindall2024confinement}. This suggests that perhaps even extremely high round QAOA circuits (e.g. where $p$ is significantly higher than what was used in this study) for these sparse heavy-hex Ising models may be easy to simulate when the number of qubits is small. It is also of interest to evaluate how well MPS simulations can be applied to these QAOA circuits when the angles are optimal (or nearly-optimal), as opposed to, for example, random QAOA angles. This is a very interesting regime to investigate since it is approaching the boundary of what is classically verifiable - we leave these high $p$ heavy-hex compatible QAOA simulation questions open for future work. Future work could also study the effects of different choices of the polynomial coefficients, besides $+1/-1$, or even different distributions of the random coefficients and how that impacts the parameter transfer. There are also interesting variants of QAOA, such as warm-start QAOA \cite{Egger_2021, Jain2022graphneuralnetwork, Tate2023warmstartedqaoa}, where parameter transfer could also be tested in future studies. 

The optimization problems studied here are computationally quite easy to solve, for example standard combinatorial optimization software can exactly solve these problems on the order of less than two seconds of CPU time. These problem instances are used specifically because they are designed to be highly hardware compatible with the heavy-hex connectivity, not because they are significantly computationally challenging for classical algorithms.

Our findings show that QAOA parameter transfer can be used in order to obtain good angles for QAOA circuits that are very high in qubit count, using a computationally efficient learning of only a single small (in this case $16$) qubit problem instance. We expect that these types of parameter transfer protocols will be useful in future implementations of QAOA. However, there is an important aspect of this which has not been studied up to this point. This is the case where the QAOA angles computed at a small problem size are so good that they reach an approximation ratio that is effectively $1$ - in other words, the QAOA performance plateaus (as a function of increasing $p$) to optimality. Once this occurs, good angles at higher $p$ can no longer be meaningfully computed for the small problem instance~\cite{pelofske2023high, golden2023numerical, golden2023qaoasat}, and thus good angles cannot be computed to be used for the larger problem instance. This is related to the question of QAOA scaling (how many $p$ rounds we need in order to obtain good approximation ratios) as a function of increasing $N$. Succinctly, the task of investigating QAOA angle parameter transfer for extremely high $p$ should be investigated in future research.


\section{Acknowledgments}
\label{section:acknowledgments}
This work was supported by the U.S. Department of Energy through the Los Alamos National Laboratory. Los Alamos National Laboratory is operated by Triad National Security, LLC, for the National Nuclear Security Administration of U.S. Department of Energy (Contract No. 89233218CNA000001). This research used resources provided by the Los Alamos National Laboratory Institutional Computing Program. We acknowledge the use of IBM Quantum services for this work. The views expressed are those of the authors, and do not reflect the official policy or position of IBM or the IBM Quantum team. The authors thank IBM Quantum Technical Support. The research presented in this article was supported by the Laboratory Directed Research and Development program of Los Alamos National Laboratory under project number 20220656ER and 20230049DR. Research presented in this article was supported by the NNSA's Advanced Simulation and Computing Beyond Moore's Law Program at Los Alamos National Laboratory. This research used resources provided by the Darwin testbed at Los Alamos National Laboratory (LANL) which is funded by the Computational Systems and Software Environments subprogram of LANL's Advanced Simulation and Computing program (NNSA/DOE). 

The figures in this article were generated using matplotlib~\cite{caswell2021matplotlib, hunter2007matplotlib}, networkx~\cite{hagberg2008exploring}, and Qiskit~\cite{Qiskit} in Python 3.

\noindent
LA-UR-23-33192

\section{Data and Code availability}
Data is publicly available on Zenodo \cite{pelofske_2024_14031608}. Code to generate the problem instances is available on Github \footnote{\url{https://github.com/lanl/QAOA_vs_QA}}.

\appendix

\section{Classical Parameter Tuning Approach: Parameter Fixing with Angle Gridsearch}
\label{section:appendix_parameter_fixing_gridsearch}

\begin{figure}[p!]
    \centering
    \includegraphics[width=0.49\textwidth]{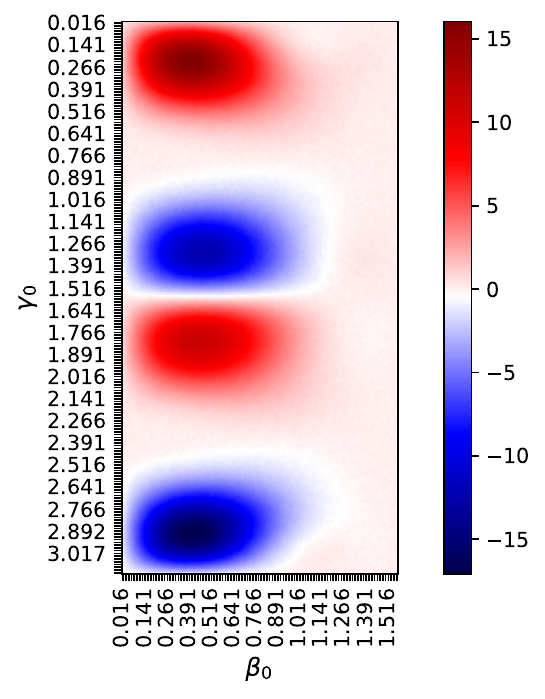}\hfill%
    \includegraphics[width=0.49\textwidth]{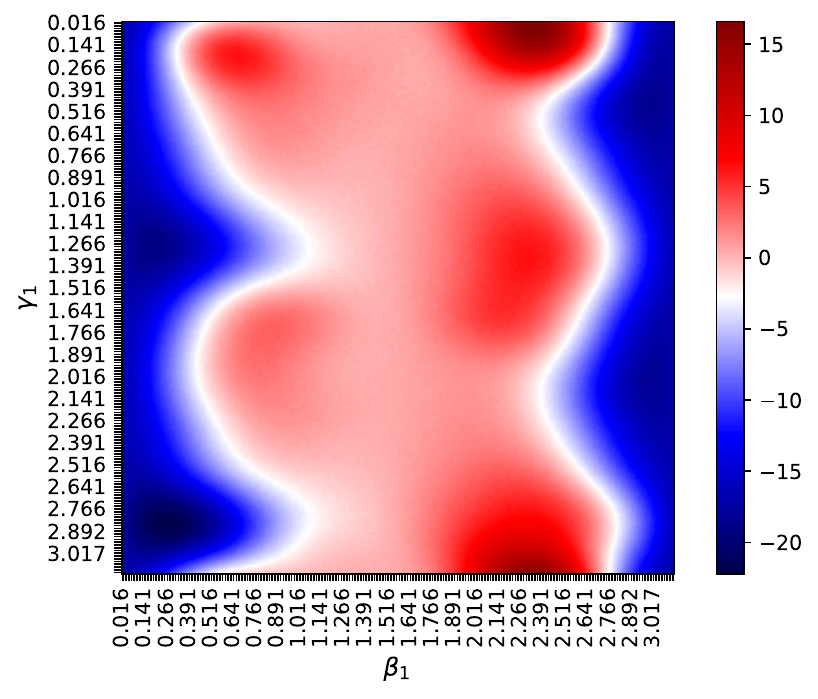}\\%
    \includegraphics[width=0.49\textwidth]{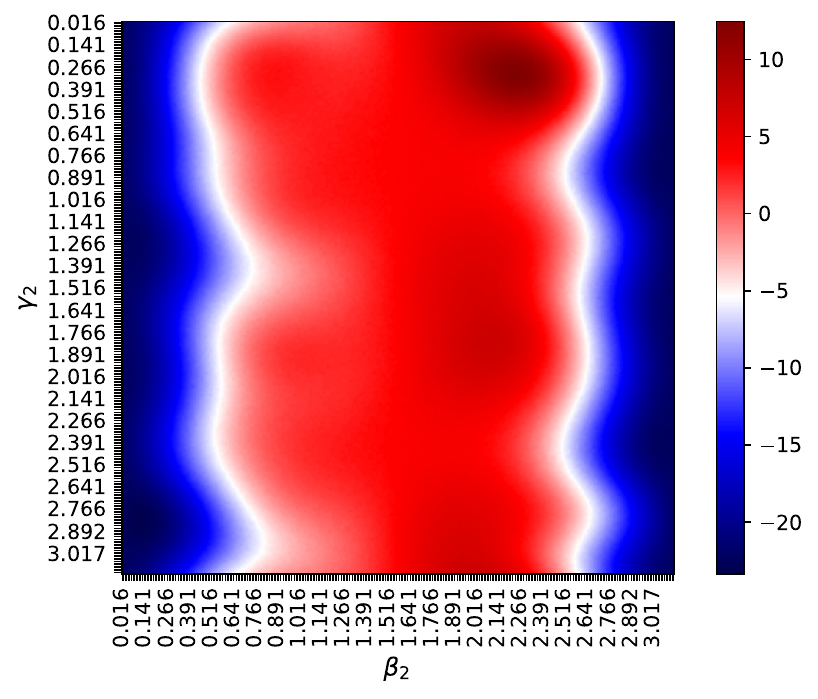}\hfill%
    \includegraphics[width=0.49\textwidth]{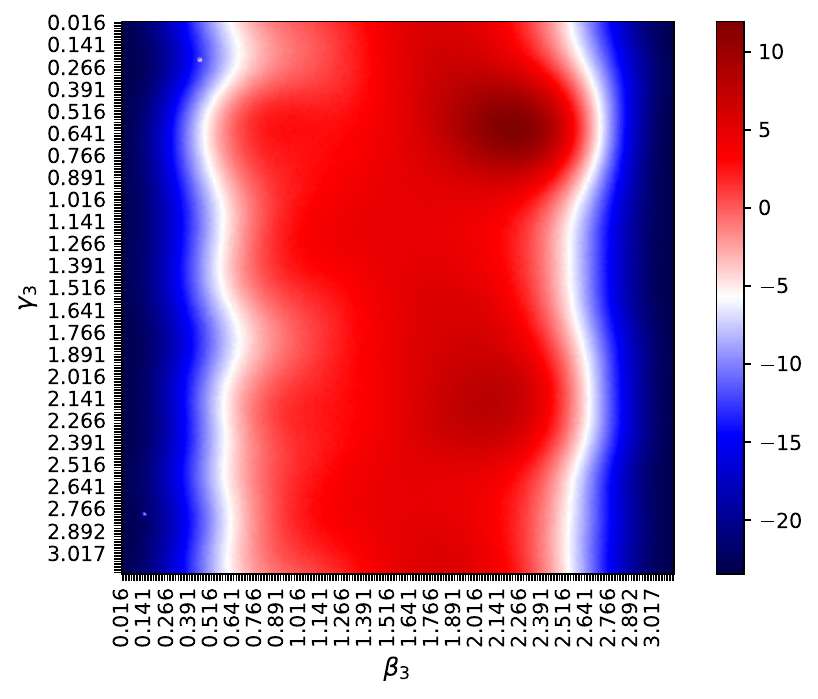}
    \caption{Heatmap showing classical simulation of the mean energy for QAOA angles as $p$ is increased, while fixing good angles found at previous rounds. The fixed angles used based on the parameter fixing gridsearch approach (e.g. the best angles found for each gridsearch, shown in the above heatmaps) is $\beta_0 = 0.422, \gamma_0 = 2.891, \beta_1 = 0.281, \gamma_1 = 2.845, \beta_2 = 0.125, \gamma_2 = 2.813, \beta_3 = 0.047, \gamma_3 = 0.203$ (indexing of $p=1$ is $0$). The problem instance being sampled here is a $27$ qubit random Ising model with cubic terms, defined on the $27$ qubit IBM Quantum processor heavy hex graph. Notably, the mean energy landscapes from $p=2, 3, 4$ are converging to similar shapes, with very little energy improvement being gained between $p=3$ and $p=4$, showing that this gridsearch and angle fixing method does not work as an angle finding approach, and therefore we utilized the better angle finding methods described in Subsection~\ref{section:methods_QAOA_angle_finding}. }
    \label{fig:increasing_p_heatmpas_angle_fixing}
\end{figure}

Figure~\ref{fig:increasing_p_heatmpas_angle_fixing} shows that iteratively performing angle gridsearches (using exact classical simulations of the $27$ qubit circuits), and fixing the best angles found at the previous round $p$, unfortunately quickly plateaus to sub-optimal angles, with very little increase from $p=3$ to $p=4$. It may be the case that random local searches around these parameter fixing search spaces yield better angles, but this direct approach of a grid search in additional to parameter fixing of the previous round did not work. Instead, we opted for learning good angles using alternative methods that were able to compute high quality QAOA angles (described in Section~\ref{section:methods_QAOA_angle_finding}). These angle fixing QAOA numerical simulations shown in Figure~\ref{fig:increasing_p_heatmpas_angle_fixing} were performed using Qiskit~\cite{Qiskit}, where the mean energy for a set of angles was computed using $10,000$ shots. The angle gridsearch was $200$ linearly spaced angles between $(0, \pi)$ (for both axis of the $\beta_p$ and $\gamma_p$ that were being varied), with the exception of $\beta_1$ (for $p=1$) where the linearly spaced gridsearch was cut half to $(0, \frac{\pi}{2})$ (see Section~\ref{section:methods_QAOA_circuits} for more details). Figure~\ref{fig:increasing_p_heatmpas_angle_fixing} shows the the pure angle fixing approach seems to cause the expectation value of the QAOA landscape to begin to converge - the mechanism behind this could be studied in future work.

\section{\texttt{ibm\_seattle} Hardware Graph}
\label{section:appendix_hardware_graphs}

\begin{figure}[t!]
    \centering
    \includegraphics[width=0.9\textwidth]{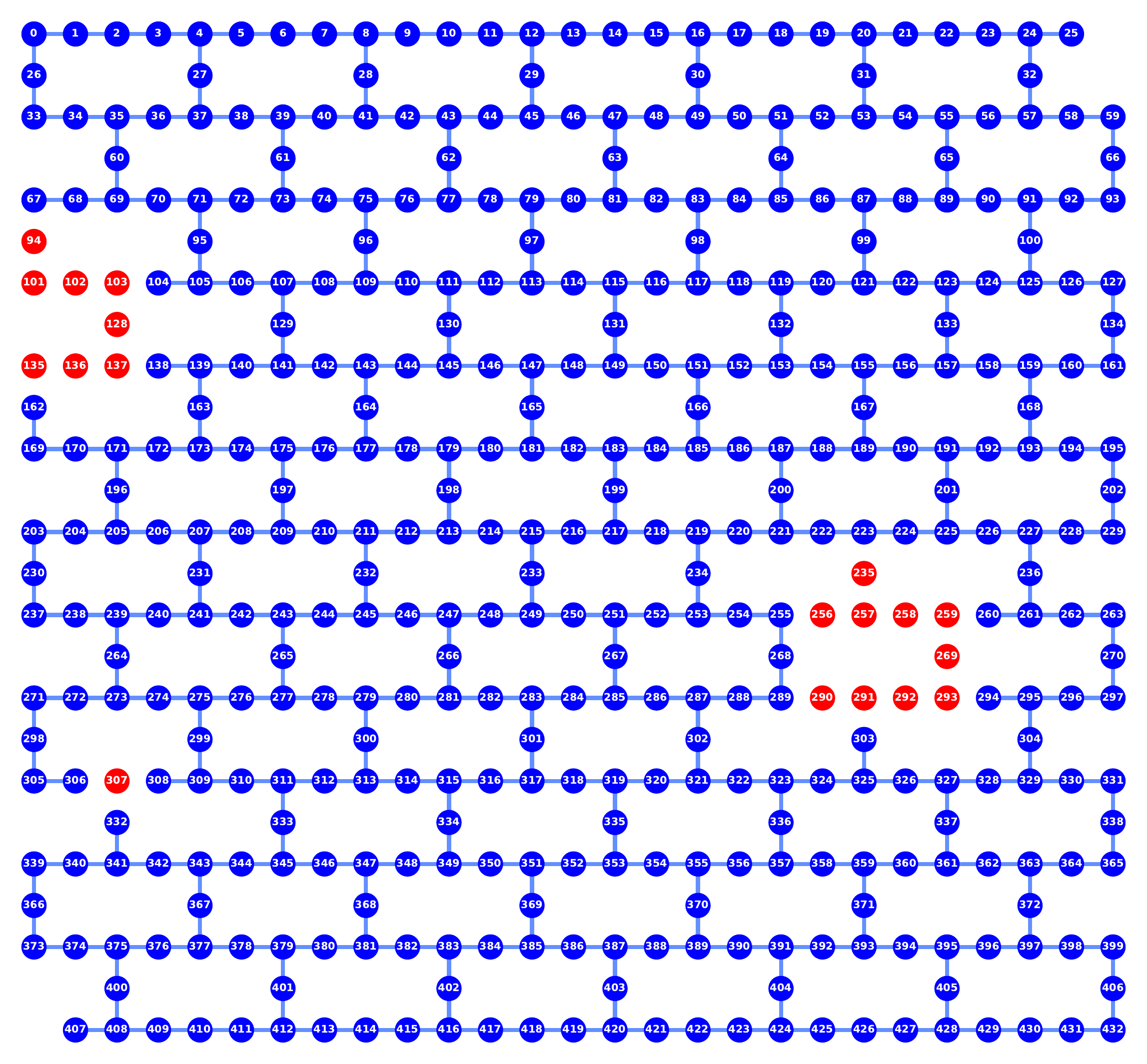}
    \caption{\texttt{ibm\_seattle} hardware graph. Red nodes indicate qubits which are hardware de-activated and not available to use at the time of these experiments. The logical hardware graph has $433$ qubits, $19$ of which are de-activated, resulting in a total of $414$ qubits which can be used for executing the short depth QAOA circuits.}
    \label{fig:hardware_graph_ibm_seattle}
\end{figure}

Figure~\ref{fig:hardware_graph_ibm_seattle} shows the heavy-hex hardware graph of \texttt{ibm\_seattle}, where the red nodes denote de-activated hardware regions that could not be used when executing circuits. 

\section{Optimal Solutions and Minimum QAOA Energy Tables}
\label{section:appendix_optimal_solution_tables}

Table~\ref{table:QAOA_minimums_127_variable} shows the optimal solution energy, along with the minimum energy sampled by the whole-chip QAOA circuits when executed on the IBM Quantum computing hardware, for all $100$ of the $127$ variable higher-order Ising model problem instances. Table~\ref{table:QAOA_minimums_16_27_variable} (left) shows the same for the $27$ variable problem instances, and Table~\ref{table:QAOA_minimums_16_27_variable} (right) shows the same for the $16$ variable problem instances. The count of samples that found the minimum energy signals the stability of that measurement. Together, these tables describe the minimum energies sampled from the set of increasing $p$ experiments described in Section~\ref{section:results_scaling_p_16_27_127_qubits}. The count of samples that found the minimum energy is out of $20000 \cdot 100 \cdot D \cdot 3 \cdot 5$ total samples, where $D$ is the number of devices that were used (for $16$ variables $D=1$, for $27$ variables this is $D=6$, and for $127$ variables this is $D=4$). 

We observe from Table~\ref{table:QAOA_minimums_127_variable} that both \texttt{ibm\_brisbane} and \texttt{ibm\_sherbrooke} actually found their best mean values at $p=3$ for a good fraction of the 100 instances (i.e., 36 and 43 instances, respectively); these two backends even found minima at $p=4$ for 9 and 1 of the instances, respectively. Thus, the higher QAOA rounds are close to paying off.

In Table~\ref{table:QAOA_minimums_127_variable}, minimum sample energies given for the specific devices are always with respect to the sample distribution \emph{for the given} $p$ and $\mathit{DD}$ parameters. 
Lower sample energies are sometimes found at non-optimal (with respect to the mean sample energy) parameters $p$, $\mathit{DD}$. Indeed, the overall QAOA minimum sample energy across all parameters
is shown for many instances strictly lower than the given minimum sample energies for the four devices at fixed parameters.

\begin{table}[t!]
    \centering
    \newlength{\deviceseplength}
    \setlength{\deviceseplength}{28pt}
    \newcommand{\devicesep}{\hspace*{\deviceseplength}}
    \renewcommand{\arraystretch}{.85}
    \scalebox{0.54}{\begin{tabular}{@{}l@{}rr@{\devicesep}c@{\devicesep}rrl@{\devicesep}rrl@{\devicesep}rrl@{\devicesep}rrl@{}}
        \toprule
        \       & \multicolumn{2}{r@{\devicesep}}{instance energies}  & overall QAOA & \multicolumn{3}{@{}l}{best \texttt{ibm\_brisbane} sample}    & \multicolumn{3}{@{}l}{best \texttt{ibm\_cusco} sample}    & \multicolumn{3}{@{}l}{best \texttt{ibm\_nazca} sample}    & \multicolumn{3}{@{}l}{best \texttt{ibm\_sherbrooke} sample}    \\
        \cmidrule(lr{\deviceseplength}){2-3}
        \cmidrule(r{\deviceseplength}){5-7}
        \cmidrule(r{\deviceseplength}){8-10}
        \cmidrule(r{\deviceseplength}){11-13}
        \cmidrule{14-16}
        \       & ground state   & maximum   & min sample energy  & mean  & min   & at params & mean  & min    & at params & mean  & min    & at params & mean  & min    & at params \\ 
        \midrule[\heavyrulewidth]
		 0	& -188	& 202	& -150 ($\times$1)	& -76.50	& -148	& p=3, DD=ALAP	& -43.30	& -120	& p=2, DD=ALAP	& -61.05	& -132	& p=2, DD=ALAP	& -67.99	& -140	& p=3, DD=none\\
		 1	& -196	& 200	& -156 ($\times$1)	& -80.18	& -154	& p=2, DD=ALAP	& -48.06	& -122	& p=2, DD=ALAP	& -67.89	& -142	& p=2, DD=ALAP	& -70.13	& -148	& p=2, DD=none\\
		 2	& -198	& 198	& -164 ($\times$1)	& -79.97	& -142	& p=2, DD=ALAP	& -42.54	& -116	& p=1, DD=ALAP	& -63.28	& -130	& p=2, DD=ALAP	& -68.75	& -138	& p=2, DD=none\\
		 3	& -198	& 196	& -156 ($\times$1)	& -81.17	& -146	& p=2, DD=ALAP	& -46.52	& -114	& p=2, DD=ALAP	& -64.09	& -136	& p=2, DD=ALAP	& -69.97	& -138	& p=2, DD=none\\
		 4	& -196	& 196	& -154 ($\times$1)	& -83.95	& -154	& p=3, DD=ASAP	& -48.46	& -124	& p=1, DD=ALAP	& -66.19	& -146	& p=3, DD=ALAP	& -75.87	& -144	& p=2, DD=none\\
		 5	& -204	& 194	& -162 ($\times$3)	& -86.86	& -162	& p=2, DD=ALAP	& -52.47	& -122	& p=2, DD=ALAP	& -67.48	& -138	& p=2, DD=ALAP	& -73.18	& -146	& p=3, DD=none\\
		 6	& -182	& 196	& -150 ($\times$1)	& -80.15	& -144	& p=2, DD=ALAP	& -45.05	& -118	& p=2, DD=ALAP	& -68.19	& -134	& p=2, DD=ALAP	& -72.39	& -136	& p=3, DD=none\\
		 7	& -204	& 192	& -158 ($\times$2)	& -84.04	& -152	& p=2, DD=ALAP	& -48.90	& -124	& p=2, DD=ALAP	& -64.89	& -142	& p=3, DD=ALAP	& -73.45	& -142	& p=2, DD=none\\
		 8	& -192	& 194	& -150 ($\times$1)	& -81.69	& -144	& p=3, DD=ALAP	& -43.72	& -116	& p=1, DD=ALAP	& -66.76	& -130	& p=2, DD=ALAP	& -68.89	& -138	& p=2, DD=none\\
		 9	& -200	& 202	& -154 ($\times$2)	& -86.35	& -152	& p=2, DD=ALAP	& -48.51	& -122	& p=1, DD=ALAP	& -67.43	& -136	& p=3, DD=ALAP	& -76.69	& -146	& p=2, DD=none\\
		\midrule
		 10	& -198	& 192	& -154 ($\times$2)	& -82.44	& -146	& p=2, DD=ALAP	& -46.11	& -110	& p=2, DD=ALAP	& -67.38	& -140	& p=2, DD=ALAP	& -73.74	& -142	& p=2, DD=none\\
		 11	& -206	& 196	& -160 ($\times$3)	& -84.46	& -156	& p=2, DD=ALAP	& -47.03	& -122	& p=2, DD=ALAP	& -68.22	& -144	& p=3, DD=ALAP	& -70.07	& -138	& p=2, DD=none\\
		 12	& -198	& 198	& -158 ($\times$1)	& -80.13	& -158	& p=2, DD=ALAP	& -45.32	& -114	& p=1, DD=ALAP	& -63.81	& -130	& p=3, DD=ASAP	& -71.95	& -140	& p=2, DD=none\\
		 13	& -202	& 200	& -158 ($\times$1)	& -79.69	& -154	& p=2, DD=ALAP	& -48.52	& -118	& p=2, DD=ALAP	& -64.90	& -140	& p=3, DD=ALAP	& -70.69	& -138	& p=3, DD=none\\
		 14	& -182	& 198	& -154 ($\times$1)	& -79.73	& -154	& p=3, DD=ALAP	& -47.68	& -116	& p=2, DD=ALAP	& -64.79	& -136	& p=3, DD=ALAP	& -69.53	& -132	& p=2, DD=none\\
		 15	& -200	& 190	& -160 ($\times$1)	& -83.22	& -160	& p=3, DD=ALAP	& -51.02	& -120	& p=2, DD=ALAP	& -67.50	& -134	& p=2, DD=ASAP	& -70.34	& -140	& p=2, DD=none\\
		 16	& -198	& 198	& -160 ($\times$2)	& -80.98	& -160	& p=3, DD=ALAP	& -48.03	& -122	& p=2, DD=ALAP	& -67.19	& -136	& p=3, DD=ALAP	& -72.10	& -142	& p=2, DD=none\\
		 17	& -188	& 182	& -152 ($\times$3)	& -78.65	& -144	& p=3, DD=ALAP	& -47.79	& -118	& p=2, DD=ALAP	& -64.56	& -138	& p=2, DD=ALAP	& -69.61	& -134	& p=3, DD=none\\
		 18	& -202	& 190	& -160 ($\times$2)	& -81.83	& -150	& p=2, DD=ALAP	& -45.69	& -126	& p=2, DD=ALAP	& -64.96	& -144	& p=3, DD=ALAP	& -73.13	& -142	& p=2, DD=none\\
		 19	& -200	& 196	& -162 ($\times$2)	& -83.27	& -158	& p=2, DD=ALAP	& -46.21	& -114	& p=1, DD=ALAP	& -61.74	& -130	& p=2, DD=ASAP	& -70.73	& -142	& p=2, DD=none\\
		\midrule
		 20	& -204	& 186	& -160 ($\times$1)	& -82.11	& -156	& p=3, DD=ALAP	& -46.46	& -118	& p=2, DD=ALAP	& -67.00	& -140	& p=3, DD=ALAP	& -73.99	& -148	& p=2, DD=none\\
		 21	& -194	& 200	& -160 ($\times$1)	& -80.87	& -142	& p=3, DD=ALAP	& -45.12	& -118	& p=2, DD=ALAP	& -65.24	& -138	& p=3, DD=ALAP	& -70.59	& -136	& p=2, DD=none\\
		 22	& -216	& 196	& -166 ($\times$1)	& -85.40	& -156	& p=3, DD=ALAP	& -44.32	& -110	& p=2, DD=ALAP	& -68.27	& -162	& p=2, DD=ALAP	& -74.13	& -152	& p=3, DD=none\\
		 23	& -198	& 204	& -158 ($\times$1)	& -81.29	& -152	& p=4, DD=ALAP	& -44.88	& -114	& p=2, DD=ALAP	& -64.74	& -136	& p=2, DD=ASAP	& -72.99	& -146	& p=3, DD=none\\
		 24	& -202	& 194	& -158 ($\times$1)	& -83.97	& -158	& p=2, DD=ALAP	& -49.08	& -118	& p=2, DD=ALAP	& -67.88	& -144	& p=3, DD=ALAP	& -74.36	& -144	& p=2, DD=none\\
		 25	& -192	& 190	& -156 ($\times$1)	& -80.85	& -156	& p=3, DD=ALAP	& -47.11	& -114	& p=2, DD=ALAP	& -63.67	& -138	& p=2, DD=ALAP	& -71.20	& -136	& p=3, DD=none\\
		 26	& -198	& 194	& -158 ($\times$2)	& -82.79	& -158	& p=2, DD=ALAP	& -47.85	& -116	& p=2, DD=ALAP	& -69.55	& -148	& p=2, DD=ASAP	& -74.40	& -148	& p=3, DD=none\\
		 27	& -198	& 208	& -150 ($\times$1)	& -78.85	& -148	& p=3, DD=ASAP	& -49.53	& -120	& p=2, DD=ALAP	& -64.84	& -138	& p=2, DD=ALAP	& -70.88	& -136	& p=2, DD=none\\
		 28	& -194	& 190	& -152 ($\times$1)	& -79.58	& -146	& p=2, DD=ALAP	& -47.34	& -114	& p=1, DD=ALAP	& -64.00	& -138	& p=2, DD=ALAP	& -71.53	& -144	& p=3, DD=none\\
		 29	& -196	& 196	& -152 ($\times$1)	& -80.95	& -144	& p=2, DD=ALAP	& -46.39	& -116	& p=1, DD=ALAP	& -65.08	& -132	& p=2, DD=ALAP	& -72.81	& -140	& p=2, DD=none\\
		\midrule
		 30	& -196	& 200	& -160 ($\times$1)	& -79.33	& -150	& p=2, DD=ALAP	& -47.08	& -114	& p=1, DD=ALAP	& -62.90	& -132	& p=2, DD=ASAP	& -70.87	& -138	& p=2, DD=none\\
		 31	& -200	& 196	& -154 ($\times$1)	& -82.09	& -150	& p=2, DD=ALAP	& -48.07	& -124	& p=2, DD=ALAP	& -64.29	& -128	& p=2, DD=ALAP	& -74.18	& -154	& p=2, DD=none\\
		 32	& -194	& 182	& -152 ($\times$2)	& -80.87	& -150	& p=2, DD=ALAP	& -50.59	& -118	& p=2, DD=ALAP	& -64.37	& -134	& p=3, DD=ALAP	& -68.28	& -140	& p=3, DD=none\\
		 33	& -206	& 192	& -170 ($\times$1)	& -81.50	& -152	& p=2, DD=ALAP	& -46.62	& -130	& p=2, DD=ALAP	& -67.57	& -150	& p=3, DD=ALAP	& -75.39	& -146	& p=3, DD=none\\
		 34	& -190	& 208	& -148 ($\times$1)	& -79.68	& -148	& p=3, DD=ALAP	& -44.12	& -116	& p=2, DD=ALAP	& -66.63	& -130	& p=3, DD=ALAP	& -71.01	& -130	& p=2, DD=none\\
		 35	& -206	& 202	& -164 ($\times$1)	& -84.09	& -154	& p=4, DD=ALAP	& -45.83	& -110	& p=2, DD=ALAP	& -64.05	& -146	& p=2, DD=ALAP	& -71.38	& -140	& p=2, DD=none\\
		 36	& -202	& 196	& -156 ($\times$6)	& -86.80	& -156	& p=2, DD=ALAP	& -49.90	& -132	& p=2, DD=ALAP	& -71.04	& -140	& p=3, DD=ASAP	& -75.27	& -140	& p=2, DD=none\\
		 37	& -190	& 204	& -154 ($\times$1)	& -78.56	& -144	& p=3, DD=ALAP	& -45.76	& -122	& p=2, DD=ALAP	& -63.80	& -132	& p=3, DD=ALAP	& -70.88	& -140	& p=2, DD=none\\
		 38	& -184	& 192	& -150 ($\times$2)	& -78.26	& -150	& p=2, DD=ALAP	& -43.86	& -114	& p=2, DD=ALAP	& -61.96	& -130	& p=2, DD=ALAP	& -71.17	& -138	& p=3, DD=none\\
		 39	& -206	& 190	& -156 ($\times$4)	& -82.67	& -156	& p=2, DD=ALAP	& -44.77	& -116	& p=2, DD=ALAP	& -65.37	& -134	& p=3, DD=ALAP	& -72.53	& -152	& p=4, DD=none\\
		\midrule
		 40	& -196	& 204	& -158 ($\times$1)	& -78.58	& -158	& p=3, DD=ALAP	& -45.80	& -120	& p=1, DD=ALAP	& -62.23	& -130	& p=2, DD=ASAP	& -70.65	& -138	& p=2, DD=none\\
		 41	& -192	& 190	& -150 ($\times$1)	& -77.36	& -144	& p=3, DD=ALAP	& -46.95	& -124	& p=1, DD=ALAP	& -61.01	& -132	& p=2, DD=ASAP	& -67.56	& -132	& p=2, DD=none\\
		 42	& -218	& 202	& -164 ($\times$1)	& -85.96	& -154	& p=2, DD=ALAP	& -51.18	& -128	& p=2, DD=ALAP	& -64.57	& -136	& p=2, DD=ASAP	& -74.02	& -146	& p=2, DD=none\\
		 43	& -198	& 198	& -156 ($\times$1)	& -83.41	& -152	& p=3, DD=ALAP	& -48.03	& -128	& p=2, DD=ALAP	& -64.65	& -132	& p=2, DD=ASAP	& -69.73	& -140	& p=2, DD=none\\
		 44	& -206	& 190	& -166 ($\times$2)	& -86.63	& -156	& p=4, DD=ALAP	& -48.53	& -120	& p=1, DD=ALAP	& -69.93	& -140	& p=2, DD=ALAP	& -71.51	& -142	& p=3, DD=none\\
		 45	& -212	& 190	& -166 ($\times$1)	& -84.47	& -154	& p=3, DD=ALAP	& -46.22	& -128	& p=1, DD=ALAP	& -66.93	& -136	& p=2, DD=ALAP	& -70.21	& -146	& p=2, DD=none\\
		 46	& -202	& 198	& -168 ($\times$1)	& -83.90	& -158	& p=4, DD=ALAP	& -45.60	& -120	& p=2, DD=ALAP	& -66.84	& -146	& p=2, DD=ALAP	& -72.27	& -142	& p=2, DD=none\\
		 47	& -192	& 190	& -156 ($\times$1)	& -82.92	& -144	& p=2, DD=ALAP	& -46.08	& -118	& p=2, DD=ALAP	& -68.31	& -132	& p=3, DD=ALAP	& -67.50	& -136	& p=3, DD=none\\
		 48	& -188	& 186	& -154 ($\times$1)	& -80.78	& -154	& p=4, DD=ALAP	& -46.01	& -124	& p=1, DD=ALAP	& -64.40	& -124	& p=2, DD=ALAP	& -71.25	& -138	& p=3, DD=none\\
		 49	& -210	& 186	& -176 ($\times$1)	& -86.88	& -168	& p=3, DD=ALAP	& -48.88	& -126	& p=2, DD=ALAP	& -68.97	& -146	& p=2, DD=ALAP	& -70.59	& -142	& p=2, DD=none\\
		\midrule
		 50	& -186	& 202	& -150 ($\times$2)	& -82.98	& -144	& p=2, DD=ALAP	& -48.21	& -124	& p=2, DD=ALAP	& -64.00	& -130	& p=2, DD=ASAP	& -69.22	& -134	& p=2, DD=none\\
		 51	& -192	& 192	& -150 ($\times$1)	& -82.73	& -144	& p=2, DD=ALAP	& -48.29	& -124	& p=2, DD=ALAP	& -61.66	& -130	& p=2, DD=ALAP	& -72.11	& -142	& p=2, DD=none\\
		 52	& -202	& 194	& -158 ($\times$2)	& -82.34	& -158	& p=4, DD=ASAP	& -48.04	& -116	& p=2, DD=ALAP	& -60.74	& -134	& p=2, DD=ASAP	& -68.28	& -144	& p=3, DD=none\\
		 53	& -194	& 214	& -154 ($\times$1)	& -81.78	& -144	& p=2, DD=ALAP	& -47.87	& -110	& p=1, DD=ALAP	& -67.16	& -136	& p=3, DD=ALAP	& -71.14	& -132	& p=3, DD=none\\
		 54	& -208	& 200	& -158 ($\times$1)	& -81.03	& -150	& p=2, DD=ALAP	& -50.15	& -114	& p=2, DD=ALAP	& -66.23	& -134	& p=2, DD=ALAP	& -71.47	& -142	& p=2, DD=none\\
		 55	& -194	& 196	& -160 ($\times$1)	& -85.15	& -154	& p=3, DD=ALAP	& -48.12	& -112	& p=2, DD=ALAP	& -66.11	& -136	& p=2, DD=ASAP	& -75.02	& -140	& p=2, DD=none\\
		 56	& -200	& 188	& -156 ($\times$1)	& -84.11	& -154	& p=3, DD=ALAP	& -48.74	& -124	& p=2, DD=ALAP	& -65.45	& -134	& p=2, DD=ALAP	& -72.34	& -138	& p=3, DD=none\\
		 57	& -192	& 188	& -156 ($\times$1)	& -79.15	& -142	& p=3, DD=ALAP	& -41.94	& -108	& p=1, DD=ALAP	& -61.65	& -140	& p=2, DD=ALAP	& -67.74	& -132	& p=2, DD=none\\
		 58	& -202	& 202	& -154 ($\times$3)	& -78.48	& -152	& p=2, DD=ALAP	& -45.46	& -116	& p=2, DD=ALAP	& -63.29	& -130	& p=2, DD=ALAP	& -70.29	& -146	& p=2, DD=none\\
		 59	& -212	& 182	& -166 ($\times$2)	& -88.49	& -158	& p=2, DD=ALAP	& -52.16	& -130	& p=2, DD=ALAP	& -66.81	& -154	& p=2, DD=ALAP	& -72.04	& -160	& p=3, DD=none\\
		\midrule
		 60	& -208	& 180	& -174 ($\times$1)	& -86.19	& -154	& p=3, DD=ALAP	& -44.94	& -122	& p=1, DD=ALAP	& -65.20	& -138	& p=2, DD=ALAP	& -69.82	& -150	& p=2, DD=none\\
		 61	& -194	& 208	& -150 ($\times$1)	& -81.90	& -142	& p=2, DD=ALAP	& -44.48	& -110	& p=1, DD=ALAP	& -64.20	& -130	& p=2, DD=ALAP	& -71.58	& -132	& p=2, DD=none\\
		 62	& -202	& 196	& -152 ($\times$2)	& -80.37	& -146	& p=2, DD=ALAP	& -44.43	& -116	& p=2, DD=ALAP	& -65.08	& -134	& p=2, DD=ALAP	& -67.74	& -140	& p=3, DD=none\\
		 63	& -208	& 192	& -166 ($\times$1)	& -84.85	& -166	& p=3, DD=ALAP	& -48.54	& -118	& p=2, DD=ALAP	& -66.42	& -152	& p=2, DD=ALAP	& -73.29	& -154	& p=2, DD=none\\
		 64	& -200	& 202	& -158 ($\times$1)	& -82.51	& -148	& p=2, DD=ALAP	& -49.25	& -118	& p=2, DD=ALAP	& -66.91	& -140	& p=2, DD=ALAP	& -73.56	& -142	& p=2, DD=none\\
		 65	& -200	& 206	& -152 ($\times$4)	& -79.76	& -140	& p=2, DD=ALAP	& -47.29	& -116	& p=2, DD=ALAP	& -67.56	& -134	& p=2, DD=ALAP	& -72.74	& -140	& p=3, DD=none\\
		 66	& -204	& 188	& -158 ($\times$2)	& -83.04	& -154	& p=2, DD=ALAP	& -45.51	& -114	& p=2, DD=ALAP	& -65.10	& -150	& p=3, DD=ALAP	& -72.54	& -144	& p=3, DD=none\\
		 67	& -196	& 202	& -158 ($\times$1)	& -83.09	& -154	& p=3, DD=ALAP	& -44.72	& -114	& p=2, DD=ALAP	& -65.67	& -132	& p=2, DD=ALAP	& -72.22	& -144	& p=3, DD=none\\
		 68	& -204	& 184	& -162 ($\times$1)	& -84.08	& -162	& p=3, DD=ALAP	& -43.25	& -116	& p=1, DD=ALAP	& -68.52	& -144	& p=3, DD=ALAP	& -71.11	& -148	& p=3, DD=none\\
		 69	& -186	& 196	& -152 ($\times$1)	& -80.54	& -148	& p=2, DD=ALAP	& -43.00	& -124	& p=2, DD=ALAP	& -62.29	& -132	& p=2, DD=ALAP	& -68.31	& -132	& p=3, DD=none\\
		\midrule
		 70	& -194	& 196	& -152 ($\times$3)	& -79.24	& -142	& p=2, DD=ALAP	& -46.98	& -110	& p=2, DD=ALAP	& -64.09	& -130	& p=2, DD=ALAP	& -72.93	& -152	& p=3, DD=none\\
		 71	& -192	& 184	& -156 ($\times$1)	& -80.06	& -150	& p=4, DD=ALAP	& -45.94	& -112	& p=2, DD=ALAP	& -66.14	& -142	& p=3, DD=ALAP	& -69.73	& -140	& p=2, DD=none\\
		 72	& -190	& 190	& -150 ($\times$1)	& -83.54	& -146	& p=2, DD=ALAP	& -49.92	& -126	& p=1, DD=ALAP	& -65.14	& -136	& p=2, DD=ALAP	& -71.05	& -134	& p=3, DD=none\\
		 73	& -194	& 194	& -160 ($\times$1)	& -82.96	& -154	& p=3, DD=ALAP	& -46.31	& -116	& p=2, DD=ALAP	& -65.35	& -136	& p=2, DD=ALAP	& -73.13	& -136	& p=2, DD=none\\
		 74	& -188	& 218	& -146 ($\times$1)	& -79.39	& -140	& p=3, DD=ALAP	& -49.73	& -118	& p=2, DD=ALAP	& -66.34	& -132	& p=2, DD=ALAP	& -68.28	& -132	& p=3, DD=none\\
		 75	& -202	& 196	& -162 ($\times$1)	& -82.93	& -158	& p=3, DD=ALAP	& -50.31	& -118	& p=2, DD=ALAP	& -68.84	& -146	& p=2, DD=ALAP	& -71.15	& -146	& p=2, DD=none\\
		 76	& -186	& 200	& -144 ($\times$1)	& -79.21	& -140	& p=2, DD=ALAP	& -46.79	& -114	& p=2, DD=ALAP	& -67.52	& -138	& p=2, DD=ALAP	& -74.88	& -136	& p=3, DD=none\\
		 77	& -204	& 196	& -152 ($\times$1)	& -83.49	& -148	& p=2, DD=ALAP	& -47.08	& -130	& p=1, DD=ALAP	& -66.10	& -138	& p=3, DD=ASAP	& -71.61	& -140	& p=3, DD=none\\
		 78	& -198	& 196	& -152 ($\times$1)	& -82.51	& -148	& p=3, DD=ALAP	& -46.89	& -112	& p=2, DD=ALAP	& -63.64	& -134	& p=2, DD=ASAP	& -72.54	& -144	& p=2, DD=none\\
		 79	& -206	& 192	& -156 ($\times$2)	& -80.16	& -156	& p=2, DD=ASAP	& -45.54	& -114	& p=1, DD=ALAP	& -66.14	& -136	& p=2, DD=ALAP	& -68.94	& -138	& p=3, DD=none\\
		\midrule
		 80	& -198	& 184	& -162 ($\times$1)	& -85.67	& -156	& p=4, DD=ALAP	& -47.32	& -112	& p=2, DD=ALAP	& -65.56	& -136	& p=3, DD=ALAP	& -73.44	& -142	& p=3, DD=none\\
		 81	& -196	& 196	& -156 ($\times$1)	& -82.98	& -152	& p=2, DD=ALAP	& -48.05	& -116	& p=2, DD=ALAP	& -63.20	& -130	& p=2, DD=ALAP	& -73.60	& -146	& p=3, DD=none\\
		 82	& -192	& 200	& -152 ($\times$2)	& -78.50	& -152	& p=2, DD=ALAP	& -44.80	& -116	& p=2, DD=ALAP	& -64.75	& -128	& p=2, DD=ALAP	& -69.70	& -130	& p=3, DD=none\\
		 83	& -192	& 194	& -154 ($\times$1)	& -85.96	& -152	& p=2, DD=ALAP	& -47.70	& -118	& p=2, DD=ALAP	& -67.02	& -146	& p=3, DD=ALAP	& -77.91	& -152	& p=2, DD=none\\
		 84	& -206	& 186	& -166 ($\times$1)	& -88.27	& -166	& p=3, DD=ASAP	& -51.70	& -122	& p=2, DD=ALAP	& -69.01	& -150	& p=3, DD=ALAP	& -77.46	& -142	& p=2, DD=none\\
		 85	& -196	& 200	& -154 ($\times$2)	& -83.82	& -154	& p=3, DD=ALAP	& -47.38	& -118	& p=2, DD=ALAP	& -64.76	& -130	& p=2, DD=ALAP	& -71.25	& -144	& p=3, DD=none\\
		 86	& -200	& 206	& -158 ($\times$2)	& -80.00	& -158	& p=4, DD=ALAP	& -44.95	& -122	& p=1, DD=ALAP	& -63.87	& -140	& p=2, DD=ALAP	& -69.19	& -134	& p=3, DD=none\\
		 87	& -204	& 178	& -160 ($\times$1)	& -86.58	& -158	& p=2, DD=ALAP	& -49.31	& -120	& p=2, DD=ALAP	& -70.03	& -142	& p=2, DD=ALAP	& -77.74	& -150	& p=2, DD=none\\
		 88	& -196	& 194	& -154 ($\times$1)	& -79.74	& -146	& p=2, DD=ALAP	& -45.30	& -112	& p=2, DD=ALAP	& -62.02	& -152	& p=3, DD=ASAP	& -73.22	& -138	& p=3, DD=none\\
		 89	& -192	& 204	& -150 ($\times$1)	& -78.42	& -146	& p=2, DD=ALAP	& -47.24	& -116	& p=2, DD=ALAP	& -61.94	& -122	& p=3, DD=ALAP	& -70.08	& -134	& p=3, DD=none\\
		\midrule
		 90	& -208	& 204	& -162 ($\times$2)	& -84.22	& -156	& p=2, DD=ALAP	& -46.32	& -120	& p=2, DD=ALAP	& -64.64	& -136	& p=2, DD=ALAP	& -68.90	& -142	& p=3, DD=none\\
		 91	& -198	& 194	& -156 ($\times$1)	& -81.15	& -156	& p=4, DD=ALAP	& -41.78	& -122	& p=2, DD=ALAP	& -63.84	& -130	& p=2, DD=ALAP	& -66.13	& -148	& p=2, DD=none\\
		 92	& -206	& 192	& -162 ($\times$1)	& -80.94	& -158	& p=3, DD=ALAP	& -44.02	& -110	& p=1, DD=ALAP	& -63.60	& -138	& p=2, DD=ASAP	& -68.01	& -156	& p=3, DD=none\\
		 93	& -188	& 192	& -150 ($\times$1)	& -78.66	& -138	& p=2, DD=ALAP	& -45.06	& -116	& p=2, DD=ALAP	& -62.83	& -126	& p=2, DD=ALAP	& -70.88	& -144	& p=2, DD=none\\
		 94	& -200	& 194	& -156 ($\times$1)	& -83.31	& -154	& p=3, DD=ALAP	& -44.73	& -116	& p=2, DD=ALAP	& -64.58	& -132	& p=2, DD=ALAP	& -73.90	& -138	& p=2, DD=none\\
		 95	& -198	& 184	& -156 ($\times$1)	& -81.58	& -150	& p=2, DD=ALAP	& -46.13	& -114	& p=2, DD=ALAP	& -65.60	& -142	& p=2, DD=ALAP	& -70.02	& -144	& p=3, DD=none\\
		 96	& -194	& 202	& -150 ($\times$2)	& -80.98	& -150	& p=3, DD=ALAP	& -44.02	& -118	& p=2, DD=ALAP	& -65.37	& -136	& p=2, DD=ASAP	& -71.41	& -134	& p=3, DD=none\\
		 97	& -198	& 192	& -162 ($\times$1)	& -82.55	& -150	& p=2, DD=ALAP	& -45.90	& -114	& p=2, DD=ALAP	& -62.76	& -132	& p=3, DD=ALAP	& -70.73	& -134	& p=2, DD=none\\
		 98	& -202	& 196	& -154 ($\times$4)	& -83.37	& -154	& p=3, DD=ALAP	& -46.38	& -116	& p=1, DD=ALAP	& -68.35	& -146	& p=2, DD=ALAP	& -71.34	& -146	& p=3, DD=none\\
		 99	& -188	& 192	& -150 ($\times$1)	& -82.46	& -146	& p=3, DD=ALAP	& -43.76	& -108	& p=2, DD=ALAP	& -64.69	& -132	& p=2, DD=ALAP	& -70.31	& -150	& p=2, DD=none\\
		\bottomrule
    \end{tabular}}
    \caption{Summary of the most successful QAOA runs for all 100 random 127-qubit instances on 4 IBMQ devices:
    We give the overall QAOA minimum sample energy across $20,000$ shots each on all $4$ devices, for each layer number $1\leq p\leq5$, and for each of the $3$ dynamical decoupling schemes $\mathit{DD}$, counting multiplicities.
    For each device, we give the \emph{parameters} $p$ and $\mathit{DD}$ that achieve the best \emph{sample mean} and also show the \emph{sample minimum} for these.
    }
    \label{table:QAOA_minimums_127_variable}
\end{table}

\begin{table}[t!]
    \centering
    \setlength{\deviceseplength}{28pt}
    \newcommand{\devicesep}{\hspace*{\deviceseplength}}
    \renewcommand{\arraystretch}{.85}
    \ \hfill
    \scalebox{0.53}{\begin{tabular}{@{}l@{\devicesep}rr@{\devicesep}r@{}r@{}}
        \toprule
        $27$-qubit       & \multicolumn{2}{r@{\devicesep}}{instance energies}  & overall\; & QAOA    \\
        \cmidrule(lr{\deviceseplength}){2-3}
        instance       & ground state   & maximum   & min sample\; & energy   \\ 
        \midrule[\heavyrulewidth]
		 0	& -42	& 32	& -42 ($\times$	&  466)	\\
		 1	& -36	& 40	& -36 ($\times$	&  946)	\\
		 2	& -38	& 40	& -38 ($\times$	&  108)	\\
		 3	& -36	& 42	& -36 ($\times$	&  146)	\\
		 4	& -36	& 44	& -36 ($\times$	&  230)	\\
		 5	& -42	& 38	& -42 ($\times$	&  342)	\\
		 6	& -40	& 38	& -40 ($\times$	&  282)	\\
		 7	& -42	& 32	& -42 ($\times$	&  150)	\\
		 8	& -42	& 38	& -42 ($\times$	&  234)	\\
		 9	& -44	& 44	& -44 ($\times$	&  138)	\\
		\midrule
		 10	& -42	& 38	& -42 ($\times$	&  285)	\\
		 11	& -40	& 40	& -40 ($\times$	&  225)	\\
		 12	& -38	& 42	& -38 ($\times$	&  164)	\\
		 13	& -36	& 38	& -36 ($\times$	&  540)	\\
		 14	& -40	& 34	& -40 ($\times$	&  143)	\\
		 15	& -34	& 34	& -34 ($\times$	& 3602)	\\
		 16	& -38	& 38	& -38 ($\times$	&  221)	\\
		 17	& -40	& 34	& -40 ($\times$	&  170)	\\
		 18	& -42	& 36	& -42 ($\times$	&  303)	\\
		 19	& -36	& 36	& -36 ($\times$	&  693)	\\
		\midrule
		 20	& -34	& 42	& -34 ($\times$	&  130)	\\
		 21	& -36	& 34	& -36 ($\times$	&  321)	\\
		 22	& -36	& 40	& -36 ($\times$	&  336)	\\
		 23	& -36	& 36	& -36 ($\times$	& 2315)	\\
		 24	& -34	& 40	& -34 ($\times$	&  817)	\\
		 25	& -38	& 38	& -38 ($\times$	&  455)	\\
		 26	& -38	& 38	& -38 ($\times$	&  512)	\\
		 27	& -46	& 38	& -46 ($\times$	&   65)	\\
		 28	& -32	& 36	& -32 ($\times$	&  636)	\\
		 29	& -34	& 40	& -34 ($\times$	&  738)	\\
		\midrule
		 30	& -36	& 50	& -36 ($\times$	&   64)	\\
		 31	& -40	& 34	& -40 ($\times$	&  592)	\\
		 32	& -44	& 38	& -44 ($\times$	&   98)	\\
		 33	& -36	& 34	& -36 ($\times$	&   50)	\\
		 34	& -32	& 38	& -32 ($\times$	& 1015)	\\
		 35	& -40	& 34	& -40 ($\times$	& 2213)	\\
		 36	& -38	& 42	& -38 ($\times$	&  444)	\\
		 37	& -32	& 36	& -32 ($\times$	& 2162)	\\
		 38	& -34	& 40	& -34 ($\times$	&  247)	\\
		 39	& -38	& 32	& -38 ($\times$	&  747)	\\
		\midrule
		 40	& -36	& 44	& -36 ($\times$	&  198)	\\
		 41	& -38	& 40	& -38 ($\times$	&  291)	\\
		 42	& -38	& 40	& -38 ($\times$	&  445)	\\
		 43	& -38	& 38	& -38 ($\times$	& 2899)	\\
		 44	& -36	& 38	& -36 ($\times$	& 1023)	\\
		 45	& -42	& 44	& -42 ($\times$	&  348)	\\
		 46	& -40	& 44	& -40 ($\times$	&   54)	\\
		 47	& -36	& 42	& -36 ($\times$	&  549)	\\
		 48	& -42	& 38	& -42 ($\times$	&  337)	\\
		 49	& -44	& 40	& -44 ($\times$	&  243)	\\
		\midrule
		 50	& -38	& 42	& -38 ($\times$	& 1175)	\\
		 51	& -38	& 38	& -38 ($\times$	&  103)	\\
		 52	& -34	& 42	& -34 ($\times$	&  521)	\\
		 53	& -40	& 46	& -40 ($\times$	&  286)	\\
		 54	& -38	& 36	& -38 ($\times$	&  234)	\\
		 55	& -40	& 46	& -40 ($\times$	&   88)	\\
		 56	& -42	& 34	& -42 ($\times$	&  647)	\\
		 57	& -40	& 40	& -40 ($\times$	&  311)	\\
		 58	& -38	& 38	& -38 ($\times$	&  487)	\\
		 59	& -40	& 40	& -40 ($\times$	&  307)	\\
		\midrule
		 60	& -34	& 40	& -34 ($\times$	&  967)	\\
		 61	& -44	& 38	& -44 ($\times$	&  250)	\\
		 62	& -36	& 34	& -36 ($\times$	& 1049)	\\
		 63	& -38	& 40	& -38 ($\times$	&  131)	\\
		 64	& -36	& 36	& -36 ($\times$	& 1999)	\\
		 65	& -38	& 36	& -38 ($\times$	& 1233)	\\
		 66	& -42	& 32	& -42 ($\times$	& 1404)	\\
		 67	& -44	& 40	& -44 ($\times$	&  510)	\\
		 68	& -42	& 34	& -42 ($\times$	&  221)	\\
		 69	& -36	& 36	& -36 ($\times$	& 1019)	\\
		\midrule
		 70	& -40	& 40	& -40 ($\times$	&  559)	\\
		 71	& -40	& 40	& -40 ($\times$	&  899)	\\
		 72	& -42	& 34	& -42 ($\times$	&  555)	\\
		 73	& -44	& 38	& -44 ($\times$	&  873)	\\
		 74	& -36	& 40	& -36 ($\times$	&  942)	\\
		 75	& -40	& 44	& -40 ($\times$	&   20)	\\
		 76	& -38	& 40	& -38 ($\times$	&  315)	\\
		 77	& -40	& 32	& -40 ($\times$	&  395)	\\
		 78	& -38	& 40	& -38 ($\times$	&  170)	\\
		 79	& -40	& 40	& -40 ($\times$	& 1190)	\\
		\midrule
		 80	& -36	& 36	& -36 ($\times$	&  949)	\\
		 81	& -40	& 38	& -40 ($\times$	& 1396)	\\
		 82	& -36	& 40	& -36 ($\times$	& 1052)	\\
		 83	& -42	& 48	& -42 ($\times$	&   96)	\\
		 84	& -42	& 36	& -42 ($\times$	&  138)	\\
		 85	& -40	& 38	& -40 ($\times$	&  600)	\\
		 86	& -36	& 38	& -36 ($\times$	& 1645)	\\
		 87	& -40	& 38	& -40 ($\times$	&  241)	\\
		 88	& -36	& 40	& -36 ($\times$	& 2150)	\\
		 89	& -36	& 36	& -36 ($\times$	&  430)	\\
		\midrule
		 90	& -34	& 40	& -34 ($\times$	&  361)	\\
		 91	& -38	& 38	& -38 ($\times$	&  406)	\\
		 92	& -34	& 40	& -34 ($\times$	& 2145)	\\
		 93	& -46	& 42	& -46 ($\times$	&   73)	\\
		 94	& -40	& 38	& -40 ($\times$	&  116)	\\
		 95	& -38	& 38	& -38 ($\times$	&  605)	\\
		 96	& -36	& 44	& -36 ($\times$	&  765)	\\
		 97	& -36	& 34	& -36 ($\times$	&  827)	\\
		 98	& -40	& 40	& -40 ($\times$	&  399)	\\
		 99	& -38	& 34	& -38 ($\times$	& 1983)	\\
		\bottomrule
    \end{tabular}}%
    \hfill%
        \scalebox{0.53}{\begin{tabular}{@{}l@{\devicesep}rr@{\devicesep}r@{}r@{}}
        \toprule
        $16$-qubit       & \multicolumn{2}{r@{\devicesep}}{instance energies}  & overall\;  & QAOA    \\
        \cmidrule(lr{\deviceseplength}){2-3}
        instance       & ground state   & maximum   & min sample\; & energy   \\ 
        \midrule[\heavyrulewidth]
		 0	& -22	& 22	& -22 ($\times$	&  493)	\\
		 1	& -20	& 24	& -20 ($\times$	& 3293)	\\
		 2	& -20	& 18	& -20 ($\times$	& 4112)	\\
		 3	& -26	& 26	& -26 ($\times$	&  529)	\\
		 4	& -20	& 20	& -20 ($\times$	& 1804)	\\
		 5	& -24	& 18	& -24 ($\times$	&  836)	\\
		 6	& -20	& 24	& -20 ($\times$	& 3213)	\\
		 7	& -22	& 22	& -22 ($\times$	& 3368)	\\
		 8	& -24	& 24	& -24 ($\times$	&  668)	\\
		 9	& -24	& 24	& -24 ($\times$	& 1071)	\\
		\midrule
		 10	& -24	& 18	& -24 ($\times$	& 1671)	\\
		 11	& -22	& 18	& -22 ($\times$	& 4183)	\\
		 12	& -22	& 20	& -22 ($\times$	& 8093)	\\
		 13	& -22	& 24	& -22 ($\times$	& 3373)	\\
		 14	& -20	& 20	& -20 ($\times$	& 6119)	\\
		 15	& -24	& 20	& -24 ($\times$	& 3854)	\\
		 16	& -24	& 24	& -24 ($\times$	&  265)	\\
		 17	& -20	& 20	& -20 ($\times$	& 1865)	\\
		 18	& -22	& 26	& -22 ($\times$	& 3470)	\\
		 19	& -22	& 26	& -22 ($\times$	&  443)	\\
		\midrule
		 20	& -24	& 20	& -24 ($\times$	& 4980)	\\
		 21	& -22	& 18	& -22 ($\times$	& 4526)	\\
		 22	& -20	& 26	& -20 ($\times$	& 2338)	\\
		 23	& -24	& 22	& -24 ($\times$	& 3674)	\\
		 24	& -20	& 24	& -20 ($\times$	&  924)	\\
		 25	& -24	& 22	& -24 ($\times$	& 3166)	\\
		 26	& -20	& 24	& -20 ($\times$	& 6678)	\\
		 27	& -22	& 26	& -22 ($\times$	&  399)	\\
		 28	& -24	& 20	& -24 ($\times$	& 3079)	\\
		 29	& -20	& 24	& -20 ($\times$	& 6886)	\\
		\midrule
		 30	& -20	& 24	& -20 ($\times$	&  647)	\\
		 31	& -24	& 20	& -24 ($\times$	& 2072)	\\
		 32	& -26	& 20	& -26 ($\times$	& 1106)	\\
		 33	& -20	& 22	& -20 ($\times$	& 2860)	\\
		 34	& -20	& 20	& -20 ($\times$	& 6171)	\\
		 35	& -24	& 20	& -24 ($\times$	& 3420)	\\
		 36	& -18	& 24	& -18 ($\times$	& 12616)	\\
		 37	& -20	& 20	& -20 ($\times$	& 4825)	\\
		 38	& -24	& 22	& -24 ($\times$	& 1269)	\\
		 39	& -22	& 20	& -22 ($\times$	& 2189)	\\
		\midrule
		 40	& -24	& 24	& -24 ($\times$	& 2063)	\\
		 41	& -24	& 22	& -24 ($\times$	& 1345)	\\
		 42	& -20	& 24	& -20 ($\times$	& 2080)	\\
		 43	& -22	& 26	& -22 ($\times$	& 1407)	\\
		 44	& -20	& 26	& -20 ($\times$	& 1615)	\\
		 45	& -24	& 20	& -24 ($\times$	& 2165)	\\
		 46	& -22	& 24	& -22 ($\times$	& 4005)	\\
		 47	& -26	& 18	& -26 ($\times$	& 1501)	\\
		 48	& -20	& 28	& -20 ($\times$	& 1276)	\\
		 49	& -24	& 20	& -24 ($\times$	& 2858)	\\
		\midrule
		 50	& -22	& 26	& -22 ($\times$	& 1069)	\\
		 51	& -20	& 20	& -20 ($\times$	& 1665)	\\
		 52	& -26	& 22	& -26 ($\times$	& 1703)	\\
		 53	& -24	& 26	& -24 ($\times$	&  316)	\\
		 54	& -26	& 22	& -26 ($\times$	&  665)	\\
		 55	& -28	& 20	& -28 ($\times$	& 2560)	\\
		 56	& -22	& 22	& -22 ($\times$	& 4350)	\\
		 57	& -26	& 22	& -26 ($\times$	&  945)	\\
		 58	& -22	& 24	& -22 ($\times$	& 2609)	\\
		 59	& -28	& 20	& -28 ($\times$	& 1511)	\\
		\midrule
		 60	& -26	& 20	& -26 ($\times$	&  509)	\\
		 61	& -24	& 20	& -24 ($\times$	&  675)	\\
		 62	& -22	& 20	& -22 ($\times$	& 1399)	\\
		 63	& -20	& 24	& -20 ($\times$	& 2315)	\\
		 64	& -26	& 22	& -26 ($\times$	&  904)	\\
		 65	& -18	& 20	& -18 ($\times$	& 10580)	\\
		 66	& -24	& 26	& -24 ($\times$	& 2729)	\\
		 67	& -24	& 24	& -24 ($\times$	& 6442)	\\
		 68	& -20	& 18	& -20 ($\times$	& 2098)	\\
		 69	& -20	& 24	& -20 ($\times$	&  725)	\\
		\midrule
		 70	& -24	& 22	& -24 ($\times$	& 1003)	\\
		 71	& -24	& 24	& -24 ($\times$	& 2151)	\\
		 72	& -20	& 24	& -20 ($\times$	&  990)	\\
		 73	& -20	& 22	& -20 ($\times$	& 2952)	\\
		 74	& -26	& 22	& -26 ($\times$	& 3743)	\\
		 75	& -20	& 26	& -20 ($\times$	& 2847)	\\
		 76	& -20	& 20	& -20 ($\times$	& 5321)	\\
		 77	& -26	& 22	& -26 ($\times$	& 1169)	\\
		 78	& -22	& 22	& -22 ($\times$	& 2885)	\\
		 79	& -20	& 22	& -20 ($\times$	& 3244)	\\
		\midrule
		 80	& -24	& 22	& -24 ($\times$	&  761)	\\
		 81	& -18	& 26	& -18 ($\times$	& 6474)	\\
		 82	& -24	& 20	& -24 ($\times$	& 1523)	\\
		 83	& -26	& 18	& -26 ($\times$	&  812)	\\
		 84	& -20	& 24	& -20 ($\times$	& 1692)	\\
		 85	& -26	& 24	& -26 ($\times$	&  634)	\\
		 86	& -20	& 22	& -20 ($\times$	& 6212)	\\
		 87	& -22	& 22	& -22 ($\times$	& 3950)	\\
		 88	& -22	& 28	& -22 ($\times$	&  334)	\\
		 89	& -24	& 20	& -24 ($\times$	& 2507)	\\
		\midrule
		 90	& -18	& 22	& -18 ($\times$	& 8225)	\\
		 91	& -26	& 20	& -26 ($\times$	& 1526)	\\
		 92	& -30	& 26	& -30 ($\times$	&  825)	\\
		 93	& -20	& 18	& -20 ($\times$	& 3139)	\\
		 94	& -22	& 22	& -22 ($\times$	& 1305)	\\
		 95	& -20	& 26	& -20 ($\times$	&  656)	\\
		 96	& -24	& 22	& -24 ($\times$	& 3101)	\\
		 97	& -22	& 20	& -22 ($\times$	& 1430)	\\
		 98	& -20	& 24	& -20 ($\times$	& 6131)	\\
		 99	& -20	& 20	& -20 ($\times$	& 4196)	\\
		\bottomrule
    \end{tabular}}\hfill\ %
    \vspace*{-1ex}
    \caption{Summary of the hardware runs of all 100 random $27$-qubit instances and all 100 random $16$-qubit instances:
    \textbf{(left)} $27$-qubit instances: We give the overall QAOA minimum sample energy across 20,000 shots each run on all 6~IBMQ devices,
    for all $1\leq p\leq 5$, and for all 3 dynamical decoupling schemes, also counting multiple appearances.
    \textbf{(right)} $16$-qubit instances: We give the same QAOA minimum sample data, with instances run on \texttt{ibm\_guadalupe}.
    For each of the 200 instances, the ground-state energy was found multiple times, though not necessarily in each of the experiments
    (e.g., for $27$-qubit instance $75$, where the ground-state was found 20 times across the $90$ experiments).
    }
    \label{table:QAOA_minimums_16_27_variable}
\end{table}

\clearpage
\setlength\bibitemsep{0pt}
\printbibliography

\end{document}